\documentclass[draft,tightenlines,nofootinbib,preprint,aps,eqsecnum,amsmath,amssymb]{revtex4}

\begin{document}

\newcommand{\sgn}{\mbox{\boldmath $\epsilon$}}

\newcommand{\beq}{\begin{equation}}
\newcommand{\eeq}{\end{equation}}
\newcommand{\bea}{\begin{eqnarray}}
\newcommand{\eea}{\end{eqnarray}}
\newcommand{\cir}{{\buildrel \circ \over =}}

\newcommand{\on}{\stackrel{\circ}{=}}
\newcommand{\byd}{\stackrel{def}{=}}
\baselineskip 20pt

\title{Quantum Mechanics in Non-Inertial Frames with a
Multi-Temporal Quantization Scheme: I) Relativistic Particles.}

\author{David Alba}

\affiliation
{Dipartimento di Fisica\\
Universita' di Firenze\\
Via G. Sansone 1\\
50019 Sesto Fiorentino (FI), Italy\\
E-mail: ALBA@FI.INFN.IT}

\author{Luca Lusanna}

\affiliation
{Sezione INFN di Firenze\\
Via G. Sansone 1\\
50019 Sesto Fiorentino (FI), Italy\\
E-mail: LUSANNA@FI.INFN.IT}

\begin{abstract}

After a review of the few attempts to define quantum mechanics in
non-inertial frames, we introduce a family of relativistic
non-rigid non-inertial frames (equal-time parallel hyper-planes
with differentially rotating 3-coordinates) as a gauge fixing of
the description of N positive energy particles in the framework of
parametrized Minkowski theories. Then we define a multi-temporal
quantization scheme in which the particles are quantized, but not
the gauge variables describing the non-inertial frames: {\it they
are considered as c-number generalized times}. We study the
coupled Schroedinger-like equations produced by the first class
constraints and we show that there is {\it a physical scalar
product independent both from time and generalized times and a
unitary evolution}. Since a path in the space of the generalized
times defines a non-rigid non-inertial frame, we can find the
associated self-adjoint effective Hamiltonian $\widehat{H}_{ni}$ for
the non-inertial evolution: it differs from the inertial energy
operator for the presence of inertial potentials and turns out to
be {\it frame-dependent} like the energy density in general
relativity. After a separation of the relativistic center of mass
from the relative variables by means of a recently developed
relativistic kinematics, inside $\widehat{H}_{ni}$ we can identify
the self-adjoint relative energy operator (the invariant mass)
${\widehat{\cal M}}$ corresponding to the inertial energy and
producing the same levels for the spectra of atoms as in inertial
frames. Instead the (in general time-dependent) effective
Hamiltonian is responsible for the interferometric effects
signalling the non-inertiality of the frame. It cannot be
interpreted as an energy (there is no relativity principle and no
kinematic group in non-inertial frames) and generically, like in
the case of time-dependent c-number external electro-magnetic
fields, it has no associated eigenvalue equation defining a
non-inertial spectrum. This formulation should help to find
relativistic Bel inequalities and to define a quantization scheme
for canonical gravity after having found a ultraviolet
regularization of the Tomonaga-Schwinger formalism in special
relativity as required by the Torre-Varadarajan no-go theorem.

\today

\end{abstract}

\maketitle

\newpage

\section{Introduction}

\subsection{Non-Rigid Non-Inertial Reference Frames.}

Till now there is no consensus on how to quantize the
gravitational field. This is connected with the fact that in
general relativity the metric tensor over space-time has a double
role: it is the potential of the gravitational field and
simultaneously describes the chrono-geometrical structure of the
space-time in a dynamical way by means of the line element. As a
consequence the light-cone in each point is dynamically varying
and this implies that the gravitational field teaches {\it
relativistic causality} to all the other fields: for instance it
selects the paths to be followed by the rays of light in the
geometrical optic approximation.

There are two main viewpoints about how to attack this problem:

i) In the weak field approximation the metric tensor over Einstein
space-times is decomposed in terms of a flat Minkowski metric (a
special solution of Einstein's equations) plus a perturbation,
$g_{\mu\nu}(x) = \eta_{\mu\nu} + h_{\mu\nu}(x)$, which, in the
family of harmonic 4-coordinates, is interpreted as a massless
spin-2 field (the graviton) over Minkowski space-time. In this way
the chrono-geometrical aspect of the gravitational field is
completely lost and, with a discontinuity, gravity is
re-formulated in the framework of special relativity with its
fixed non-dynamical chrono-geometrical structure and the residual
coordinate gauge freedom is replaced with the gauge freedom of a
spin 2 field theory. Effective quantum field theories and string
theories follow this approach, in which there is no conceptual
difference between gravitons, photons, gluons..: all propagate on
directrices of the fixed background light-cones.

ii) Remaining in the framework of Einstein's general relativity,
one tries to study both the classical and the quantum theory in a
background- and coordinate-independent way. The more advanced
approach of this type is loop quantum gravity with its quantum
geometry. However, since it is defined in spatially compact
space-times not admitting an action of the Poincare' group and,
moreover, since it does not lead to a Fock space, its main
drawback is the absence of a working prescription for
incorporating electro-magnetic fields and the standard model of
elementary particles.

\bigskip

One of the conceptual problems in general relativity is the {\it
absence of global rigid inertial systems}. According to Riemannian
geometry and to the equivalence principle {\it only observers in
free fall can define a local inertial frame}, which is strictly
valid only if restricted to the observer world-line. As a
consequence, only non-rigid non-inertial frames can be defined in
a finite region of space-time. As shown in Ref.\cite{1}, after a
discussion of how to individuate the points of the mathematical
space-time as point-events of a non spatially compact space-time
with suitable boundary conditions such that the asymptotic
symmetries are reduced to the ADM Poincare' group, we need the
Hamiltonian formalism to separate the predictable observable
degrees of freedom of the gravitational field (the {\it Dirac
observables describing its generalized tidal effects}) from the
arbitrary gauge variables ({\it describing generalized inertial
effects due to the absence of rigid inertial frames}). The
canonical formalism presupposes the introduction of arbitrary
admissible 3+1 splittings of the globally hyperbolic,
asymptotically flat at spatial infinity, space-time. Namely the
space-time is foliated with space-like leaves $\Sigma_{\tau}$
\footnote{$\tau$ is the mathematical time labeling the leaves. On
each $\Sigma_{\tau}$ curvilinear 3-coordinates $\sigma^r$ are
introduced, whose origin is an arbitrary centroid, namely the
world-line of a time-like observer. The observer proper time is a
good candidate  for $\tau$. These observer-dependent scalar
coordinates $(\tau ; \sigma^r)$ are generalized {\it radar
4-coordinates} (see Ref.\cite{2} for their use in special
relativity). } with a double role:\medskip

i) they are simultaneity instantaneous 3-spaces (a conventional
{\it present}) corresponding to a convention on the
synchronization of distant clocks (see Ref.\cite{2} for a complete
discussion of this aspect in special relativity) with certain
admissibility conditions to avoid the coordinate-singularities of
the rotating frames;

ii) they are Cauchy surfaces for the initial value problem.
\medskip

Given the embedding $x^{\mu} = z^{\mu}(\tau ,\vec \sigma )$ of its
leaves in the space-time, every admissible foliation allows to
define \cite{2} two congruences of time-like (in general
accelerated) observers, which are the {\it natural} ones for
describing the phenomena with the chosen notion of simultaneity.
For one congruence the 4-velocity field is the field of normals to
the leaves of the foliation. For the other (in general rotating,
i.e. non-surface forming) congruence the 4-velocity field is
determined by the $\tau$-derivative of the embedding describing
the leaves of the foliation. These local non-inertial observers
(endowed with tetrads, whose triads are arbitrary but whose
time-like vector is the 4-velocity of the observer) replace the
inertial ones (with their inertial reference frames). The pair
consisting of an arbitrary accelerated observer and an admissible
3+1 splitting parametrized by observer-dependent radar
4-coordinates defines a non-rigid non-inertial reference frame
(i.e. an extended physical laboratory) having the observer
world-line as  time axis \cite{2}.

\medskip

The main consequence of the general covariance of general
relativity is that the descriptions associated to different
admissible notions of simultaneity are {\it gauge equivalent} when
one uses the Hamiltonian formulation of metric and tetrad gravity
developed in Refs.\cite{3}. In this canonical framework each
solution of Einstein's equations determines an Einstein space-time
{\it and also} a well defined associated family of admissible 3+1
splittings of it \footnote{Each solution of Hamilton equations in
a completely fixed gauge with admissible Cauchy data allows to
determine not only a 4-metric but also the extrinsic curvature
$K_{rs}$ of the foliation and the lapse and shift functions, from
which we can evaluate the embedding of the simultaneity leaves
associated to that solution.}, which individuates its
chrono-geometrical structure and its natural associated observers.
Therefore there is a {\it dynamical determination} of the
synchronization convention and of the associated  notions of
simultaneous 3-space, one-way velocity of light, spatial distance,
variable light-cone, locally varying inertial effects.

\bigskip

Since the absence of rigid global inertial systems is at the basis
of the obstruction to a background-independent quantization of
gravity, one can hope to gain some insight in these problem {\it
studying the description of physical systems in non-inertial
frames in special relativity in absence of gravity}.\medskip

As shown in Ref.\cite{2}, also in special relativity one has to
introduce admissible 3+1 splittings of Minkowski space-time and
choose an arbitrary observer with the associated
observer-dependent radar coordinates to define all the previous
notions (instantaneous 3-space, clock synchronization
convention,..) and to treat physics in rotating frames (think to
the rotating disk and the Sagnac effect). Only in inertial frames
with Cartesian 4-coordinates Einstein's convention for the
synchronization of clocks determines the notion of simultaneity
given by the space-like hyper-planes of constant time. It cannot
be used in non-inertial either linearly accelerated or rotating
frames and we must use different conventions (for instance {\it
rigidly rotating frames are not allowed}: only differentially
rotating ones are admissible \cite{2}). In any case, in special
relativity the chrono-geometrical structure of Minkowski
space-time is absolute (i.e. {\it non-dynamical}) and {\it every}
admissible 3+1 splitting is allowed. When isolated systems are
reformulated in the framework of parametrized Minkowski theories
(see later on), again it can be shown that all the admissible
notions of simultaneity are {\it gauge equivalent} like in
canonical gravity, since these theories are reparametrization
invariant under frame preserving diffeomorphisms (the {\it
restricted general  covariance of special relativity}).
\bigskip

However, even if there is no more the problem of quantizing
gravity, still there is {\it no consensus on how to define quantum
mechanics in non-inertial frames not only at the relativistic
level but also in the non-relativistic limit}.

\subsection{Problems with the Definition of Quantum Mechanics in
Non-Inertial Frames.}

The postulates of non-relativistic quantum mechanics are
formulated in {\it global inertial reference frames}, connected by
the transformations of the  kinematical (extended) Galilei group,
which, due to the {\it Galilei relativity principle}, connect the
observations of an inertial observer to those of another one. The
self-adjoint operators on the Hilbert space, in particular the
{\it Hamiltonian operator} (governing the time-evolution in the
Schroedinger equation and identified with the {\it energy
operator} in the projective representation of the quantum Galilei
group associated to the system), correspond to the quantization of
classical quantities defined in these frames. The resulting
quantum theory is extremely successful {\it both for isolated and
open systems} (viewed as sub-systems of isolated systems), except
for the problem of the theory of measurement, i.e. of how to
realize the transition from {\it potentialities} to {\it
realities} in such a way to {\it avoid entangled states of
macroscopic classical bodies} (collapse of the wave function?
decoherence?).
\medskip

At the relativistic level conceptually nothing changes: we have
the {\it relativity principle} stating the impossibility to
distinguish special relativistic inertial frames and the
kinematical Poincare' group replacing the Galilei one. Again the
{\it energy} is one of the generators of the kinematical group and
is identified with the {\it canonical Hamiltonian} governing the
evolution of a relativistic Schroedinger equation.

\bigskip

In this framework, with a semi-relativistic treatment of the
electro-magnetic field \footnote{Strictly speaking we would need
relativistic quantum mechanics, since in the limit $c \rightarrow
\infty$ we get two non communicating (no induction) {\it electric}
and {\it magnetic} worlds as shown by LeBellac and Levy-Leblond
\cite{4}.} we get an extremely successful theory of atomic spectra
in inertial reference frames both for isolated inertial atoms
(closed systems) and for accelerated ones in presence of external
forces (open systems). The following cases are an elementary list
of possibilities.\medskip

a) {\it Isolated atom} - From the time-dependent Schroedinger
equation $i\, {{\partial}\over {\partial t}}\, \psi = H_o\, \psi$,
through the position $\psi = e^{i\, E_n\, t/\hbar }\, \psi_n$ we
get the time-independent Schroedinger equation $H_o\, \psi_n =
E_n\, \psi_n$ for the stationary levels and the energy spectrum
$E_n$ with its degenerations. Being isolated the atom can decay
only through spontaneous emission.

b) {\it Atom in an external c-number, maybe time-dependent,
electro-magnetic field} - Now the (energy) Hamiltonian operator is
in general non conserved (open system). Only for time-independent
external fields it is clear how to define the time-independent
Schroedinger equation for the stationary states and the
corresponding (modified) spectrum. Time-independent external
electro-magnetic fields lead to removal of degeneracies (Zeeman
effect) and/or shift of the levels (Stark effect). With
time-dependent external fields we get the Schroedinger equation
$i\, {{\partial}\over {\partial t}}\, \psi = H(t)\, \psi$ with
$H(t) = H_o + V(t)$. Therefore at each instant $t$ the
self-adjoint operator $H(t)$ defines a different basis of the
Hilbert space with its spectrum, but, since in general we have
$[H(t_1), H(t_2)] \not= 0$, it is not possible to define a unique
associated  eigenvalue equation and an associated spectrum varying
continously in $t$. Only when we have $[H(t_1), H(t_2)] = 0$ we
can write $H(t)\, \psi_n(t) = E_n(t)\, \psi_n(t)$ with
time-dependent eigenvalues $E_n(t)$ and a visualization of the
spectrum as a continuous function of time. In any case, when
$V(t)$ can be considered a perturbation, time-dependent
perturbation theory with suitable approximations can be used to
find the transition amplitudes among the levels of the unperturbed
Hamiltonian $H_o$. Now the atom can decay both for spontaneous or
stimulated emission and be excited through absorption.

c) {\it Atom plus an external c-number "mechanical" potential
inducing, for instance, the rotational motion of an atom fixed to
a rotating platform} (see the Moessbauer effect \cite{5}) - If the
c-number potential is $V(t)$, i.e. it is only time-dependent, we
have $i\, {{\partial}\over {\partial t}}\, \psi = [H_o + V(t)]\,
\psi = H(t)\, \psi$ with $[H(t_1), H(t_2)] = 0$ and the position
$\psi = e^{i\, \int_o^t\, V(t_1)\, dt_1/\hbar}\, \psi_1$ leads to
$i\, {{\partial}\over {\partial t}}\, \psi_1 = H_o\, \psi_1$, so
that the energy levels are $E_{1n} = E_n + \int_o^t dt_1\,
V(t_1)$. The addition of a c-number external time-dependent
electro-magnetic field leads again to the problems of case b).

d) At the relativistic level we can consider the {\it isolated
system atom + electro-magnetic field} as an approximation to the
theory of bound states in quantum electrodynamics. Both the atom
and the electro-magnetic field are separately accelerated open
subsystems described in an inertial frame.

\medskip

In any case the modifications of the energy spectrum of the
isolated atom is induced by {\it physical force fields} present in
the inertial frame of the observer.

\bigskip

In case c) we can consider an accelerated observer carrying a
measuring apparatus and rotating with the atom. In this case the
theory of measurement is based on the {\it locality hypothesis}
\cite{6,2} according to which at each instant the measurements of
the accelerated apparatus coincide with those of an identical
comoving inertial one. As a consequence the observer will detect
the same spectrum as an inertial observer.

Let us consider the description of the previous cases from the
point of view of a {\it non-inertial observer} carrying a
measuring apparatus by doing a {\it passive coordinate
transformation} adapted to the motion of the observer. Since,
already at the non-relativistic level, there is  {\it no
relativity principle for non-inertial frames}, there is {\it no
kinematical group} (larger than the Galilei group) whose
transformations connect the non-inertial measurements to the
inertial ones:  given the non-inertial frame with its linear and
rotational accelerations with respect to a standard inertial
frame, we can only define the {\it succession} of time-dependent
Galilei transformations identifying at each instant the {\it
comoving inertial observers}, with the same measurements of the
non-inertial observer if the locality hypothesis holds.

\medskip

Since we are considering a {\it purely passive viewpoint}, there
is no physical reason to expect that the atom spectra will change:
there are no physical either external or internal forces but only
a different viewpoint which changes the appearances and introduces
the fictitious (or inertial) mass-proportional forces to describe
these changes.

\medskip

At the special relativistic level the natural framework to
describe non-inertial (mathematical) observers is given by {\it
parametrized Minkowski theories} \cite{7,8,9} (see also the
Appendix of the first paper in Ref.\cite{3}). In them, one makes
an arbitrary 3+1 splitting of Minkowski space-time with a global
foliation of space-like hyper-surfaces (Cauchy simultaneity
surfaces) described by an embedding $x^{\mu} = z^{\mu}(\tau ,\vec
\sigma )$ with respect to an arbitrary inertial observer with his
associated inertial reference frame. In this approach, besides the
configuration variables of the isolated system, there are the
embeddings $z^{\mu}(\tau ,\vec \sigma )$ as extra {\it gauge
configuration} variables in a suitable Lagrangian determined in
the following way. Given the Lagrangian of the isolated system in
the Cartesian 4-coordinates of an inertial system, one makes the
coupling to an external gravitational field and then replaces the
external 4-metric with $g_{AB}(\tau ,\vec \sigma ) = [z^{\mu}_A\,
\eta_{\mu\nu}\, z^{\nu}_B](\tau ,\vec \sigma )$. Therefore the
resulting Lagrangian depends on the embedding through the
associated metric $g_{AB}$. As already said, the presence of {\it
the special relativistic type of general covariance} [the action
is invariant under frame-preserving diffeomorphisms \footnote{The
analogous sub-group $x^{{'}\, i} = f^i(x^r)$, $x^{{'}\, o} =
f^o(x^o, x^r)$ of the general coordinate transformations (passive
4-diffeomorphisms) of Einstein's general relativity, which leave a
frame of reference unchanged,  are used as a starting point by
Schmutzer and Plebanski \cite{10} in their treatment of quantum
mechanics in non-inertial reference frames, after having
considered the non-relativistic limit. } $\tau^{'} = a(\tau ,\vec
\sigma )$, ${\vec \sigma}^{'} = \vec b(\vec \sigma )$], the
transition from a foliation to another one (i.e. a change of the
notion of simultaneity) is a {\it gauge transformation} of the
theory generated by its first class constraints. Like in general
relativity, these passive 4-diffeomorphisms on Minkowski
space-time imply general covariance, but {\it do not form} a
kinematical group (extending the Poincare' group), because there
is {\it no relativity principle} for non-inertial observers.
Therefore, there is {\it no kinematical generator interpretable as
a non-inertial energy}.

\medskip

The $c \rightarrow \infty$ limit of parametrized Minkowski
theories allows to define {\it parametrized Galilei theories} and
to describe non-relativistic congruences of non-inertial
observers. Again there is no relativity principle for such
observers, no kinematical group extending the Galilei one and,
therefore, no kinematical generator to be identified as a
non-inertial energy.
\medskip

Let us remark that in Ref.\cite{11} Newtonian gravity was
rephrased as a gauge theory of the extended Galilei group by
taking the non-relativistic limit of the ADM action for metric
gravity: local non-rigid non-inertial reference frames are
introduced in this way. Instead Kuchar \cite{12} introduced a
quasi-Galilean gauge group connecting Galilean (rigid,
non-rotating) frames in the presence of Newtonian gravitational
fields, in the framework of Cartan's geometrical description of
Newtonian space-times, to build a geometrical description of the
quantum mechanics of a single non-relativistic particle freely
falling in an external Newtonian gravitational field. This
definition of non-inertiality is used to try to show how the
equivalence principle works for non-relativistic quantum systems.
Klink \cite{13} generalized the kinematical Galilei group to the
Euclidean line group [$\vec x \mapsto {\vec x}^{'} = R(t)\, \vec x
+ \vec a(t)$] to describe global rigid, but time-dependent,
rotations and translations to be implemented  as time-dependent
unitary transformation on the Hilbert space. Schmutzer and
Plebanski  \cite{10} ask for the form invariance of the equations
of motion in rigid non-inertial frames under the non-relativistic
limit of the frame invariant coordinate transformations. Then they
postulate that non-inertial quantum mechanics must be form
invariant under time-dependent unitary transformations connecting
such frames. They study also the Dirac equation in this framework
with the statement that inertial effects change the spectrum. In
Ref.\cite{14} Greenberger and Overhauser consider non-relativistic
rigid time-dependent translations and uniform rotations to extend
the equivalence principle to quantum physics (neutron
interferometry), where the transformations are unitarily
implemented; they also consider uniform 4-acceleration (hyperbolic
motion) for the Klein-Gordon equation. Finally another approach
\cite{15}, oriented to atomic physics and matter-waves
interferometers, considers the limit to either Minkowski or
Galilei space-time of the Dirac equation coupled to an external
gravitational field.

\bigskip

In conclusion, all the existing attempts \cite{10,13,14} to extend
the standard formulation of quantum mechanics from global rigid
inertial frames to  {\it special global rigid non-inertial
reference frames} carried by observers with either linear (usually
{\it constant}) acceleration or rotational (usually {\it
constant}) angular velocity are equivalent to the definition of
suitable {\it time-dependent unitary transformations} acting in
the Hilbert space associated with inertial frames. \medskip

While in inertial frames the generator of the time evolution,
namely the Hamiltonian operator $H$ appearing in the Schroedinger
equation $i\, {{\partial}\over {\partial t}}\, \psi = H\, \psi$,
also describes the energy of the system, after a time-dependent
unitary transformation $U(t)$ the {\it generator $\tilde H(t) =
U(t)\, H\, U^{-1}(t) + i \dot U(t)\, U^{-1}(t)$ of the time
evolution} \footnote{This {\it non-inertial Hamiltonian}
containing the potential $i \dot U\, U^{-1}$ of the {\it
fictitious or inertial forces is not a generator of any
kinematical group}.} in the transformed Schroedinger equation $i\,
{{\partial}\over {\partial t}}\, \tilde \psi = \tilde H(t)\,
\tilde \psi$, with $\tilde \psi = U(t)\, \psi$, {\it differs from
the energy operator} $H^{'} = U(t)\, H\, U^{-1}(t)$. And also in
this case like in example b), only if we have $[\tilde H(t_1),
\tilde H(t_2)] = 0$ it is possible to define a unique stationary
equation with time-dependent eigenvalues for $\tilde H(t)$.

The situation is analogous to the Foldy-Wouthuysen transformation
\cite{16}, which is a time-dependent unitary transformation when
it exists: in this framework $H^{'}$ is the energy, while $\tilde
H(t)$ is the Hamiltonian for the new Schroedinger equation and the
associated S-matrix theory (theoretical treatment of
semi-relativistic high-energy experiments, $\pi\, N$,..).

\medskip

Since in general the self-adjoint operator $\tilde H(t)$ does not
admit a unique associated eigenvalue equation \footnote{Even when
it does admit such an equation, we have $< \psi | H | \psi
> \not= < \tilde \psi | \tilde H | \tilde\psi
>$ and different stationary states are connected, following the
treatment of the time-independent examples,  in which it is
possible to find the spectrum of both of them, of Kuchar
\cite{12}, by a generalized transform.} and, moreover, since the
two self-adjoint operators $\tilde H(t)$ and $H^{'}$ are in
general (except in the static cases) non commuting, there is {\it
no consensus} about the results of measurements in non-inertial
frames: {\it does a non-inertial observer see a variation of the
emission spectra of atoms}? Which is the spectrum of the hydrogen
atom seen by a non-inertial observer? Since for constant rotation
we get $\tilde H = H^{'} + \vec \Omega \cdot \vec J$, does the
uniformly rotating observer see the inertial spectra or are they
modified by a Zeeman effect? If an accelerated observer would
actually measure the Zeeman levels with an energy measurement,
this would mean that the stationary states of $\tilde H$ (and not
those of the inertial energy operator $H^{'}$) are the relevant
ones. Proposals for an experimental check of this possibility are
presented in Ref.\cite{19}. Usually one says that a possible
non-inertial Zeeman effect from constant rotation is either too
small to be detected or masked by physical magnetic fields, so
that the distinction between $\tilde H$ and $H^{'}$ is irrelevant
from the experimental point of view.
\medskip

Here we have exactly the same problem like in the case of an atom
interacting with a time-dependent external field: the atom is
defined by its inertial spectrum, the only one  unambiguously
defined when $[\tilde H(t_1), \tilde H(t_2)] \not= 0$. When
possible, time-dependent perturbation theory is used to find the
transition amplitudes among the inertial levels. Again only in
special cases (for instance time-independent $\tilde H$) a
spectrum for $\tilde H$ may be evaluated and usually, except in
special cases like the Zeeman effect, it has no relation with the
inertial spectrum (see Ref.\cite{12} for an example). Moreover,
also in these special cases the two operators may not commute so
that the two properties described by these operators cannot in
general be measured simultaneously.\medskip

As a consequence of these problems the description of measurements
in non-inertial frames is often replaced by an explanation of how
to correlate the phenomena to the results of measurements of the
energy spectra in inertial frames. For instance in the Moessbauer
effect \cite{5} one  only considers the correction for Doppler
effect (evaluated by the instantaneous comoving inertial observer)
of unmodified spectra. Regarding the spectra of stars in
astrophysics, only correction for gravitational red-shift of
unmodified spectra are considered. After these corrections,
notwithstanding the quoted complex theoretical situation the
inertial effects connected to the emission in non-inertial frames
manifest themselves only in a {\it broadening} of the inertial
spectral lines. See Hughes \cite{17} for the red-shift
interpretation based upon the equivalence principle in atomic,
nuclear and particle physics in special relativity, relying on all
Einstein's statements, which, however, are explicitly  referred to
{\it static constant gravitational field} \footnote{According to
Synge \cite{18}, it does not exists in general relativity due to
tidal effects: the free fall exists only along a time-like
geodesics and approximately in its neighborhood till when tidal
effects are negligible.}. In conclusion {\it atoms are always
identified through their inertial spectra in absence of external
fields}. The non-inertial effects, precluding the unique existence
of a spectrum continuous in time, are usually small and appear as
a noise over-imposed to the continuous spectrum of the center of
mass.

\bigskip

An apparatus for measuring $\tilde H$ can be an {\it
interferometer} measuring the variation $\triangle\, \phi$ of the
phase of the wave-function describing the two wave-packets
propagating, in accord with the non-inertial Schroedinger equation
(one uses the Dirac-Feynman path integral with $\tilde H$ to
evaluate $\triangle\, \phi$) along the two arms of the
interferometer. However, {\it the results of the interferometer
only reveal the eventual non-inertial nature of the reference
frame}, namely  {\it they amount to a detection of the
non-inertiality of the frame of reference}, as remarked in
Ref.\cite{2}. In this connection see Ref.\cite{20} on neutron
interferometry, where there is a full account of the following
topics: a) the effect of the Earth rotation in the generation of
the Sagnac effect, b) the detection of the Coriolis force, c) the
detection of linear acceleration, d) a treatment of propagating
light and of its dragging by a moving medium (Fizeau effect and
neutron Fizeau effect), e) the connection of the Sagnac effect
with the spin-rotation coupling to detect the rotational inertia
of an intrinsic spin, f) a neutron Aharonov-Bohm analogue, g) the
confinement and gravity quantized phases, h) the Anandan
acceleration.

\bigskip

Now in the non-relativistic literature there is an {\it active
re-interpretation in terms of gravitational potentials of the
previous passive view} according to a certain reading of the {\it
non-relativistic limit of the classical (weak or strong)
equivalence principle} (universality of free fall or identity of
inertial and gravitational asses) and to its extrapolation to
quantum mechanics (see for instance Hughes \cite{17} for its use
done by Einstein). According to this interpretation, at the
classical level the {\it passive fictitious forces} seen by the
accelerated observer are interpreted as an {\it active external
Newtonian gravitational force}  acting in an inertial frame, so
that at the quantum level $\tilde H$ is interpreted as the energy
operator in an inertial frame in presence of an external quantum
gravitational potential $\tilde H - H^{'} = i\, \dot U U^{-1}$.
Therefore the shift from the levels of $H^{'}$ to those of $\tilde
H$ is justified and expected. However this interpretation and use
of the equivalence principle is subject to criticism already at
the classical level.

A first objection is that a physical external gravitational field
(without any connection with non-inertial observers) leads to the
Schroedinger equation $i\, {{\partial}\over {\partial t}}\, \psi =
[H + V_{grav}]\, \psi$ and not to $i\, {{\partial}\over {\partial
t}}\, \tilde \psi = [UHU^{-1} + V_{grav}]\, \tilde \psi$, $\tilde
\psi = U\, \psi$. \medskip

Moreover, since we are going to define a quantization scheme in
non-inertial frames directly in Minkowski space-time, and then the
non-relativistic limit $c\, \rightarrow\, \infty$ will restrict it
to the Galileo space-time, and since there is no
action-at-a-distance formulation of gravity in Minkowski
space-time (Newtonian gravity is obtained as the $c\,
\rightarrow\, \infty$ limit of general relativity), we do not
think that the equivalence principle is playing any role in this
type of quantization. Therefore we shall not consider it any more
and we shift to the Conclusions any comment on it.

\subsection{Parametrized Minkowski Theories and the Quantization
of First Class Constraints.}

In this paper  we consider a system of N relativistic
positive-energy scalar particles (either free or with mutual
action-at-a-distance interaction) in the framework of parametrized
Minkowski theories. In Refs.\cite{7,9} such an isolated system has
been described on the special Wigner hyper-planes orthogonal to
the total 4-momentum of the system: this defines the {\it
Wigner-covariant rest-frame instant form} and the two associated
congruences of time-like observers reduce to a unique congruence
of inertial observers. At the classical level the first task, by
using results from Ref.\cite{2}, will be to extend these results
to more general foliations, whose associated congruences describe
non-inertial observers. We will study a special family of 3+1
splittings, whose leaves are {\it hyper-planes with differentially
rotating 3-coordinates}, and we will identify the first class
constraints and the effective non-inertial Hamiltonian ruling the
evolution in these non-inertial non-rigid reference frames.
Moreover, we identify a time-dependent canonical transformation
connecting the effective non-inertial Hamiltonian to the inertial
one. This allows to introduce action-at-a-distance relativistic
interactions among the particles in a consistent way. The
non-relativistic limit of parametrized Minkowski theories leads to
parametrized Galilei theories (they will be discussed in a second
paper, referred to as  II), to their first class constraints and
to a formulation of the Newtonian N-body problem in non-inertial
(in general non-rigid) reference frames. The restriction to rigid
non-inertial frames allows to recover the quoted existing
formulations in these frames. \medskip

The main problem is how to  quantize the system of first class
constraints resulting from  the restriction of parametrized
Minkowski (Galilei) theories to this special family of foliations.
Since there are the gauge variables describing the class of
embeddings belonging to the family with their associated inertial
effects, the quantization is not trivial.

\bigskip

Let us remark that quantization at the relativistic level is
always complicated by the fact that all isolated relativistic
systems are described by singular Lagrangians \cite{8}, whose
Hamiltonian formulation requires Dirac-Bergmann theory of
constraints \cite{21} (see also Refs.\cite{8,22}). The
quantization of systems with first class constraints
\footnote{Here we ignore the extra complications arising when
second class constraints are present.} is a complicate affair.
There are two main viewpoints:

A) {\it First quantize, then reduce} - This program was originated
from  Dirac's quantization procedure \cite{21} of systems with
first class constraints, starting from a non-physical Hilbert
space in which also the gauge variables are quantized and then
projecting to a physical one containing only gauge-invariant
physical observables. There are many (in general inequivalent)
ways to implement it and its main drawback is the {\it ordering
problem}, namely the Groenewold-Van Hove no-go theorem \cite{23}
stating the absence of a unique quantization rule for observables
more than quadratic in the canonical variables even using notions
as Weyl correspondence and deformation quantization techniques.
Since in many models the classical phase space is a symplectic
manifold but not a cotangent bundle, in these cases there is the
extra problem of which coordinate system, if any, is more suitable
for the quantization of theories invariant under diffeomorphisms
\cite{24}. Many points of view, geometric and/or algebraic
quantization \cite{25} and group-theoretical quantization
\cite{27}, have been developed. However, since in all these
approaches the physical Hilbert space is some quotient of the
non-physical one with respect to the group of gauge
transformations, every approach has to find a strategy to identify
the physical scalar product \cite{24,27}. The most developed and
used scheme is the BRS quantization procedure \cite{22}, both in
canonical and path integral quantization, but it too has the
problem of how to identify the physical scalar product \cite{28}.

B) {\it First reduce, then quantize} - In this program the idea is
to utilize differential geometry to identify the classical reduced
phase space, containing only physical observable degrees of
freedom, of every model with first class constraints and then to
quantize only the gauge-invariant observables. However, canonical
reduction is usually very complicated and the reduced phase space
is in general a very complicated topological space \cite{29}
\footnote{The program A) often is not able to treat many of these
topological problems in the correct way for the absence of
suitable mathematical tools.}. As a consequence it is usually not
known how to arrive to a global quantization. In any case, when it
is possible to quantize both in this way and in the way A), the
two quantized models are in general inequivalent.

\bigskip
We shall introduce a new viewpoint: the {\it multi-temporal
quantization}, in which the gauge variables (one for each first
class constraint) are treated as {\it extra $c$-number generalized
times and only the gauge invariant Dirac observables are
quantized} \footnote{See Ref.\cite{30} for the general theory,
Ref.\cite{31} for an explicit example of quantization of an
interacting relativistic two-particle system with the
determination of the physical scalar product and Ref.\cite{32} for
non-relativistic examples.}. Besides the ordinary Schroedinger
equation with the canonical Hamiltonian operator, there are as
many other generalized Schroedinger-like equations as first class
constraints. The wave function will depend on a space of
parameters parametrized by the time and the gauge variables: each
line in this parametric space corresponds to a classical gauge
\footnote{The topological problems of the program B) are replaced
by the global properties in the large of the parametric space of
the generalized times. Our quantization is defined only locally
around the origin of this parametric space.}. If we find an
ordering such that we get a correct quantum algebra of the
constraints, the system of coupled equations is formally
integrable, solutions corresponding to different classical gauge
(different non-inertial reference frames in our case) are
unitarily equivalent and there is no problem in finding a physical
scalar product independent from all the times and in showing that
the evolution is unitary.

Therefore our philosophy is {\it not to quantize inertial effects}
(describing only the {\it appearances} of the phenomena), i.e. the
embedding gauge variables, at every level: i) general relativity
(yet to be developed); ii) special relativity (either as a limit
of general relativity or as an autonomous theory); iii)
Newton-Galilei non-relativistic theories. As we will see in this
paper and in II, in the cases ii) and iii) this leads to coupled
generalized Schroedinger-like equations with the wave-functions
depending on a parametric space. Each curve in the parametric
space may be put in correspondence with some non-inertial frame
and we can find the effective non-inertial Hamiltonian governing
the evolution in that frame. We will show that both in the
non-relativistic (see II) and in the relativistic case the
solutions of the effective Schroedinger equation valid in a given
non-inertial frame are connected to the solutions of the standard
inertial Schroedinger equation by {\it time-dependent unitary
transformations}. As it will be shown in II, the previous
non-relativistic attempts of Refs.\cite{10,12,13,14} can be
recovered as special cases of our construction.\medskip

Then we can show that the time-dependent canonical transformation
connecting the effective non-inertial Hamiltonian to the inertial
one can be used, in both the  relativistic and non-relativistic
cases, to define a further canonical transformation (of the type
of those studied in Refs.\cite{33}) realizing a separation of the
center of mass from the relative variables in non-inertial frames.
The effective non-inertial Hamiltonian turns  out to be the sum of
a term containing the relative energy and the center-of-mass
kinetic energy plus a term with the inertial potentials. As it
will be discussed in Subsection IVB, in the relativistic case a
satisfactory definition of bound states on {\it equal-time} Cauchy
surfaces can be achieved {\it only if we apply the multi-temporal
quantization scheme after a separation of the relativistic center
of mass from the relativistic relative variables}. At the
non-relativistic  quantum level we have that: A) in rigid
non-inertial frames the non-inertial wave function for the N-body
problem can be factorized as the product of a center-of-mass term
(a free decoupled particle) and of a bound-state wave function
(depending on $N-1$ relative variables); B) in non-rigid
non-inertial frames the unitary evolution operator does not
factorize and the factorization of the wave function on the Cauchy
surface is lost at later times: this is also the general situation
in the relativistic case.

\medskip

Our main result is that  {\it  in non-rigid non-inertial either
relativistic or non-relativistic frames the relative energy
operator (the invariant mass of the system), depending only on
relative variables, remains a self-adjoint operator ${\widehat {\cal
M}}$ with the same spectrum for bound states  of the energy
operator $\widehat{H}_{inertial}$ in inertial frames} (here atoms are
approximated with N-body bound states with an effective potential
extracted from quantum field theory). Over-imposed to this
discrete spectrum there is the continuum spectrum of the decoupled
center of mass of the atom. Instead the self-adjoint effective
Hamiltonian operator $\widehat{H}_{ni}$ for the non-inertial unitary
evolution has the structure $\sqrt{{\widehat {\cal M}}^2 + {\vec k}^2}
+ (inertial\, potentials)$ at the relativistic level (in the
momentum representation where $\vec k$ is the momentum of the
decoupled center of mass), which becomes ${\widehat {\cal M}} +
{{{\vec k}^2}\over {2\, {\widehat {\cal M}}}} + (inertial\,
potentials)$ in the non-relativistic limit. In both cases the (in
general time-dependent) potentials for the inertial forces are
such that $\widehat{H}_{ni}$ is time-dependent, does not in general
admit a unique associated spectrum of eigenvalues and is only
relevant for the interferometric experiments signalling the
non-inertiality of the frame. Finally, since the potentials for
the inertial forces are frame-dependent, we have a frame-dependent
effective Hamiltonian like it happens with the energy density in
general relativity where only non-inertial frames are allowed.

\subsection{Content of the  Paper.}

In Section IIA we review some notions on non-inertial observers
and on the synchronization of clocks in Minkowski space-time.
Then, after a review of parametrized Minkowski theories for a
system of N free positive-energy particle in Section IIB, in
Subsection IIC we study the description of such a system in a
family of foliations with parallel hyper-planes but with
differentially rotating 3-coordinates, which defines a class of
non-rigid non-inertial frames. This allows to find the effective
frame-dependent Hamiltonian for the non-inertial evolution of the
particles. In Section IIIA we define our multi-temporal
quantization of the first class constraints for a system of N free
particles in such a class of non-inertial frames. In Section IIIB,
after the definition of a suitable ordering, we introduce a
frame-dependent physical Hilbert space, whose wave functions
depend on time and on the generalized times (the gauge variables
describing inertial effects) and satisfy an integrable set of
coupled Schroedinger-like equations, some of which have a
non-self-adjoint Hamiltonian. However, the frame dependence of the
measure of the scalar product implies that there is an isometric
evolution and that the scalar  product is independent from all the
times. This allows to reformulate the theory in a
frame-independent Hilbert space with a standard scalar product: it
amounts to a change of the ordering making all the Hamiltonians
self-adjoint due to the introduction of extra inertial potentials.
After selecting a non-inertial-frame by choosing a path in the
parametric space of the generalized times, we identify the
effective self-adjoint Hamiltonian operator for the non-inertial
evolution. After the introduction of action-at-a-distance
interactions in Section IVA, in Section IVB we make the separation
of the relativistic center of mass from the relative variables
with a recently developed relativistic kinematics \cite{33,34} and
we show that the term in the effective Hamiltonian operator
describing the relative energy operator (the rest-frame invariant
mass of the system) is self-adjoint. This allows to define the
same bound states in non-inertial frames as it is done in the
inertial ones. Then in Section V there are some final remarks and
a sketch of the non-relativistic limit studied in paper II.

Appendix A contains some relativistic kinematics connected with
Wigner boosts and rotations. Appendix B contains some calculations
for Section IIC, while Appendix C is devoted to the study of a
pseudo-differential operator. Finally Appendix D contains the
extension of our results to positive-energy spinning particles.

\newpage

\section{Classical Relativistic Positive-Energy Particles.}

In this Section, after a review about non-inertial observers and
admissible 3+1 splittings of Minkowski space-time (Subsection A),
we introduce the description of positive-energy relativistic
scalar particles \cite{7,8,9} by means of a parametrized Minkowski
theory (Subsection B) and we make a comment on its restriction to
Wigner hyper-planes, leading to the rest-frame instant form with
the associated inertial observers. Then, in Subsection C, we study
in detail the class of admissible embeddings corresponding to
space-like hyper-planes with differentially rotating 3-coordinates
(defined at the end of Subsection A), since it will be needed to
get the description of the particles from the point of view of a
non-inertial observer without an explicit breaking of manifest
Lorentz covariance.

\subsection{ Non-Inertial Observers and Synchronization of Clocks
in  Minkowski Space-Time.}

In Refs.\cite{2} we studied how to describe non-rigid non-inertial
references frames (admissible 3+1 splittings, extended physical
laboratories) in Minkowski space-time and how an arbitrary
accelerated observer can use them to establish an
observer-dependent radar 4-coordinate system, whose time
coordinate is the  observer proper time $\tau$.

\bigskip

The starting point, given an inertial system with Cartesian
coordinates $x^{\mu}$ in Minkowski space-time \footnote{The
Minkowski metric has the signature $\sgn\, (+ - - -)$ with $\sgn =
\pm 1$ according to the particle physics or general relativity
convention. As a consequence, for a spatial vector we have $V^r =
- \sgn\, V_r$.}, are M$\o$ller \cite{35} (Chapter VIII, Section
88) {\it admissible coordinates transformations} $x^{\mu}\,
\mapsto\, y^{\mu} = f^{\mu}(x)$ [with inverse transformation
$y^{\mu}\, \mapsto\, x^{\mu} = h^{\mu}(y)$]: they are those
transformations whose associated metric tensor $g_{\mu\nu}(y) =
{{\partial h^{\alpha}(y)}\over {\partial y^{\mu}}}\, {{\partial
h^{\beta}(y)}\over {\partial y^{\nu}}}\, \eta_{\alpha\beta}$
satisfies the following conditions

\bea
 && \sgn\, g_{oo}(y) > 0,\nonumber \\
 &&{}\nonumber \\
 && \sgn\, g_{ii}(y) < 0,\qquad \begin{array}{|ll|} g_{ii}(y)
 & g_{ij}(y) \\ g_{ji}(y) & g_{jj}(y) \end{array}\, > 0, \qquad
 \sgn\, \det\, [g_{ij}(y)]\, < 0,\nonumber \\
 &&{}\nonumber \\
 &&\Rightarrow \det\, [g_{\mu\nu}(y)]\, < 0.
 \label{II1}
 \eea

These are the necessary and sufficient conditions for having
${{\partial h^{\mu}(y)}\over {\partial y^o}}$ behaving as the
velocity field of a relativistic fluid, whose integral curves, the
fluid flux lines, are the world-lines of time-like observers.
Eqs.(\ref{II1}) say:

i) the observers are time-like because $\sgn g_{oo} > 0$;

ii) that the hyper-surfaces $y^o = f^{o}(x) = const.$ are good
space-like simultaneity surfaces.

\medskip

Moreover we must ask that $g_{\mu\nu}(y)$ tends to a finite limit
at spatial infinity on each of the hyper-surfaces $y^o = f^{o}(x)
= const.$ As shown in Ref.\cite{2} this implies that {\it the
simultaneity surfaces must tend to space-like hyper-planes at
spatial infinity} and that an important sub-group of the
admissible coordinate transformations are the frame-preserving
diffeomorphisms $x^o \mapsto y^o = f^o(x^o, \vec x)$, $\vec  x
\mapsto \vec y = \vec f(\vec x)$. Let us remark that admissible
coordinate transformations $x^{\mu} \mapsto y^{\mu} = f^{\mu}(x)$
constitute the most general extension of the Poincare'
transformations $x^{\mu} \mapsto y^{\mu} = a^{\mu} +
\Lambda^{\mu}{}_{\nu}\, x^{\nu}$ compatible with special
relativity. However they do not form a kinematical group due to
the absence of a relativity principle for non-inertial frames.

\bigskip

It is then convenient to describe \cite{7,8,9} the simultaneity
surfaces of an admissible foliation (3+1 splitting of Minkowski
space-time) with {\it adapted Lorentz scalar admissible
coordinates} $x^{\mu}\, \mapsto \sigma^A = (\tau ,\vec \sigma ) =
f^A(x)$ [with inverse $\sigma^A\, \mapsto\, x^{\mu} =
z^{\mu}(\sigma ) = z^{\mu}(\tau ,\vec \sigma )$] such that:

i) the scalar time coordinate $\tau$ labels the leaves
$\Sigma_{\tau}$ of the foliation ($\Sigma_{\tau} \approx R^3$);

ii) the scalar curvilinear 3-coordinates $\vec \sigma = \{
\sigma^r \}$ on each $\Sigma_{\tau}$ are defined with respect to
an arbitrary time-like centroid $x^{\mu}(\tau )$ chosen as their
origin.

\medskip

The use of these Lorentz-scalar adapted coordinates allows to make
statements depending only {\it on the foliation} but not on the
4-coordinates $y^{\mu}$ used for Minkowski space-time.

\medskip

{\it If we identify the centroid $x^{\mu}(\tau )$ with the
world-line $\gamma$ of an arbitrary time-like observer and $\tau$
with the observer proper time, we obtain as many globally defined
observer-dependent Lorentz-scalar radar 4-coordinates for an
accelerated observer as admissible 3+1 splittings of Minkowski
space-time and each 3+1 splitting can be viewed as a conventional
choice of an instantaneous 3-space and of a synchronization
prescription for distant clocks. The world-line $\gamma$ is not
orthogonal to the simultaneity leaves and Einstein ${1\over 2}$
convention is suitably generalized.}

\bigskip

The simultaneity hyper-surfaces $\Sigma_{\tau}$ are described by
their embedding $x^{\mu} = z^{\mu}(\tau ,\vec \sigma )$ in
Minkowski space-time [$(\tau ,\vec \sigma )\, \mapsto\,
z^{\mu}(\tau ,\vec \sigma )$, $R^3\, \mapsto \, \Sigma_{\tau}
\subset M^4$] and the induced metric is $g_{AB}(\tau ,\vec \sigma
) = z^{\mu}_A(\tau ,\vec \sigma )\, z^{\nu}_B(\tau ,\vec \sigma
)\, \eta_{\mu\nu}$ with $z^{\mu}_A = \partial z^{\mu} / \partial
\sigma^A$ (they are flat cotetrad fields over Minkowski
space-time) and $g(\tau ,\vec \sigma ) = \det\, (g_{AB}(\tau ,\vec
\sigma )) \not= 0$. Since the vector fields $z^{\mu}_r(\tau ,\vec
\sigma )$ are tangent to the surfaces $\Sigma_{\tau}$, the
time-like vector field of normals is $l^{\mu}(\tau ,\vec \sigma )
= {1\over {\gamma (\tau ,\vec \sigma )}}\,
\epsilon^{\mu}{}_{\alpha\beta\gamma}\, z^{\alpha}_1(\tau ,\vec
\sigma )\, z^{\beta}_2(\tau ,\vec \sigma )\, z^{\gamma}_3(\tau
,\vec \sigma )$ \footnote{Here $\gamma (\tau ,\vec \sigma ) = -
\det\, (g_{rs}(\tau ,\vec \sigma ))$. The inverse metric of the
3-dimensional part $g_{rs}(\tau ,\vec \sigma  ) = - \sgn\,
h_{rs}(\tau ,\vec \sigma )$ [the 3-metric $h_{rs}(\tau ,\vec
\sigma )$ has signature $(+++)$] of the induced metric is
$\gamma^{rs}(\tau ,\vec \sigma ) = - \sgn\, h^{rs}(\tau ,\vec
\sigma )$, $\gamma^{ru}(\tau ,\vec \sigma )\, g_{us}(\tau ,\vec
\sigma ) = \delta^r_s$.} We have $l^2(\tau ,\vec \sigma ) = \sgn$
and $\eta^{\mu\nu} = \sgn\, l^{\mu}(\tau ,\vec \sigma )\,
l^{\nu}(\tau ,\vec \sigma ) + \gamma^{rs}(\tau ,\vec \sigma )\,
z^{\mu}_r(\tau ,\vec \sigma )\, z^{\nu}_s(\tau ,\vec \sigma )$.
Instead the time-like evolution vector field is
$z^{\mu}_{\tau}(\tau ,\vec \sigma ) = N_{[z]}(\tau ,\vec \sigma
)\, l^{\mu}(\tau ,\vec \sigma ) + N^r_{[z]}(\tau ,\vec \sigma )\,
z^{\mu}_r(\tau ,\vec \sigma )$ [$N_{[z]} = \sqrt{{g\over
{\gamma}}}$, $N^r_{[z]} = g_{\tau s}\, \gamma^{sr}$ are the flat
lapse and shift functions, which now, differently from metric
gravity, are not independent variables but functionals of the
embedding].

\medskip

Therefore the accelerated observer plus one admissible 3+1
splitting with the observer-dependent radar 4-coordinates define a
{\it non-rigid non-inertial reference frame} whose time axis is
the world-line $\gamma$ of the observer and whose instantaneous
3-spaces are the simultaneity hyper-surfaces $\Sigma_{\tau}$.

\bigskip

The main property of {\it each foliation with simultaneity
surfaces} associated to an admissible 4-coordinate transformation
is that the embedding of the leaves of the foliation automatically
determine two time-like vector fields and therefore {\it two
congruences of (in general) non-inertial time-like observers}:

i) The time-like vector field $l^{\mu}(\tau ,\vec \sigma )$ of the
normals to the simultaneity surfaces $\Sigma_{\tau}$ (by
construction surface-forming, i.e. irrotational), whose flux lines
are the world-lines of the so-called (in general non-inertial)
Eulerian observers. The simultaneity surfaces $\Sigma_{\tau}$ are
(in general non-flat) Riemannian instantaneous 3-spaces in which
the physical system is visualized and in each point the tangent
space to $\Sigma_{\tau}$ is the local observer rest frame
$R_{\tilde l(\tau_{\gamma})}$ of the Eulerian observer through
that point. This 3+1 viewpoint is called {\it hyper-surface 3+1
splitting}.

ii) The time-like evolution vector field $z^{\mu}_{\tau}(\tau
,\vec \sigma ) / \sqrt{\sgn\, g_{\tau\tau}(\tau ,\vec \sigma ) }$,
which in general is not surface-forming (i.e. it has non-zero
vorticity like in the case of the rotating disk). The observers
associated to its flux lines have the local observer rest frames
$R_{\tilde u(\tau_{\gamma})}$ not tangent to $\Sigma_{\tau}$:
there is no notion of instantaneous 3-space for these observers
(1+3 point of view or {\it threading splitting}) and no
visualization of the physical system in large. However these
observers can use the notion of simultaneity associated to the
embedding $z^{\mu}(\tau ,\vec \sigma )$, which determines their
4-velocity. This 3+1 viewpoint is called {\it slicing 3+1
splitting}.

\bigskip

As shown in Ref.\cite{2} the 3+1 point of view allows to arrive at
the following results:

i) The main  byproduct of the restrictions (\ref{II1}) is that
there exist admissible 4-coordinate transformations interpretable
as {\it rigid systems of reference with arbitrary translational
acceleration}. However there is {\it no admissible 4-coordinate
transformation corresponding to a rigid system of reference with
rotational motion}. When rotations are present, the admissible
4-coordinate transformations give rise to a continuum of local
systems of reference like it happens in general relativity ({\it
differential rotations}). This leads to a new treatment of
problems like the rotating disk, the Sagnac effect and the
one-way time delay for signals from an Earth station to a
satellite.

ii) The simplest foliation of the previous class, whose
simultaneity surfaces are space-like hyper-planes with
differentially rotating 3-coordinates is given by the embedding

\bea
  &&z^{\mu}(\tau ,\vec \sigma ) = x^{\mu}(\tau ) + \epsilon^{\mu}_r\,
R^r{}_s(\tau , \sigma )\, \sigma^s\,
 {\buildrel {def}\over =}\, x^{\mu}(\tau ) + b^{\mu}_r(\tau
 ,\sigma )\, \sigma^r,\nonumber \\
 &&{}\nonumber \\
 &&R^r{}_s(\tau ,\sigma ) {\rightarrow}_{\sigma \rightarrow
 \infty} \delta^r_s,\qquad \partial_A\, R^r{}_s(\tau
 ,\sigma )\, {\rightarrow}_{\sigma \rightarrow
 \infty}\, 0,\nonumber \\
 &&{}\nonumber \\
 &&b^{\mu}_s(\tau ,\sigma ) = \epsilon^{\mu}_r\, R^r{}_s(\tau
 ,\sigma )\, {\rightarrow}_{\sigma \rightarrow
 \infty}\, \epsilon^{\mu}_s,\quad [b^{\mu}_r\, \eta_{\mu\nu}\, b^{\nu}_s](\tau ,\sigma )
 = - \sgn\, \delta_{rs},\nonumber \\
 &&{}\nonumber \\
 &&R = R(\alpha ,\beta ,\gamma ),\quad \mbox{with Euler angles
 satisfying}\nonumber \\
 &&{}\nonumber \\
 && \alpha (\tau ,\sigma) =F(\sigma )\, \tilde \alpha (\tau
 ),\qquad
 \beta (\tau ,\sigma ) = F(\sigma )\, \tilde \beta (\tau
 ),\qquad
 \gamma (\tau ,\sigma )=F(\sigma )\, \tilde \gamma (\tau
 ),\nonumber \\
 &&{}\nonumber \\
  &&0< F(\sigma ) <
 {m\over {2\, K\, M_1\, \sigma}}\,(K-1)=\frac{1}{M\,\sigma},\qquad
 {{d F(\sigma )}\over {d \sigma }} \not= 0,\nonumber \\
 &&{}\nonumber \\
 \mbox{ or }\qquad&&| \partial_{\tau} \alpha (\tau ,\sigma )|,
  | \partial_{\tau} \beta (\tau ,\sigma )|,
   | \partial_{\tau} \gamma (\tau ,\sigma )| <
   {{m}\over {2\, K\, \sigma}}\,(K-1).
 \label{II2}
 \eea

\bigskip

Let us now consider an isolated system restricted to hyper-planes
with constant unit normal $l^{\mu}$: $z^{\mu}(\tau ,\vec \sigma )
= x^{\mu}(\tau ) + l^{\mu}\, \tau + \epsilon^{\mu}_r(\tau )\,
\sigma^r$, different from the Wigner hyper-planes orthogonal to
the conserved 4-momentum of the field configuration (the {\it
rest-frame instant form}).

\noindent In both cases the two associated congruences of
time-like observers include {\it only} inertial observers.
However, while in the rest-frame instant form there is a built-in
Wigner covariance of the quantities defined inside the Wigner
hyper-planes, in the case of hyper-planes with constant normal in
a give reference inertial system there is an {\it explicit
breaking of the action of Lorentz boosts}. Therefore, in Appendix
A of the second paper in Refs.\cite{2} we defined a general class
of admissible embeddings containing a given foliation with
hyper-planes with unit normal $l^{\mu}$ and with admissible
differentially rotating 3-coordinates of the type of
Eq.(\ref{II2}). If, as we will show, we allow $l^{\mu}$ to become
a dynamical variable, then all the foliations whose hyper-planes
have unit normal $\Lambda^{\mu}{}_{\nu}\, l^{\nu}$ for every
Lorentz transformation $\Lambda$ can be obtained without breaking
manifest Lorentz covariance. This material will be reviewed in
Subsection C, where we discuss the description of free
relativistic positive-energy scalar particles in non-inertial
frames.

Let us first review the parametrized Minkowski theory for a N
particle system.

\subsection{Parametrized Minkowski Theories: the $N$-Body Problem.}

As shown in Refs.\cite{7,8,9}, given an admissible 3+1 splitting
with embedding $z^{\mu}(\tau , \vec \sigma )$ and a set of $N$
massive positive energy particles with time-like world line
$x^\mu_i(\tau)$, $i=1,...,N$, we describe the particles with the
$\Sigma_\tau$-adapted Lorentz-scalar 3-coordinates
$\vec{\eta}_i(\tau)$ defining their intersection with the
hyper-surface $\Sigma_\tau$

\begin{equation}
x^\mu_i(\tau)=z^\mu(\tau,\vec{\eta}_i(\tau))\;\; \Rightarrow\;\;
\dot{x}^\mu(\tau)=z^\mu_\tau(\tau,\vec{\eta}_i(\tau))+
z^\mu_r(\tau,\vec{\eta}_i(\tau))\,\dot{\eta}^r(\tau).
 \label{II3}
\end{equation}

The Lagrangian density of parametrized Minkowski theories for $N$
free positive energy particles is

\bea
 {\cal L}(\tau,\vec{\sigma}) &=& -\sum_i
m_i\,\delta^3(\vec{\sigma}-\vec{\eta}_i(\tau)) \sqrt{\sgn\, [
g_{\tau\tau}(\tau,\vec{\sigma}) +2g_{\tau
{r}}(\tau,\vec{\sigma})\dot{\eta}^{{r}}_i(\tau)
+g_{{r}{s}}(\tau,\vec{\sigma})
\dot{\eta}^{{r}}_i(\tau)\dot{\eta}^{{s}}_i(\tau)]}, \nonumber \\
&&\nonumber\\
 S &=& \int d\tau\, L(\tau ) = \int d\tau d^3\sigma\, {\cal
 L}(\tau ,\vec \sigma ),
  \label{II4}
 \end{eqnarray}

From this Lagrangian we can obtain the following momenta:

\begin{eqnarray*}
 \rho_{\mu}(\tau ,\vec \sigma ) &=& - {{\partial {\cal L}(\tau
 ,\vec \sigma )}\over {\partial z^{\mu}_{\tau}(\tau ,\vec \sigma
 )}}=\nonumber\\
&&\nonumber\\
&=&\sgn\, \sum_{i=1}^N\, \delta^3(\vec \sigma - {\vec
 \eta}_i(\tau ))
 m_i\, {{z_{\tau \mu}(\tau ,\vec \sigma ) + z_{r\mu}(\tau
 ,\vec \sigma )\, {\dot \eta}_i^r(\tau )}\over {\sqrt{\sgn\, [
g_{\tau\tau}(\tau,\vec{\sigma}) +2g_{\tau
\check{r}}(\tau,\vec{\sigma})\dot{\eta}^{\check{r}}_i(\tau)
+g_{\check{r}\check{s}}(\tau,\vec{\sigma})
\dot{\eta}^{\check{r}}_i(\tau)\dot{\eta}^{\check{s}}_i(\tau)]}}},
 \end{eqnarray*}

\bea
 \kappa_{ir}(\tau ) &=& - {{\partial L(\tau )}\over {\partial
 {\dot \eta}_i^r(\tau )}} = \nonumber\\
&&\nonumber\\
&=&\sgn\, m_i\, {{g_{\tau r}(\tau ,{\vec
 \eta}_i(\tau )) + g_{rs}(\tau ,{\vec \eta}_i(\tau ))\, {\dot
 \eta}_i^r(\tau )}\over {\sqrt{\sgn\, [
g_{\tau\tau}(\tau,{\vec \eta}_i(\tau )) +2g_{\tau {r}}(\tau,{\vec
\eta}_i(\tau ) )\, \dot{\eta}^{{r}}_i(\tau) +g_{{r}{s}}(\tau,{\vec
\eta }_i(\tau ))
\dot{\eta}^{{r}}_i(\tau)\dot{\eta}^{{s}}_i(\tau)]}}}.
 \label{II5}
\eea

The action $S$ is invariant under frame-preserving
reparametrizations. This special relativistic general covariance
implies that, as already said, the embedding
$z^\mu(\tau,\vec{\sigma})$ are arbitrary {\em gauge variables} not
determined by the Euler-Lagrange equations.

\bigskip

In  the  Hamiltonian formulation we have the following pairs of
canonical variables:

i) the {\em external} variables associated to the {\em embedding}
degrees of freedom

\begin{equation}
z^\mu(\tau,\vec{\sigma}),\;\;\rho_{\mu}(\tau,\vec{\sigma}):\;\;\;\;
\{z^\mu(\tau,\vec{\sigma}),\rho_{\nu}(\tau,\vec{\sigma}')\}=
-\eta^{\mu}_{\nu}\,\delta(\vec{\sigma}-\vec{\sigma}'),
 \label{II6}
\end{equation}

ii) the {\em internal} variables associated to the degrees of
freedom of the physical system living on the hyper-surface, which
in the $N$-particles case are the 3-position and the conjugate
momentum

\begin{equation}
\eta^r_i(\tau),\;\;\kappa_{i\,r}(\tau):\;\;\;\;
\{\eta^r_i(\tau),\kappa_{j\,s}(\tau)\}=-\delta^r_s\delta_{ij}.
 \label{II7}
\end{equation}

\medskip

The ordinary momenta $p^{\mu}_i(\tau )$ are derived variables
given by the following positive-energy solutions of the mass-shell
constraints $\sgn\, p^2_i - m^2_i = 0$

\begin{equation}
p^{\mu}_i(\tau ) = \sqrt{m^2_i + h^{rs}(\tau ,{\vec \eta}_i(\tau
))\, \kappa_{ir}(\tau )\, \kappa_{is}(\tau )}\, l^{\mu}(\tau ,
{\vec \eta}_i(\tau )) + \sgn\, z^{\mu}_r(\tau ,{\vec \eta}_i(\tau
))\, h^{rs}(\tau ,{\vec \eta}_i(\tau ))\, \kappa_{is}(\tau ).
\label{II8}
\end{equation}
\medskip

The canonical variables are not independent but there are the
following first class constraints on the phase space
($T_{\tau\tau}$ and $T_{\tau r}$ are the energy- and
momentum-density components of the energy-momentum tensor written
in coordinates adapted to the leaves $\Sigma_{\tau}$)

\begin{eqnarray*}
 {\cal H}_{\mu}(\tau ,\vec \sigma ) &=& \rho_{\mu}(\tau ,\vec
 \sigma ) - \sgn\, l_{\mu}(\tau ,\vec \sigma )\, T_{\tau\tau}(\tau
 ,\vec \sigma ) - \sgn\, z_{r\mu}(\tau ,\vec \sigma )\,
 h^{rs}(\tau ,\vec \sigma )\, T_{\tau r}(\tau ,\vec \sigma )
 =\nonumber \\
&&\nonumber\\
 &=& \rho_{\mu}(\tau ,\vec
 \sigma ) - \sgn\, l_{\mu}(\tau ,\vec \sigma )\, \sum_{i=1}^N\,
 \delta^3(\vec \sigma - {\vec \eta}_i(\tau ))\, \sqrt{m^2_i +
 h^{rs}(\tau ,\vec \sigma )\, \kappa_{ir}(\tau )\, \kappa_{is}(\tau )}
  +\end{eqnarray*}

\bea
 &+& \sgn\, z_{r\mu}(\tau ,\vec \sigma )\, h^{rs}(\tau ,\vec \sigma )\,
  \sum_{i=1}^N\, \delta^3(\vec \sigma - {\vec \eta}_i(\tau ))\,
  \kappa_{is}(\tau ) \approx 0,\nonumber \\
 &&{}\nonumber \\
 &&or\nonumber \\
 &&{}\nonumber \\
 {\cal H}_\perp(\tau,\vec{\sigma})&=&
\rho_\mu(\tau,\vec{\sigma})\,l^\mu(\tau,\vec{\sigma})-
\sum_i\delta^3(\vec{\sigma}-\eta_i(\tau)) \sqrt{m_i^2\,c^2 + h
^{rs}(\tau,\vec{\sigma})\kappa_{i\,r}(\tau) \kappa_{i\,s}(\tau)}
\approx 0,\nonumber\\
 &&\nonumber\\
  {\cal H}_r(\tau,\vec{\sigma})&=& \rho_\mu(\tau,\vec{\sigma})\,
z^\mu_r(\tau,\vec{\sigma})-
\sum_i\delta^3(\vec{\sigma}-\eta_i(\tau)) \kappa_{i\,r}(\tau)
\approx 0.
 \label{II9}
\end{eqnarray}

\noindent While we have $\{ {\cal H}_{\mu}(\tau ,\vec \sigma ),
{\cal H}_{\nu}(\tau ,{\vec \sigma}^{'}\} = 0$, the Poisson bracket
algebra \cite{7,8,9} of the constraints ${\cal H}_{\perp}$ and
${\cal H}_r$ is the universal Dirac Algebra like the
super-Hamiltonian and the super-momentum constraints of ADM
canonical metric gravity \cite{3}, with the ${\cal H}_r$
generating the 3-diffeomorphisms on $\Sigma_{\tau}$. The
Hamiltonian gauge transformations generated by these constraints
change the form and the coordinatization of the space-like
hyper-surface $\Sigma_\tau$, showing explicitly that the
embeddings $z^\mu(\tau,\vec{\sigma})$ are {\it gauge variables}.
\medskip

Since the {\em canonical Hamiltonian} $H_c$, obtained by the
Legendre transformation of the original Lagrangian, is null
($H_c\equiv 0$), we have to use the {\em Dirac Hamiltonian}

\begin{equation}
H_D(\tau)= \int d^3\sigma\, \lambda^{\mu}(\tau ,\vec \sigma )\,
{\cal H}_{\mu}(\tau ,\vec \sigma ) = \int d^3\sigma\,
\Big[\lambda_\perp(\tau,\vec{\sigma})\,{\cal
H}_\perp(\tau,\vec{\sigma})+ \lambda^r(\tau,\vec{\sigma})\,{\cal
H}_r(\tau,\vec{\sigma})\Big],
 \label{II10}
\end{equation}

\noindent where $\lambda_\perp(\tau,\vec{\sigma}),
\lambda^r(\tau,\vec{\sigma})$ are arbitrary Dirac multipliers.
These arbitrary Dirac multipliers can be used as new  flat lapse
and shift functions $N(\tau ,\vec \sigma ) = \lambda_\perp(\tau
,\vec \sigma )$, $N^ r(\tau ,\vec \sigma ) = \lambda^r(\tau ,\vec
\sigma )$. In Minkowski space-time they are quite distinct from
the previous lapse and shift functions $N_{[z]}$, $N^r_{[z]}$,
defined starting from the metric. Only with the use of the
Hamilton equations $z^{\mu}_{\tau}(\tau ,\vec \sigma )\,
{\buildrel \circ \over =}\, \{ z^{\mu}(\tau ,\vec \sigma ), H_D
\}$ we get $N_{[z](flat)}\, {\buildrel \circ \over =}\,
N_{(flat)}$, $N_{[z](flat)\,r}\, {\buildrel \circ \over =}\,
N_{(flat)\,r}$.
\medskip

Since only the embedding functions and their momenta carry
Minkowski indices and the Lagrangian formulation is {\em
manifestly covariant}, the canonical generators of the Poincar\'e
transformations are

\begin{eqnarray}
p_\mu(\tau)&=&\int d^3\sigma\, \rho_\mu(\tau,\vec{\sigma}),
\nonumber \\
 &&\nonumber\\
  J^{\mu\nu}(\tau)&=&\int
d^3\sigma\, \left[
z^\mu(\tau,\vec{\sigma})\rho^\nu(\tau,\vec{\sigma})-
z^\nu(\tau,\vec{\sigma})\rho^\mu(\tau,\vec{\sigma})\right].
 \label{II11}
\end{eqnarray}

In Refs.\cite{12,14} there is the study of the restriction of the
embedding to space-like hyper-planes by means of the gauge fixings

\beq
 z^{\mu}(\tau ,\vec \sigma ) - x^{\mu}(\tau ) - b^{\mu}_r(\tau )\,
 \sigma^r \approx 0,
 \label{II12}
 \eeq

 \noindent where the $b^{\mu}_r(\tau )$ are three space-like
 ortho-normal vectors, forming an ortho-normal tetrad with the
 normal $b^{\mu}_{\tau}(\tau ) = l^{\mu}(\tau )$ to the
 hyper-planes. Then from Eqs.(\ref{II11}) we get $J^{\mu\nu} =
 x^{\mu}\, p^{\nu} - x^{\nu}\, p^{\mu} + S^{\mu\nu}$. In the case
 $l^{\mu} = const.$ (see also Appendix A of the first paper in Ref.\cite{3}) it can
 be shown that a non-Darboux canonical basis of the reduced phase
 space is $x^{\mu}$, $p^{\mu}$, $b^{\mu}_r$, $S^{\mu\nu}$ [the
 remaining degrees of freedom of the embedding], ${\vec
 \eta}_i(\tau )$, ${\vec \kappa}_i(\tau )$, and that only seven
 first class constraints survive

 \bea
  H^{\mu}(\tau ) &=& p^{\mu} - l^{\mu}\, \sum_{i=1}^N\, \sqrt{m^2_i +
  {\vec \kappa}^2_i(\tau )} + b^{\mu}_r(\tau )\, \sum_{i=1}^N\,
  \kappa_{ir}(\tau ) \approx 0,\nonumber \\
  &&{}\nonumber \\
  \vec H(\tau ) &=& \vec S - \sum_{i=1}^N\, {\vec \eta}_i(\tau )
  \times {\vec \kappa}_i(\tau ) \approx 0.
  \label{II13}
  \eea

However, with $l^{\mu} = const.$ in the given inertial system we
have a breaking of the action  of the Lorentz boosts  and this
induces a breaking of the Lorentz-scalar nature of the particle
coordinates ${\vec \eta}_i(\tau )$, ${\vec \kappa}_i(\tau )$: they
transform in a complex non-tensorial way. Moreover in the quoted
papers there was no study of the admissibility, in the sense of
Subsection A, of the rotating 3-coordinates when the $b^{\mu}_r$
are $\tau$-dependent: only $b^{\mu}_r = const.$ is admissible.

\bigskip

Instead in Refs.\cite{7,8,9} there is a detailed study of the {\it
rest-frame instant form}, which corresponds to the limiting case
$l^{\mu} = p^{\mu} / \sqrt{\sgn\, p^2} = u^{\mu}(p)$. In it the
leaves $\Sigma_{\tau}$ are orthogonal to the conserved 4-momentum
of the isolated system: they are the {\it Wigner hyper-planes}
with Wigner covariance for all the quantities defined on them. It
can be shown that  after suitable gauge fixings only the variables
${\tilde x}^{\mu}$, $p^{\mu}$, ${\vec \eta}_i$, ${\vec \kappa}_i$
survive with ${\tilde x}^{\mu} = x^{\mu} + (spin\, terms)$ not
being a 4-vector and with the particle positions given by the
equations $x^{\mu}_i(\tau ) = x^{\mu}(\tau ) +
\epsilon^{\mu}_r(u(p))\, \eta^r_i(\tau )$. Only the following four
first class constraints survive [$u^{\mu}(p)$ and
$\epsilon^{\mu}_r(u(p))$ are the columns of the Wigner boost
(\ref{a1})]

\begin{eqnarray*}
 {\cal H} &=& \sqrt{\sgn\, p^2} - \sum_{i=1}^N\, \sqrt{m^2_i +
 {\vec \kappa}^2_i} = \sqrt{\sgn\, p^2} - M \approx 0,
 \end{eqnarray*}

\bea
 {\vec {\cal H}} &=& \sum_{i=1}^N\, {\vec \kappa}_i \approx 0.
 \label{II14}
 \eea

After the gauge fixing $\tau - p \cdot \tilde x /\sqrt{\sgn\, p^2}
= \tau - p \cdot x /\sqrt{\sgn\, p^2} \approx 0$ to ${\cal H}
\approx 0$, it can be shown that the $\tau$-evolution is ruled by
the Hamiltonian ($M$ is the invariant mass of the isolated system)

\beq
 H = M + \vec \lambda (\tau ) \cdot \sum_{i=1}^N\, {\vec \kappa}_i
 (\tau ),
 \label{II15}
 \eeq

\noindent and only the rest-frame conditions $\sum_{i=1}^N\, {\vec
\kappa}_i \approx 0$ remain. It is possible to decouple the center
of mass of the isolated system (see  Ref.\cite{33}) and to study
only the (Wigner covariant) relative motions on the Wigner
hyper-planes, where there is a degenerate {\it internal} Poincare'
algebra with $M$ as energy and $\vec S = \sum_{i=1}^N\, {\vec
\eta}_i \times {\vec \kappa}_i {|}_{\sum_i\, {\vec \kappa}_i = 0}$
as angular momentum.

\subsection{A Family of Admissible Foliations with Parallel Hyper-Planes
and Differentially Rotating 3-Coordinates.}

Let us now consider the following special embeddings (see
Eq.(\ref{a5}) for the vector decompositions)

\begin{eqnarray*}
 z^{\mu}_{U}(\tau ,\vec \sigma ) &=& x^{\mu}(0) + {\hat U}^{\mu}\, x_U(\tau ) +
 \epsilon^{\mu}_a({\hat U})\, \xi^a_U(\tau ,\vec \sigma ) =\nonumber \\
 &=& x^{\mu}_U(\tau ) + F^{\mu}_U(\tau ,\vec \sigma ),\nonumber \\
 &&\nonumber\\
 x^{\mu}_U(\tau ) &=& x^{\mu}(0) + {\hat U}^{\mu}\, x_U(\tau ) +
 \epsilon^{\mu}_a({\hat U})\, x^a_U(\tau ) = z^{\mu}_{U}(\tau ,\vec 0) =\nonumber \\
 &=& [x_U(\tau ) + \sgn\, x_{\nu}(0)\, {\hat U}^{\nu}]\, {\hat
 U}^{\mu} + [x^a_U(\tau ) + \sgn\, x^{\nu}(0)\,
 \epsilon^a_{\nu}(\hat U)]\, \epsilon^{\mu}_a(\hat U),\nonumber \\
&&\nonumber\\
 x^{\mu}(0) &=& \sgn\, [x_{\nu}(0)\, {\hat U}^{\nu}\, {\hat
 U}^{\mu} + x^{\nu}(0)\, \epsilon^a_{\mu}(\hat U)\,
 \epsilon^{\mu}_a(\hat U)],
 \end{eqnarray*}

\bea
 &&\xi^a_U(\tau ,\vec \sigma ) = x^a_U(\tau ) + \zeta^a(\tau ,\vec
 \sigma ),\nonumber \\
&&\nonumber\\
 &&F^{\mu}_U(\tau ,\vec \sigma ) =
 \epsilon^{\mu}_a({\hat U})\, \zeta^a(\tau ,\vec \sigma
 ),\qquad F^{\mu}_U(\tau ,\vec 0) = \zeta^a(\tau ,\vec 0) =
 0,
 \label{II16}
 \eea

\medskip

\noindent where ${\hat U}^\mu$ is the unit normal to the
hyper-surface. The time-like vector  ${\hat U}^{\mu} =
\epsilon^{\mu}_o(\hat U)$ and the triad of space-like vectors
$\epsilon^{\mu}_a({\hat U})$ are the columns of the standard
Wigner boost defined in Eq.(\ref{a1}) of Appendix A
[$\epsilon^A_{\mu}(\hat U)$ are the cotetrads associated to the
tetrads $\epsilon^{\mu}_A(\hat U)$]. As a consequence the
hyper-surfaces $\Sigma_{\tau}$ of this foliation are {\it parallel
hyper-planes orthogonal to ${\hat U}^{\mu}$} with arbitrary
admissible 3-coordinates described by the functions $\zeta^a(\tau
,\vec \sigma )$ [for instance the embedding (\ref{II2}) is
recovered with $\zeta^a(\tau ,\vec \sigma ) = \sigma^s\, {\cal
R}_s{}^a(\tau ,\vec \sigma )$, ${\cal R}^{-1}(\tau ,\vec \sigma )
= {\cal R}^T(\tau ,\vec \sigma )$]. The world-line $x^{\mu}_U(\tau
)$ of an arbitrary non-inertial time-like observer is the
time-axis of a {\it non-inertial reference frame} centered on this
observer \footnote{In general $\tau$ is not the proper time of
this observer.} with the hyper-planes $\Sigma_{\tau}$ as
instantaneous 3-spaces. While ${\ddot x}^{\mu}_U(\tau )$ describes
the {\it translational} 4-acceleration of the non-inertial frame
(both the freedom in the choice of the mathematical time $\tau$
and the linear 3-acceleration), the functions $\zeta^a(\tau ,\vec
\sigma )$ describe its {\it rotational} properties [${\cal R}(\tau
,\vec \sigma )$ are rotation matrices].

\bigskip

To avoid problems with manifest Lorentz covariance, we shall
enlarge our framework so that the normal ${\hat U}^{\mu}$ becomes
a dynamical 4-vector. To this end  let us add a free relativistic
particle $X^{\mu}(\tau )$ of unit mass to the Lagrangian
(\ref{II5}), which is replaced by

\begin{equation}
 L(\tau ) \mapsto L^{'}(\tau ) = \int d^3\sigma\, {\cal L}(\tau
 ,\vec \sigma ) - \sqrt{\sgn\, {\dot X}^2(\tau )},
\label{II17}
\end{equation}

With this new Lagrangian we have  the extra momentum

\begin{equation}
U^{\mu}(\tau ) = - {{\partial L^{'}(\tau )}\over {\partial
 {\dot X}^{\mu}(\tau )}} = {{{\dot X}^{\mu}(\tau )}\over {\sqrt{\sgn\,
 {\dot X}^2(\tau )}}},\qquad
 \{X^\mu(\tau), U_\nu(\tau)\} = - \eta^\mu_\nu.
 \label{II18}
 \end{equation}

Then we get the extra first class constraint

\begin{equation}
\chi(\tau)= \sgn \, U^2(\tau)-1\approx 0,\quad \Rightarrow\quad
{\hat U}^{\mu}(\tau ) = \frac{U^{\mu}(\tau )}{\sqrt{\sgn \,
U^2(\tau )}} \approx U^{\mu}(\tau ),
 \label{II19}
\end{equation}

\noindent and  the new Dirac Hamiltonian [see Eq.(\ref{II10})]
with the extra Dirac multiplier $\kappa (\tau )$

 \begin{equation}
H_D(\tau)=\int d^3\sigma \left[{\tilde \lambda}_\perp
(\tau,\vec{\sigma}) {\cal H}_\perp (\tau,\vec{\sigma})+ {\tilde
\lambda}^r(\tau,\vec{\sigma}){\cal H}_r(\tau,\vec{\sigma})
\right]+\kappa(\tau)\,\chi(\tau).
 \label{II20}
\end{equation}

\medskip

The configurational variable ${\hat U}^{\mu}(\tau )$ is  a
constant of motion, since the Hamiltonian of Eq.(\ref{II20})
implies ${{d {\hat U}^{\mu}(\tau )}\over {d \tau}} = 0$ and ${{d
X^{\mu}(\tau )}\over {d \tau}} \approx - 2\, \sgn\, \kappa (\tau
)\, {\hat U}^{\mu}(\tau )$.

\bigskip

The canonical generators (\ref{II11}) of the Poincar\'e group are
replaced by

\begin{eqnarray}
p^\mu(\tau)&=& U^\mu(\tau)+\int
d^3\sigma\,\rho^\mu(\tau,\vec{\sigma}),  \label{II21}\\
 &&\nonumber\\
 J^{\mu\nu}(\tau) &=& X^\mu(\tau)U^\nu(\tau)-
X^\nu(\tau)U^\mu(\tau)+ \int d^3\sigma\,[
z^\mu(\tau,\vec{\sigma})\rho^\nu(\tau,\vec{\sigma})-
z^\nu(\tau,\vec{\sigma})\rho^\mu(\tau,\vec{\sigma})]. \nonumber
\end{eqnarray}

Then, let us restrict the arbitrary embeddings $z^{\mu}(\tau ,\vec
\sigma )$ with the following gauge fixing constraint

\begin{equation}
S(\tau,\vec{\sigma})={\hat U}^\mu(\tau)\,\left[
z_\mu(\tau,\vec{\sigma})-z_\mu(\tau,0)\right]\approx 0.
 \label{II22}
\end{equation}

\noindent The surviving family of embeddings admits the following
parametrization

\begin{eqnarray*}
 z^{\mu}(\tau ,\vec \sigma ) &\approx& \theta (\tau )\, {\hat
 U}^{\mu}(\tau ) + \epsilon^{\mu}_a(\hat U(\tau ))\, {\cal
 A}^a(\tau ,\vec \sigma ) =\nonumber \\
&&\nonumber\\
 &=& z^{\mu}(\tau ,\vec 0) +
\epsilon^{\mu}_a(\hat U(\tau ))\, \Big[{\cal A}^a(\tau ,\vec
\sigma ) - {\cal A}^a(\tau ,\vec 0)\Big] ,\nonumber \\
 &&{}\nonumber \\
 &&z^{\mu}(\tau ,\vec 0) = \theta (\tau )\, {\hat U}^{\mu}(\tau )
 + \epsilon^{\mu}_a(\hat U(\tau ))\, {\cal A}^a(\tau ,\vec
 0) = x^{\mu}_U(\tau ),
  \end{eqnarray*}

 \bea
 &&\theta (\tau ) = \sgn\, {\hat U}^{\mu}(\tau )\, z_{\mu}(\tau
,\vec
 0),\qquad {\cal A}^a(\tau ,\vec \sigma ) = \sgn\,
 \epsilon^a_{\mu}(\hat U(\tau ))\,
 z^{\mu}(\tau ,\vec \sigma ),\nonumber \\
 &&{}\nonumber \\
 &&z^{\mu}_r(\tau ,\vec \sigma ) \approx \epsilon^{\mu}_a(\hat U(\tau
 ))\, {{\partial {\cal A}^a(\tau ,\vec \sigma )}\over {\partial
 \sigma^r}},\quad \Rightarrow \quad l^{\mu}(\tau ,\vec \sigma )
 \approx {\hat U}^{\mu}(\tau ),\nonumber \\
 &&{}\nonumber \\
 \Rightarrow&& g_{rs}(\tau ,\vec \sigma ) \approx -\sgn\, \sum_a\,
 {{\partial  {\cal A}^a(\tau ,\vec \sigma )}\over {\partial \sigma^r}}\,
 {{\partial {\cal A}^a(\tau ,\vec \sigma )}\over {\partial \sigma^s}}
 = - \sgn\, h_{rs}(\tau ,\vec \sigma ).
 \label{II23}
 \eea

Therefore the gauge fixing (\ref{II22}) implies that the
simultaneity surfaces $\Sigma_{\tau}$ are hyper-planes orthogonal
to the arbitrary time-like unit vector ${\hat U}^{\mu}(\tau )$. We
see that  $\theta (\tau )$ {\it describes the freedom in the
choice of the mathematical time $\tau$, ${\ddot {\cal A}}^a(\tau
,\vec 0)$ the linear 3-acceleration of the non-inertial frame and
${{\partial {\cal A}^a(\tau ,\vec \sigma )}\over {\partial \tau}}
- {\dot {\cal A}}^a(\tau ,\vec 0)$ its angular velocity,
describing its rotational properties. As a consequence, a gauge
fixing for $\theta (\tau )$ and ${\cal A}^a(\tau ,\vec \sigma )$
realizes the choice of a well defined non-inertial frame}. The
embedding $z^{\mu}_U(\tau , \vec \sigma )$ of Eqs.(\ref{II16}) is
recovered if $\theta (\tau ) = x_U(\tau ) + \sgn\, x_{\nu}(0)\,
{\hat U}^{\nu}(\tau )$, ${\cal A}^a(\tau ,\vec \sigma ) =
x^a_U(\tau ) + \zeta^a(\tau ,\vec \sigma ) + \sgn\, x^{\nu}(0)\,
\epsilon^a_{\nu}(\hat U(\tau ))$.

\medskip

The time preservation of the gauge fixing (\ref{II22}) implies

\begin{equation}
\frac{d}{d\tau}\,S(\tau,\vec{\sigma})\approx 0\,
\Rightarrow\,{\tilde \lambda}_\perp(\tau,\vec{\sigma}) \approx
{\tilde \lambda}_\perp(\tau,\vec 0)\, {\buildrel {def}\over =}\,
\mu(\tau),
 \label{II24}
\end{equation}

\noindent and then in the reduced theory we have the Dirac
Hamiltonian

 \begin{equation}
H_D(\tau)=\mu (\tau) {\cal H}_\perp (\tau)+\int d^3\sigma\,
{\tilde \lambda}^r(\tau,\vec{\sigma}){\cal H}_r(\tau,\vec{\sigma})
+\kappa(\tau)\,\chi(\tau),
 \label{II25}
\end{equation}

\noindent where

\begin{equation}
H_\perp (\tau)=\int d^3\sigma\, {\cal H}_\perp
(\tau,\vec{\sigma})\approx 0.
 \label{II26}
\end{equation}
\medskip

Since  we  have

\bea
  \rho^{\mu}(\tau ,\vec \sigma ) &=& \sgn\, \Big[
 \rho_U(\tau ,\vec \sigma)\, {\hat U}^{\mu}(\tau ) - \sum_a\,
 \epsilon^{\mu}_a(\hat U(\tau ))\,
 \rho_{U\,a}(\tau ,\vec \sigma )\Big],\nonumber \\
 &&{}\nonumber \\
 \rho_U(\tau ,\vec \sigma )
 &\approx& \sgn\, \sum_{i=1}^N\,
 \delta^3(\vec \sigma - {\vec \eta}_i(\tau ))\, \sqrt{m^2_i +
 h^{rs}(\tau ,\vec \sigma )\, \kappa_{ir}(\tau )\, \kappa_{is}(\tau
 )},\nonumber \\
 &&{}\nonumber \\
 \rho_{Ua}(\tau ,\vec \sigma ) &=& - \sgn\, \rho^a_U(\tau ,\vec
\sigma ) = \epsilon^{\mu}_a(\hat U(\tau ))\, \rho_{\mu}(\tau ,\vec
\sigma ),\nonumber \\
 &&{}\nonumber \\
 \rho_{Ua}(\tau ,\vec \sigma )&\approx& {{\partial {\cal A}^a(\tau ,\vec
 \sigma )}\over {\partial \sigma^r}} \, \sum_{i=1}^N\,
 \delta^3(\vec \sigma - {\vec \eta}_i(\tau ))\, \kappa_{ir}(\tau ).
 \label{II27}
\eea

\medskip

Since we have ${\cal A}^a(\tau ,\vec \sigma ) =  \sgn\,
\epsilon^a_{\mu}(\hat U(\tau ))\, z^{\mu}(\tau ,\vec  \sigma )$,
Eqs.(\ref{II6}) imply

 \beq
  \{ {\cal A}^a(\tau ,\vec \sigma ), \rho_{U\,b}(\tau ,{\vec
\sigma}^{'}) \} = - \sgn\,
 \delta^a_b\, \delta^3(\vec \sigma - {\vec \sigma}^{'}),
\label{II29}
 \eeq

\medskip

\noindent Eqs.(\ref{II9}) can be rewritten in the following form
[Eqs.(\ref{II23}) are used]

\begin{eqnarray*}
{\cal H}_\perp (\tau,\vec{\sigma}) &=& l^{\mu}(\tau ,\vec \sigma
)\, {\cal H}_{\mu}(\tau ,\vec \sigma ) \approx {\cal H}_U(\tau
,\vec \sigma ) = {\hat U}^{\mu}(\tau
)\, {\cal H}_{\mu}(\tau ,\vec \sigma ) \approx \nonumber \\
&&\nonumber\\
 &\approx& \rho_U(\tau,\vec{\sigma})- \sgn\,
\sum_{i=1}^N\, \delta^3(\vec \sigma - {\vec \eta}_i(\tau ))\,
\sqrt{m^2_i + h^{rs}(\tau ,\vec \sigma )\, \kappa_{ir}(\tau )\,
\kappa_{is}(\tau )} \approx 0,
 \end{eqnarray*}

 \bea
  {\cal H}_r(\tau,\vec{\sigma}) &=&
z^\mu_r(\tau,\vec{\sigma})\rho_\mu(\tau,\vec{\sigma})- \sgn\,
\sum_{i=1}^N\, \delta^3(\vec \sigma - {\vec \eta}_i(\tau ))\,
\kappa_{ir}(\tau ) \approx \nonumber \\
&&\nonumber\\
 &\approx& {{\partial {\cal A}^a(\tau ,\vec \sigma )}\over
 {\partial \sigma^r}}\, \rho_{U\,a}(\tau ,\vec \sigma ) - \sgn\, \sum_{i=1}^N\,
 \delta^3(\vec \sigma - {\vec \eta}_i(\tau ))\, \kappa_{ir}(\tau ) \approx 0.
 \label{II28}
\end{eqnarray}

\medskip

Introducing the  {\it internal mass} of the N-body system on the
simultaneity surface $\Sigma_{\tau}$

\beq
 M_U(\tau) = \int d^3\sigma\, \rho_U(\tau,\vec{\sigma}), \qquad
   \{\theta (\tau ), M_U(\tau ) \} = -\sgn,
 \label{II30}
\eeq

\noindent the constraint (\ref{II26}) can be written in the form

\begin{eqnarray}
H_\perp (\tau)&=&M_U(\tau)- {\cal E}[{\cal
A}^a, {\vec \eta}_i, {\vec \kappa}_i]\approx 0,\nonumber\\
 &&\nonumber\\
  {\cal E}[{\cal A}^a, {\vec \eta}_i, {\vec \kappa}_i]&=& \sgn\, \sum_{i=1}^N\,
  \sqrt{m^2_i + h^{rs}(\tau ,{\vec \eta}_i(\tau ))\,
   \kappa_{ir}(\tau )\, \kappa_{is}(\tau )}.
 \label{II31}
\end{eqnarray}

\medskip

Eqs.(\ref{II31}) show that the gauge fixing (\ref{II22}) and the
constraints

\bea
 \chi (\tau ,\vec \sigma ) &=& {\hat U}^{\mu}(\tau )\,
 [{\cal H}_{\mu}(\tau ,\vec \sigma ) -
\delta^3(\vec \sigma )\, \int d^3\sigma_1\, {\cal H}_{\mu}(\tau
,{\vec \sigma}_1)] \approx {\hat U}^{\mu}(\tau )\, {\cal
H}_{\mu}(\tau ,\vec \sigma ) - H_{\perp}(\tau )\, \delta^3(\vec
\sigma ) \approx\nonumber \\
&&\nonumber\\
 &\approx& {\tilde \rho}_U(\tau ,\vec \sigma ) -
\sgn\, \sum_{i=1}^N\, [\delta^3(\vec \sigma - {\vec \eta}_i(\tau
))\, \sqrt{m^2_i + h^{rs}(\tau ,\vec \sigma )\, \kappa_{ir}(\tau
)\, \kappa_{is}(\tau )} -\nonumber \\
&&\nonumber\\
 &-& \delta^3(\vec \sigma )\,
\sqrt{m^2_i + h^{rs}(\tau ,{\vec \eta}_i(\tau ))\,
\kappa_{ir}(\tau )\, \kappa_{is}(\tau )}] \approx 0,
 \label{II32}
 \eea

\noindent with ${\tilde \rho}_U(\tau ,\vec \sigma ) = \rho_U(\tau
,\vec \sigma ) - \sgn\, M_U(\tau )$, form a pair of second class
constraints, so that the only surviving first class constraints
are $H_{\perp}(\tau ) \approx 0$ and ${\cal H}_r(\tau ,\vec \sigma
) \approx 0$.

\medskip

In the second paper of Refs.\cite{2} it is shown that the Dirac
brackets associated to the gauge fixing (\ref{II22}) are

\begin{eqnarray}
&&\{A(\tau),B(\tau)\}^*= \{A(\tau),B(\tau)\}+\nonumber\\
 &&\nonumber\\
 &+&\int d^3\sigma[ \{A(\tau),S(\tau,\vec{\sigma})\}
\{{\cal H}_U (\tau,\vec{\sigma}),B(\tau)\}-
\{B(\tau),S(\tau,\vec{\sigma})\} \{{\cal H}_U
(\tau,\vec{\sigma}),A(\tau)\}].
 \label{II33}
\end{eqnarray}

\bigskip

After the gauge fixing (\ref{II22}), a set of independent
variables for the reduced embedding, the particles and the extra
particle are $\theta (\tau )$, $M_U(\tau )$, ${\cal A}^r(\tau
,\vec \sigma )$, $\rho_{Ur}(\tau ,\vec \sigma )$, ${\vec
\eta}_i(\tau )$, ${\vec  \kappa}_i(\tau )$, $X^{\mu}(\tau )$,
$U^{\mu}(\tau )$. However this is not a Darboux basis, because the
reduced embedding variables have non-zero Poisson brackets with
$X^{\mu}(\tau )$. To find a canonical basis let us introduce the
following variables

\bea
 &&{\cal A}^{A}(\tau,\vec{\sigma}) =  \sgn\,
\epsilon^{A}_\mu(\hat U(\tau ))\, z^\mu(\tau,\vec{\sigma}) =
\Big({\cal A}^o(\tau,\vec \sigma ) \approx \theta (\tau ); {\cal
A}^a(\tau ,\vec \sigma )\Big), \nonumber \\
&&\nonumber\\
&& R^{A}(\tau,\vec{\sigma}) =\epsilon^{A}_\mu(\hat
 U(\tau ))\, \rho^\mu(\tau,\vec{\sigma}) = \Big(R^o = \rho_U;
 R^a = \rho_U^a\Big)(\tau ,\vec
 \sigma ),\nonumber \\
 &&\nonumber\\
 &&\{{\cal A}^{A}(\tau,\vec{\sigma}), {\cal
A}^{B}(\tau,\vec{\sigma}')\}= \{R^{A}(\tau,\vec{\sigma}),
R^{B}(\tau,\vec{\sigma})'\} = 0,
\nonumber\\
&&\nonumber\\
 &&\{{\cal A}^{A}(\tau,\vec{\sigma}),
R^{B}(\tau,\vec{\sigma}')\}= -\eta^{AB}\,
\delta^3(\vec{\sigma}-\vec{\sigma}').
  \label{II34}
 \eea

\noindent Then, following the construction of the
Newton-Wigner-like canonical non-covariant variable of
Ref.\cite{7}, we define the pseudo-vector

\bea
 \widetilde{X}^\mu(\tau) &=&  X^\mu(\tau)-
\frac{\partial\epsilon_{A}^\sigma(\hat U(\tau ))}{\partial {\hat
U}_\mu}\, \epsilon_{\sigma{B}}(\hat U(\tau ))\, \int
d^3\sigma\,{\cal A}^{A}(\tau,\vec{\sigma})\,
R^{B}(\tau,\vec{\sigma}),\nonumber \\
 &&{}\nonumber \\
  &=& X^{\mu}(\tau ) + \frac{\epsilon^\mu_a(\hat U(\tau ))}{\sqrt{\sgn\, U^2(\tau)}}\,\int
d^3\sigma\,\left(\theta(\tau)\,  \rho^a_U(\tau,\vec{\sigma})-
{\cal A}^a(\tau,\vec{\sigma})\, \rho_U(\tau,\vec{\sigma}) \right)
+\nonumber \\
&&\nonumber\\
 &+& \frac{\partial \epsilon^\alpha_a(\hat U(\tau ))}{\partial
{\hat U}_\mu} \,\epsilon_{b\alpha}(\hat U(\tau )) \int
d^3\sigma\,{\cal A}
^a(\tau,\vec{\sigma})\,\rho^b_U(\tau,\vec{\sigma}).
 \label{II35}
\eea

\medskip

Eqs.(\ref{II6}) and (\ref{II7}) imply the following Poisson
brackets

\begin{eqnarray}
&&\{\widetilde{X}^\mu(\tau),\widetilde{X}^\nu(\tau)\}=0,
 \{\widetilde{X}^\mu(\tau),U^\nu(\tau)\}=-\eta^{\mu\nu},\nonumber \\
 &&{}\nonumber \\
 &&\{ {\tilde X}^{\mu}(\tau ), {\cal A}^A(\tau ,\vec \sigma )\}
 =0, \qquad \{ {\tilde X}^{\mu}(\tau ), {\cal R}^A(\tau ,\vec
 \sigma )\} =0.
 \label{II36}
\end{eqnarray}

\medskip

As a consequence, the canonical variables $\theta (\tau )$,
$M_U(\tau )$, ${\cal A}^a(\tau ,\vec \sigma )$, $\rho_{Ur}(\tau
,\vec \sigma )$, ${\vec \eta}_i(\tau )$, ${\vec \kappa}_i(\tau )$,
${\tilde X}^{\mu}(\tau )$, $U^{\mu}(\tau )$ give a Darboux basis
for the reduced phase space, which, as shown in Appendix B,
remains a canonical basis also at the level of the Dirac brackets.
Let us remark that the gauge fixing (\ref{II22}) can now be
rewritten as $S(\tau ,\vec \sigma ) = {\cal A}^o(\tau ,\vec \sigma
) - {\cal A}^o(\tau ,\vec 0) \approx 0$.

\bigskip

Let us now study the Lorentz covariance of the new variables in
the reduced phase space. Let us first observe that substituting
for $X^{\mu}$ its expression implied by Eq.(\ref{II35}) in
Eq.(\ref{II21}) we obtain the following expression of the momentum
and the following splitting of the angular momentum

\begin{eqnarray*}
 p^{\mu}(\tau ) &=& [1 +  M_U(\tau )]\, {\hat U}^{\mu}(\tau ) - \sgn\,
 \epsilon^{\mu}_a(\hat U(\tau ))\, \int d^3\sigma\, \rho_{U\,a}(\tau
 ,\vec \sigma )
\approx\nonumber \\
 &\approx& \Big[ \sqrt{\sgn\, U^2(\tau )} +
 \sum_{i=1}^N\, \sqrt{m^2_i + h^{rs}(\tau ,{\vec \eta}_i(\tau ))\,
 \kappa_{ir}(\tau )\, \kappa_{is}(\tau )}\Big]\, {\hat U}^{\mu}(\tau )
 -\nonumber \\
 &-& \sgn\, \sum_a\,
 \epsilon^{\mu}_a(\hat U(\tau ))\, \int d^3\sigma\, \rho_{U\,a}(\tau
 ,\vec \sigma ),\nonumber\\
&&\nonumber\\
 J^{\mu\nu} &=& {\tilde L}^{\mu\nu} + {\tilde S}^{\mu\nu},
\nonumber \\
 &&{}\nonumber \\
 {\tilde S}^{\mu\nu} &=& D_{ab}{}^{\mu\nu}(\hat U)\,
 \int d^3\sigma\, [{\cal A}^a\, \rho^b_U - {\cal A}^b\,
 \rho^a_U](\tau ,\vec \sigma ), \nonumber\\
&&\nonumber\\
 {\tilde L}^{\mu\nu} &=&
{\tilde X}^{\mu}(\tau )\, U^{\nu}(\tau ) - {\tilde X}^{\nu}(\tau
)\, U^{\mu}(\tau ), \nonumber \\
&&\nonumber\\
 &&\{ {\tilde L}^{\mu\nu}, {\tilde S}^{\alpha\beta}\} \not=
 0,\nonumber \\
 &&{}\nonumber \\
 D^{\alpha\beta}_{ab}(\hat U) &=& \frac{1}{2}\left[
\epsilon^\alpha_a(\hat U)\epsilon^\beta_b(\hat U)-
\epsilon^\alpha_b(\hat U)\epsilon^\beta_b(\hat U)-\left( {\hat
U}^\alpha\frac{\partial \epsilon^\mu_a(\hat U)}{\partial {\hat
U}_\beta}- {\hat U}^\beta\frac{\partial \epsilon^\mu_a(\hat
U)}{\partial {\hat U}_\alpha} \right)\,\epsilon_{b\mu}(\hat
U)\right],\end{eqnarray*}

\[  {\tilde X}^\mu(\tau)
 =(\hat{U}^\sigma(\tau)\,X_\sigma(\tau))\,\hat{U}^\mu(\tau)+
J^{\mu\rho}(\tau)\hat{U}_\rho(\tau)\frac{1}{\sqrt{\sgn\,
U^2(\tau)}}
  -\frac{\partial
\epsilon^\alpha_a(\hat U(\tau )}{\partial {\hat U}_\nu}
\,\epsilon_{b\alpha}(\hat U(\tau ))\, S^{ab}(\tau ),
\]

\bea
  \{ {\tilde S}^{\mu\nu}, {\tilde S}^{\alpha\beta} \} &=&
 C^{\mu\nu\alpha\beta}_{\rho\sigma}\, {\tilde S}^{\rho\sigma} +
 \Big({{\partial D_{ab}{}^{\mu\nu}(\hat U)}\over {\partial {\hat U}_{\beta}}}\,
 U^{\alpha} - {{\partial D_{ab}{}^{\mu\nu}(\hat U)}\over {\partial {\hat U}_{\alpha}}}\,
 U^{\beta} -\nonumber \\
 &&\quad - {{\partial D_{ab}{}^{\alpha\beta}(\hat U)}\over {\partial {\hat U}_{\nu}}}\,
 U^{\mu} + {{\partial D_{ab}{}^{\alpha\beta}(\hat U)}\over {\partial {\hat U}_{\mu}}}\,
 U^{\nu}\Big)\, S^{ab},\nonumber \\
 &&{}\nonumber \\
 S^{ab}(\tau ) &=& \int d^3\sigma\, ({\cal A}^a\, \rho^b_U - {\cal A}^b\,
 \rho^a_U)(\tau ,\vec \sigma ).
 \label{II37}
 \eea

This decomposition of $J^{\mu\nu}$ is a direct consequence of
Eqs.(\ref{II34}) and it is left unchanged by the gauge fixing
(\ref{II22}).

 Moreover we get

\begin{equation}
\{J^{\mu\nu}(\tau),S(\tau,\vec{\sigma})\}=
\{J^{\mu\nu}(\tau),{\cal H}_\perp(\tau,\vec{\sigma})\}=0,
 \label{II38}
\end{equation}

\noindent  so that the Dirac brackets  (\ref{II33}) preserve the
Poincare' algebra

\begin{eqnarray}
 \{ p^{\mu}, p^{\nu}\}^* &=& 0,\qquad \{ p^{\mu}, J^{\rho\sigma}\}^*
 = \eta^{\mu\rho}\, p^{\sigma} - \eta^{\mu\sigma}\,
 p^{\rho},\nonumber \\
&&\nonumber\\
\{J^{\mu\nu}(\tau), J^{\sigma\rho}(\tau)\}^*&=&\{J^{\mu\nu}(\tau),
J^{\sigma\rho}(\tau)\}=
C^{\mu\nu\sigma\rho}_{\alpha\beta}J^{\alpha\beta}(\tau),\nonumber\\
 &&\nonumber\\
  C^{\mu\nu\rho\sigma}_{\alpha\beta}&=&
\eta^\nu_\alpha\eta^\rho_\beta\eta^{\mu\sigma}+
\eta^\mu_\alpha\eta^\sigma_\beta\eta^{\nu\rho}-
\eta^\nu_\alpha\eta^\sigma_\beta\eta^{\mu\rho}-
\eta^\mu_\alpha\eta^\rho_\beta\eta^{\nu\sigma}.
 \label{II39}
\end{eqnarray}

\medskip

As shown in Appendix B, at the level of these Dirac brackets the
variables $F({\cal I})$, $M_U(\tau )$, $\theta (\tau )$ are
Lorentz scalars, ${\cal A}^a(\tau ,\vec \sigma )$, $\rho^a_U(\tau
,\vec \sigma ) = - \sgn\, \rho_{Ua}(\tau ,\vec \sigma )$ are
Wigner spin 1 3-vectors, $U^{\mu}(\tau )$ is a 4-vector but
${\tilde X}^{\mu}(\tau )$ is not a 4-vector. Therefore, since
Eq.(\ref{b10}) remains true, we still have that under a Lorentz
transformation $\Lambda$ we get $U^{\mu} \mapsto
\Lambda^{\mu}{}_{\nu}\, U^{\nu}$.

\bigskip

Following Refs.\cite{7,31} we can make the following canonical
transformation from ${\tilde X}^{\mu}$, $U^{\mu}$ to the canonical
basis ($\vec z$ is a Newton-Wigner-like non-covariant 3-vector)

\bea
 {\cal U}(\tau)=\sqrt{\sgn\, U^2(\tau)} \approx 1,&\qquad& {\cal
W}(\tau)=\frac{U(\tau)\cdot \tilde X(\tau)}{\sqrt{\sgn\,
U^2(\tau)}} = \frac{U(\tau)\cdot  X(\tau)}{\sqrt{\sgn\,
U^2(\tau)}}, \nonumber\\
 &&\nonumber\\
 k^i(\tau)=\frac{U^i(\tau)}{\sqrt{\sgn\, U^2(\tau)}} ,&\qquad&
z^i(\tau)=\sqrt{\sgn\, U^2(\tau)}\,\left(
 {\tilde X}^i(\tau)-{\tilde X}^o(\tau)\,\frac{U^i(\tau)}{U^o(\tau)} \right),\nonumber
\eea
\beq
\{{\cal U}(\tau),{\cal W}(\tau)\}^*=1,\;\;\;\;
\{z^i(\tau),k^i(\tau)\}^*=\delta^{ij}.
 \label{II40}
  \eeq

\medskip

Eqs.(\ref{II25}) and (\ref{II35}) imply ${{d k^i(\tau )}\over {d
\tau}} = 0$, ${{d z^i(\tau )}\over {d \tau}} \approx {\dot {\tilde
X}}^i(\tau ) - {\dot {\tilde X}}^o(\tau )\, {{U^i(\tau )}\over
{U^o(\tau )}} \approx - 2\, \sgn\, \kappa (\tau )\, \Big( {\hat
U}^i(\tau ) - {\hat U}^o(\tau )\, {{U^i(\tau )}\over {U^o(\tau
)}}\Big) \approx 0$, namely $\vec z$ and $\vec k$ are Jacobi
non-evolving initial data.
\medskip

Let us remark that Eq.(\ref{II23}) implies that the centroid
origin of the 3-coordinates, namely the non-inertial observer
$x^{\mu}_U(\tau )$ and ${\tilde X}^{\mu}(\tau )$, have  different
4-velocities: ${\dot x}^{\mu}_U(\tau ) \approx \dot \theta (\tau
)\, {\hat U}^{\mu}(\tau ) + \epsilon^{\mu}_a(\hat U(\tau ))\,
{\dot {\cal A}}^a(\tau ,\vec 0) \not= {\dot {\tilde X}}^{\mu}(\tau
) = - 2\, \sgn\, \kappa (\tau )\, {\hat U}^{\mu}(\tau )$. Let us
remark that ${\dot x}^{\mu}_U(\tau )$ is not orthogonal to the
hyper-planes $\Sigma_{\tau}$.

\bigskip

 If we want to eliminate the constraint $\chi (\tau ) = \sgn\, U^2(\tau ) - 1
 \approx 0$, we must add the gauge fixing

\bea
 &&{\cal W}(\tau ) - \sgn\, \theta (\tau )
  \approx 0,\quad \Rightarrow\quad \kappa (\tau ) =
  - {{\sgn}\over 2}\, \dot \theta (\tau ),\nonumber \\
 &&{}\nonumber \\
 &&\Downarrow\nonumber \\
 &&{}\nonumber \\
 &&{\tilde X}^{\mu}(\tau ) = z^{\mu}(\tau , {\vec \sigma}_{\tilde
 X}(\tau )),\quad for\, some\quad {\vec \sigma}_{\tilde X}(\tau
 ),\nonumber \\
 &&X^{\mu}(\tau ) = z^{\mu}(\tau ,{\vec \sigma}_X(\tau )),\quad
 for\, some\quad {\vec \sigma}_X(\tau ),\nonumber \\
 &&{}\nonumber \\
 U^{\mu}(\tau ) &=& \Big( \sqrt{1 + {\vec k}^2}; k^i(\tau )
 \Big) = {\hat U}^{\mu}(\vec k),\nonumber \\
&&\nonumber\\
 {\tilde X}^{\mu}(\tau ) &=& \Big( \sqrt{1 + {\vec k}^2}\, [\sgn\,
 \theta (\tau ) + \vec k(\tau ) \cdot \vec z(\tau )];
 z^i(\tau ) + k^i(\tau )\, [\sgn\, \theta (\tau ) + \vec k(\tau
 ) \cdot \vec z(\tau )] \Big)=\nonumber \\
&=&
z^{\mu}(\tau ,{\vec
 \sigma}_{\tilde X}(\tau )),\nonumber \\
 &&{}\nonumber \\
 &&L^{ij} = z^i\, k^j - z^j\, k^i,\qquad L^{oi} = - L^{io} = -
 z^i\, \sqrt{1 + {\vec k}^2}.
 \label{II45}
 \eea
 \bigskip

After having introduced new Dirac brackets, the extra added point
particle of unit mass is reduced to the decoupled non-evolving
variables $\vec z$, $\vec k$ and the not yet determined ${\vec
\sigma}_{\tilde X}(\tau )$ and ${\vec \sigma}_X(\tau )$ give the
3-location on $\Sigma_{\tau}$ of ${\tilde X}^{\mu}(\tau )$ and
$X^{\mu}(\tau )$, respectively, which do not coincide with the
world-line $x^{\mu}_U(\tau )$ of the non-inertial observer. Now we
get ${\dot {\tilde X}}^{\mu}(\tau ) = \dot \theta (\tau )\, {\hat
U}^{\mu}(\tau )$ [$\dot \theta (\tau ) = {\dot x}_U(\tau )$] and
this determines ${\vec \sigma}_{\tilde X}(\tau )$ as solution of
the equation ${{\partial {\cal A}^a(\tau ,{\vec \sigma}_{\tilde
X}(\tau ))}\over {\partial \tau}} + {{\partial {\cal A}^a(\tau
,\vec \sigma )}\over {\partial \sigma^s}}{|}_{\vec \sigma = {\vec
\sigma}_{\tilde X}(\tau ) }\, {\dot \sigma}^s_{\tilde X}(\tau ) =
0$.

\bigskip

In the reduced phase space we have the following constraints'
algebra  (the constraints ${\cal H}_r$ still satisfy the algebra
of 3-diffeomorphisms)

\begin{eqnarray}
&& \{{\cal H}_r(\tau,\vec{\sigma}) ,{\cal
H}_s(\tau,\vec{\sigma}')\}^*= {\cal H}_r(\tau,\vec{\sigma}')
\frac{\partial}{\partial\sigma^{\prime\,s}}
\delta^3(\vec{\sigma}-\vec{\sigma}')- {\cal
H}_s(\tau,\vec{\sigma}) \frac{\partial}{\partial\sigma^{s}}
\delta^3(\vec{\sigma}-\vec{\sigma}'),\nonumber\\
 &&\nonumber\\
  &&\{H_\perp(\tau),{\cal H}_r(\tau,\vec{\sigma})\}^*=0.
   \label{II41}
\end{eqnarray}

\bigskip

Finally, the embedding  whose hyper-planes have a fixed unit
normal $l^{\mu}$, implying the breaking of the action of Lorentz
boosts, is obtained by adding by hand the first class constraints
(only three are independent)

\beq
 {\hat U}^{\mu}(\vec k) - l^{\mu} \approx 0,
 \label{II46}
 \eeq

\noindent which determine the non-evolving constant $\vec k$. The
conjugate constant $\vec z$ can be eliminated with the
non-covariant gauge fixing

\beq
 \vec z \approx 0, \qquad \Rightarrow\quad {\tilde X}^{\mu}(\tau )
 \approx {\tilde X}^{\mu}(0) + \sgn\, \theta (\tau )\, l^{\mu}.
 \label{II47}
 \eeq

\medskip

The constraints (\ref{II46}) and (\ref{II7}) eliminate the extra
non-evolving degrees of freedom $\vec k$ and $\vec z$ of the added
decoupled point particle, respectively.

\bigskip

If we want to recover the embedding (\ref{II16}) with
$\zeta^a(\tau ,\vec \sigma ) = \sigma^s\, {\cal R}_s{}^a(\tau
,\vec \sigma )$, ${\cal R}^{-1} = {\cal R}^T$, namely to choose a
well defined {\it non-inertial non-rigid reference frame}, we must
add the following gauge fixings to the first class constraints
$H_{\perp}(\tau ) \approx 0$ and ${\cal H}_r(\tau ,\vec \sigma )
\approx 0$ [here we do not eliminate the constraint $\chi (\tau )
\approx 0$ with the gauge fixing (\ref{II45})]

\begin{eqnarray*}
 &&\theta (\tau ) - x_U(\tau ) - {\hat U}_{\mu}(\tau )\,
 x^{\mu}(0) \approx 0,\nonumber \\
 &&{}\nonumber \\
 &&{\cal A}^a(\tau ,\vec \sigma ) - \xi^a_U(\tau ,\vec \sigma ) -
 \epsilon^a_{\mu}(\hat U(\tau ))\, x^{\mu}(0) \approx 0,\nonumber \\
 &&{}\nonumber \\
 &&\xi^a_U(\tau ,\vec \sigma ) + \epsilon^a_{\mu}(\hat U(\tau ))\,
 x^{\mu}(0) = \epsilon^a_{\mu}(\hat U(\tau ))\, x^{\mu}(0) +
 x^a_U(\tau ) + \zeta^a(\tau ,\vec \sigma ) =\nonumber \\
 &&\qquad = {\tilde x}^a(\tau ) + \sigma^a\, {\cal R}_s{}^a(\tau
 ,\vec \sigma ),\end{eqnarray*}

\begin{eqnarray*}
 &&\Downarrow\\
&&\\
  &&{\cal A}^s_a(\tau ,\vec \sigma )\, \mbox{ inverse of }\,
 {{\partial {\cal A}^a(\tau ,\vec \sigma )}\over {\partial \sigma^s}} \approx {{\partial
 \zeta^a(\tau ,\vec \sigma )}\over {\partial
 \sigma^s}} = [{\cal R}_s{}^a + \sigma^u\, {{\partial {\cal
 R}_u{}^a}\over {\partial \sigma^s}}](\tau ,\vec \sigma ),
\nonumber \\
 &&{}\nonumber \\
 &&z^{\mu}(\tau ,\vec 0) = x^{\mu}_U(\tau ),\quad
 F^{\mu}_U(\tau ,\vec \sigma ) = \epsilon^{\mu}_a(\hat U(\tau ))\,
 [{\cal A}^a(\tau ,\vec \sigma ) - {\cal A}^a(\tau ,\vec 0)] =
 \epsilon^{\mu}_a(\hat U(\tau ))\, \zeta^a(\tau ,\vec \sigma ),\nonumber \\
 &&{}\nonumber \\
  &&\qquad h_{rs}(\tau ,\vec \sigma ) = \sum_a\, {{\partial
 \zeta^a(\tau ,\vec \sigma )}\over {\partial
 \sigma^r}}\, {{\partial \zeta^a(\tau ,\vec \sigma )}\over
 {\partial \sigma^s}},\qquad h^{rs}(\tau ,\vec \sigma ) = \sum_a\,
 {\cal A}^r_a(\tau ,\vec \sigma )\, {\cal A}^s_a(\tau ,\vec \sigma ),
 \end{eqnarray*}

 \bea
  &&\rho_{Ua}(\tau ,\vec \sigma ) \approx \sgn\, {\cal A}^s_a(\tau ,\vec
 \sigma )\, \sum_{i=1}^N\, \delta^3(\vec \sigma - {\vec
 \eta}_i(\tau ))\, \kappa_{is}(\tau ),\nonumber \\
 &&{}\nonumber \\
 &&S^{ab} = \sum_{i=1}^N\, \Big[ (\xi^a_U\, {\cal A}^{bv} - \xi^b_U\,
 {\cal A}^{av})(\tau ,{\vec \eta}_i(\tau ))\, \kappa_{iv}(\tau ) +\nonumber \\
 &&\qquad + x^{\mu}(0)\, \Big(\epsilon^a_{\mu}(\hat U(\tau ))\,
 {\cal A}^{bv}(\tau ,{\vec \eta}_i(\tau ) ) - \epsilon^b_{\mu}(\hat U(\tau ))\,
 {\cal A}^{av}(\tau ,{\vec \eta}_i(\tau ) )\Big) \kappa_{iv}(\tau )\Big],
 \label{II42}
 \eea

\bea
 p^{\mu}(\tau )  &\approx& \Big[1 + \sum_{i=1}^N\,
 \sqrt{m^2_i + h^{rs}(\tau ,{\vec \eta}_i(\tau )\,
 \kappa_{ir}(\tau )\, \kappa_{is}(\tau )} \Big]\, {\hat U}^{\mu}(\tau )
 -\nonumber \\
 &-& \epsilon^{\mu}_a(\hat U(\tau ))\, \sum_{i=1}^N\,
 {\cal A}^s_a(\tau ,{\vec \eta}_i(\tau ) )\, \kappa_{is}(\tau
 ),\nonumber \\
&&\nonumber\\
 J^{\mu\nu}(\tau ) &\approx& {\tilde X}^{\mu}\, {\hat
 U}^{\nu}(\tau ) - {\tilde X}^{\nu}\, {\hat U}^{\mu}(\tau ) +
 D^{\mu\nu}_{ab}(\hat U(\tau ))\, S^{ab}(\tau ),\nonumber \\
&&\nonumber\\
 {\tilde X}^{\mu}(\tau ) &\approx& {\cal W}(\tau )\, {\hat
 U}^{\mu}(\tau ) + J^{\mu\rho}(\tau )\, {\hat U}_{\rho}(\tau ) -
 {{\partial \epsilon^{\alpha}_a(\hat U(\tau ))}\over {\partial
 {\hat U}_{\mu}}}\, \epsilon_{b\alpha}(\hat U(\tau ))\,
 S^{ab}(\tau ).
 \label{II43}
 \eea

\bigskip

The preservation in time of the gauge fixings (\ref{II42}) and
${{d {\hat U}^{\mu}(\tau )}\over {d \tau}} = 0$ imply $\mu (\tau )
= - \dot \theta (\tau ) = - {\dot x}_U(\tau ) = - {\dot
x}^{\mu}_U(\tau )\, {\hat U}_{\mu}(\tau )$, $\lambda^r(\tau ,\vec
\sigma ) = -\sgn\, {\cal A}^r_a(\tau ,\vec \sigma )\, {{\partial
{\cal A}^a(\tau ,\vec \sigma )}\over {\partial \tau}} = -\sgn\,
{\cal A}^r_a(\tau ,\vec \sigma )\, \Big({\dot x}^{\mu}_U(\tau )\,
\epsilon^a_{\mu}(\hat U(\tau )) + {{\partial \zeta^a(\tau ,\vec
\sigma )}\over {\partial \tau}}\Big)$ for the Dirac multipliers
appearing in the Dirac Hamiltonian (\ref{II25}) and in the
associated Hamilton equations.

\bigskip

Eqs.(\ref{II3}) and (\ref{II16}) show that the particle
world-lines are given by $x^{\mu}_i(\tau ) = x^{\mu}_U(\tau ) +
\epsilon^{\mu}_a(\hat U)\, \zeta^a(\tau ,{\vec \eta}_i(\tau ))$ $=
x^{\mu}_U(\tau ) + \epsilon^{\mu}_a(\hat U)\, \eta^s_i(\tau )\,
{\cal R}_s{}^a(\tau ,{\vec \eta}_i(\tau ))$ and the particle
4-momenta, satisfying $\sgn\, p^2_i = m^2_i$, are $p^{\mu}_i(\tau
) = \sqrt{m^2_i + h^{rs}(\tau ,{\vec \eta}_i(\tau ))\,
\kappa_{ir}(\tau )\, \kappa_{is}(\tau ) }\, {\hat U}^{\mu} -
\epsilon^{\mu}_a(\hat U)\, {\cal A}_a^s(\tau ,{\vec \eta}_i(\tau
))\, \kappa_{is}(\tau )$. Eqs.(\ref{II43}) imply that the
Poincare' generators can be written in the form  $p^{\mu} = {\hat
U}^{\mu} + \sum_{i=1}^N\, p^{\mu}_i$, $J^{\mu\nu} =
D^{\mu\nu}_{ab}(\hat U)\, S^{ab}$ with $S^{ab}$ of
Eq.(\ref{II42}).

\bigskip

If we go to new Dirac brackets, in the new reduced phase space we
get $H_D = \kappa (\tau )\, \chi (\tau )$ and this Dirac
Hamiltonian does not reproduce the just mentioned Hamilton
equations after their restriction to Eqs.(\ref{II42}) due to the
explicit $\tau$-dependence of the gauge fixings. As a consequence,
we have to find the correct Hamiltonian ruling the evolution in
the reduced phase space. As shown in the second paper of
Refs.\cite{2} this effective Hamiltonian for the non-inertial
evolution is

\begin{eqnarray*}
 H_{ni} &=& - \mu (\tau )\,  \sum_{i=1}^N\, \sqrt{m^2_i +
 h^{rs}(\tau ,{\vec \eta}_i(\tau ))\,
 \kappa_{ir}(\tau )\, \kappa_{is}(\tau )}
  - \sum_{i=1}^N\, \lambda^r(\tau ,{\vec \eta}_i(\tau ) )\,
  \kappa_{ir}(\tau ) +\nonumber \\
 &+& \kappa (\tau )\, \chi (\tau ) =\nonumber \\
 & &{}\nonumber \\
 &=& \dot \theta (\tau )\, \sum_{i=1}^N\, \sqrt{m^2_i +
 {\cal A}^r_a(\tau ,{\vec \eta}_i(\tau ))\,
 \kappa_{ir}(\tau )\, \delta^{ab}\,
 {\cal A}^s_b(\tau ,{\vec \eta}_i(\tau ))\, \kappa_{is}(\tau )} +\nonumber \\
 &+& \sgn\, \sum_{i=1}^N\,
 {\cal A}^r_a(\tau ,{\vec \eta}_i(\tau ))\, {{\partial {\cal
 A}^a(\tau ,{\vec \eta}_i(\tau ))}\over {\partial \tau}}\,
 \kappa_{ir}(\tau ) + \kappa (\tau )\, \chi (\tau ) =\end{eqnarray*}

 \bea
 &=& {\dot x}^{\mu}_U(\tau )\, \Big[ {\hat U}_{\mu}(\tau )\,
 \sum_{i=1}^N\, \sqrt{m^2_i + h^{rs}(\tau ,{\vec \eta}_i(\tau ))\,
 \kappa_{ir}(\tau )\, \kappa_{is}(\tau )} -\nonumber \\
 &-& \epsilon_{a\mu}(\hat U(\tau ))\, \sum_{i=1}^N\,
 {\cal A}^r_a(\tau ,{\vec \eta}_i(\tau ))\, \kappa_{ir}(\tau )  \Big] +
  \sum_{i=1}^N\, {\cal A}^r_a(\tau ,{\vec \eta}_i(\tau ))\,
  {{\partial \zeta^a(\tau ,{\vec \eta}_i(\tau ) )}\over
 {\partial \tau}}\, \kappa_{ir}(\tau ) +\nonumber \\
 &+& \kappa (\tau )\, \chi (\tau ) =\nonumber \\
 &&{}\nonumber \\
 &=& {\dot x}^{\mu}_U(\tau )\, \Big[p_{\mu} - {\hat U}_{\mu}(\tau
 )\Big] +
  \sum_{i=1}^N\, {\cal A}^r_a(\tau ,{\vec \eta}_i(\tau ))\,
  {{\partial \zeta^a(\tau ,{\vec \eta}_i(\tau ) )}\over
 {\partial \tau}}\, \kappa_{ir}(\tau ) + \kappa (\tau )\, \chi (\tau ).
 \label{II44}
  \eea

\noindent where ${\dot x}^{\mu}_U(\tau ) = \dot \theta (\tau )\,
{\hat U}^{\mu}(\tau ) + {\dot {\cal A}}^a(\tau ,\vec 0)\,
\epsilon^{\mu}_a(\hat U(\tau ))$ from Eq.(\ref{II23}) and
Eq.(\ref{c3}) has been used in the second line.

\medskip

We find that, apart from the contribution of the remaining first
class constraint, the {\it effective non-inertial Hamiltonian}
ruling the $\tau$-evolution seen by the (in general non-inertial)
observer $x^{\mu}_U(\tau )$ (the centroid origin of the
3-coordinates) is the sum of the projection of the total
4-momentum along the 4-velocity of the observer (without the term
pertaining to the decoupled unit mass particle) plus a term
induced by the differential rotation of the 3-coordinate system
around the world-line of the observer. Instead the asymptotic
observers at spatial infinity see an evolution ruled only by $
{\dot x}^{\mu}_U(\tau )\, \Big[p_{\mu} - {\hat U}_{\mu}(\tau
 )\Big]$.

\medskip

It is important to remark that in non-inertial systems the
effective Hamiltonian depends on the gauge variables $\theta$ and
${\cal A}^a$ describing the inertial effects, like it happens for
every notion of energy density in general relativity (see for
instance the integrand of the weak ADM energy in Refs.\cite{3}).

\bigskip

As it is clear from the second line of Eq.(\ref{II44}), the {\it
generalized inertial (or fictitious) forces} in a non-rigid
non-inertial frame of this type are generated by the potential
$\sum_{i=1}^N\, {{\partial {\cal A}^a(\tau ,{\vec \eta}_i(\tau
))}\over {\partial \tau}}\, {\cal A}^r_a(\tau , {\vec \eta}_i(\tau
))\, \kappa_{ir}(\tau )$.

In non-inertial frames it is not clear if there is a non-inertial
analogue of the internal Poincare' group of the rest-frame instant
form.

\medskip

To recover the rest-frame instant form, having the Wigner
hyper-planes orthogonal to the total 4-momentum as simultaneity
surfaces, we must require ${\hat U}^{\mu}(\tau ) - p^{\mu} /
\sqrt{\sgn\, p^2} \approx 0$ instead of Eq.(\ref{II46}) and put
$\zeta^{a=r}(\tau ,\vec \sigma ) = \sigma^r$. Then from
Eq.(\ref{II43}) we get the rest-frame conditions
$\epsilon^a_{\mu}(\hat U)\, p^{\mu} \approx 0$ (whose gauge fixing
is the vanishing  of the internal center of mass, see the second
paper in Ref.\cite{33}) and the invariant mass ${\cal E} + 1$,
which is the correct one if we neglect the constant extra mass
$1$.

\vfill\eject

\section{A Multi-Temporal Quantization Scheme for Positive-Energy Relativistic Particles
in Non-Inertial Frames.}

In this Section we define a new quantization scheme for systems
with first class constraints and we  apply it to the quantization
of positive-energy relativistic particles described in the
framework of parametrized Minkowski theories,  namely on the
arbitrary space-like hyper-surface of the Subsection IIIA.
Subsequently, due to ordering problems,  we will restrict
ourselves to the space-like hyper-planes of Subsection IIIB.

\subsection{A Multi-Temporal Quantization Scheme.}

Let us first remind the method of Dirac quantization applied to
the system described by the Lagrangian density (\ref{II11}),
namely to N free positive-energy particles on arbitrary admissible
3-surfaces $\Sigma_{\tau}$. The phase space of this system has the
canonical basis $z^{\mu}(\tau ,\vec \sigma )$, $\rho_{\mu}(\tau
,\vec \sigma )$, ${\vec \eta}_i(\tau )$, ${\vec \kappa}_i(\tau )$,
$i=1,..,N$, and the dynamics is restricted to the sub-manifold
defined by the first class constraints in strong involution
(\ref{II14})

\begin{eqnarray}
{\cal H}_\mu(\tau,\vec{\sigma})&=&l_\mu(\tau,\vec{\sigma}) {\cal
H}_\perp(\tau,\vec{\sigma}) - z_{r\,\mu}(\tau,\vec{\sigma})\,
h^{rs}(\tau,\vec{\sigma}) \,{\cal
H}_s(\tau,\vec{\sigma})=\nonumber\\ &&\nonumber\\ &\byd&
\rho_\mu(\tau,\vec{\sigma})-{\cal G}_\mu\Big(
z(\tau,\vec{\sigma}),\eta^r_i(\tau),\kappa_{i\,r}(\tau)\Big)\approx
0,\nonumber \\
 &&{}\nonumber \\
 &&\{{\cal H}_\mu(\tau,\vec{\sigma}),{\cal
H}_\nu(\tau,\vec{\sigma}')\}=0,
 \label{III1}
\eea

\noindent Since the canonical Hamiltonian is identically zero, the
evolution is ruled by the Dirac Hamiltonian (\ref{II16}).

Dirac's quantization prescription implies the following steps:

i) The definition of a suitable {\it non-physical Hilbert space}
${\bf H_{NP}}$.

ii) The replacement of the canonical basis with {\it self-adjoint
operators} ${\hat z}^{\mu}(\tau ,\vec \sigma )$, ${\hat
\rho}_{\mu}(\tau ,\vec \sigma )$, ${\hat {\vec \eta}}_i(\tau )$,
${\hat {\vec \kappa}}_i(\tau )$ acting on ${\bf H_{NP}}$ and
satisfying the canonical commutation relations

\beq
 [{\hat z}^{\mu}(\tau ,\vec \sigma ), {\hat \rho}_{\nu}(\tau
 ,{\vec \sigma }^{'}) ] = i\, \hbar\, \delta^3(\vec \sigma - {\vec
 \sigma}^{'}),\qquad [ {\hat \eta}^r_i(\tau ), {\hat
 \kappa}_{js}(\tau ) ] = i\, \hbar\, \delta^r_s\, \delta_{ij}.
 \label{III2}
 \eeq

iii) The search of an {\it operator ordering} such that the
resulting constraint operators ${\hat {\cal H}}_{\mu}(\tau ,\vec
\sigma )$ satisfy a commutator algebra of the type (${\hat {\cal
C}}^{\alpha}_{\mu\nu}$ may be operators)

\begin{equation}
[\widehat{\cal H}_\mu(\tau,\vec{\sigma}_1), \widehat{\cal
H}_\nu(\tau,\vec{\sigma}_2)]=\int d^3\sigma \; {\hat {\cal
C}}^\alpha_{\mu\nu} (\vec{\sigma}_1,\vec{\sigma}_2,\vec{\sigma})
\widehat{\cal H}_\alpha(\tau,\vec{\sigma}).
 \label{III3}
\end{equation}

In most cases, in particular in ADM gravity, this is an open
problem. When Eq.(\ref{III3}) holds and the quantum constraints
are self-adjoint operators, then they are the generators of {\it
quantum unitary gauge transformations} in ${\bf H_{NP}}$.

iv) The replacement of the classical Hamilton equations, ruled by
the Dirac Hamiltonian $H_D$, with a Schroedinger equation, ruled
by the quantum Dirac Hamiltonian ${\hat H}_D$, in the Schroedinger
coordinate representation where ${\hat z}^{\mu}(\tau ,\vec \sigma
) = z^{\mu}(\tau ,\vec \sigma )$ and ${\hat \eta}^r_i(\tau ) =
\eta^r_i(\tau )$ are c-number multiplicative operators

\beq
 i\, \hbar\, {{\partial}\over {\partial \tau}}\, \psi_{NP}(\tau | \lambda^{\mu}|
 z^{\mu}, {\vec \eta}_i] = {\hat H}_D\, \psi_{NP}(\tau | \lambda^{\mu} |z^{\mu},
 {\vec \eta}_i].
 \label{III4}
 \eeq

The non-physical wave functions depend on the c-number Dirac
multipliers $\lambda^{\mu}(\tau ,\vec \sigma )$.

\medskip

v) The choice of the {\it non-physical scalar product} $<
\psi_{NP\, 1}, \psi_{NP\, 2} >$ induced by this Schroedinger
equation. With a suitable behavior of $z^{\mu}(\tau ,\vec \sigma
)$ at spatial infinity, this scalar product is {\it
$\tau$-independent}.

vi) The {\it selection of the physical} (gauge invariant and
$\lambda^{\mu}$-independent) {\it states} $\psi$ through the
conditions (Eqs. (\ref{III3}) are necessary for their formal
consistency)

\bea
 &&{\hat {\cal H}}_{\mu}(\tau ,\vec \sigma )\, \psi (\tau  |z^{\mu},
 {\vec \eta}_i] = 0,\nonumber \\
 &&{}\nonumber \\
 &&\Downarrow\nonumber \\
 &&{}\nonumber \\
 &&i\, \hbar\, {{\partial}\over {\partial \tau}}\, \psi (\tau |
 \lambda^{\mu} |z^{\mu}, {\vec \eta}_i] = 0,\quad \Rightarrow\quad \psi = \psi
 [z^{\mu}, {\vec \eta}_i].
 \label{III5}
  \eea

\noindent However, since the zero eigenvalue of the operators
${\hat {\cal H}}_{\mu}$ lies usually in the continuum spectrum,
usually the states $\psi$ are {\it not normalizable} in ${\bf
H_{NP}}$. These states live in the {\it quotient space} ${\bf
H_{NP}} / \{ group\, of\, gauge\, transformations\}$ and the hard
task is {\it to find a physical scalar product $(\psi_1, \psi_2)$
such that the quotient space becomes a Hilbert space}.
\medskip

The BRS approach, modulo the  physical scalar product problem
\cite{28}, is the most developed formalization of this way to do
the quantization, taking also into account the cohomology of the
constraint manifold. In this way one also gets the {\em
Tomonaga-Schwinger} approach to manifestly covariant relativistic
quantum theory \cite{36}; in fact Eqs.(\ref{III5}) are nothing
else that the {\em Tomonaga-Schwinger equations} for $N$
relativistic particles.

\medskip

However, besides the physical scalar product problem, usually
there are formal obstructions to realize the previous scheme of
quantization. In Ref.\cite{37}, it is shown that already in the
case of free quantum fields  the evolution between two 3-surfaces
$\Sigma_{\tau}$ governed by  {\em Tomonaga-Schwinger equations} of
the type of Eq.(\ref{III5}), {\it is in general not unitary due to
an ultraviolet problem}. To avoid this problem, which will be
studied elsewhere, in this paper we concentrate only on particles.
\bigskip

The new quantization scheme is based on the {\it multi-temporal
approach} of Refs.\cite{30,31}. In it the arbitrary Dirac
multipliers appearing in the non-physical Schroedinger equation
(\ref{III4}) and describing the arbitrary gauge aspects of the
description are re-interpreted as new {\it generalized times}
${\cal T}^{\mu}(\tau ,\vec \sigma )$ \footnote{When in
Eq.(\ref{III3}) there is ${\hat C}^{\alpha}_{\mu\nu} = 0$, the
identification is done through the 1-forms $\theta^{\mu}(\tau
){|}_{\vec \sigma = const.} = d {\cal T}^{\mu}(\tau ,\vec \sigma )
= \lambda^{\mu}(\tau ,\vec \sigma )\, d\tau$. Otherwise we have
$\theta^{\mu} = A^{\mu}_{\nu}({\cal T}^{\alpha})\, d{\cal T}^{\nu}
= \lambda^{\mu}(\tau ,\vec \sigma = const.)\, d\tau$ with the
matrix $A$ determined by the structure functions appearing in
Eqs.(\ref{III3}). When in Eqs.(\ref{III3}) there are the structure
constants of a Lie algebra, namely when there is an action of this
Lie algebra on the constraint sub-manifold, the 1-forms $\theta$
are the Maurer-Cartan 1-forms. },the non-physical wave functions
are re-written in the form $\psi_{NP}(\tau | \lambda^{\mu} |
z^{\mu}, {\vec \eta}_i] = {\tilde \psi}_{NP}(\tau , {\cal T}^{\mu}
| z^{\mu}, {\vec \eta}_i]$ and Eq.(\ref{III4}) is replaced by the
following set of coupled Schroedinger-like equations
\footnote{When ${\hat C}^{\alpha}_{\mu\nu} = 0$ in
Eqs.(\ref{III3}) we have ${\hat Y}_{\mu}({\cal T}^{\alpha}) =
\hbar\, {{\delta}\over {\delta {\cal T}^{\mu}(\tau ,\vec \sigma
)}}$. Otherwise we have ${\hat Y}_{\mu}({\cal T}^{\alpha}) =
\hbar\, A^{-1,\, \mu}_{\nu}({\cal T}^{\alpha})\, {{\delta}\over
{\delta {\cal T}^{\nu}(\tau ,\vec \sigma )}}$. } (with
Eqs.(\ref{III3}) as formal integrability conditions)

\bea
 &&i\, \hbar\, {{\partial}\over {\partial \tau}}\, {\tilde \psi}
 _{NP}(\tau , {\cal T}^{\mu} | z^{\mu}, {\vec \eta}_i] =
 0,\nonumber \\
 &&{}\nonumber \\
 &&i\, {\hat Y}_{\mu}({\cal T}^{\alpha})\, {\tilde \psi}_{NP}(\tau ,
 {\cal T}^{\mu} | z^{\mu}, {\vec \eta}_i] = {\hat {\cal
 H}}_{\mu}(\tau ,\vec \sigma )\, {\tilde \psi}_{NP}(\tau , {\cal T}^{\mu} |
z^{\mu}, {\vec \eta}_i].
 \label{III6}
 \eea

The {\it generalized times} ${\cal T}^{\mu}(\tau ,\vec \sigma )$
are nothing else that the {\it Abelianized gauge degrees of
freedom} of the description and, when Eqs.(\ref{III3}) hold, the
second half of Eqs.(\ref{III6}) shows that, in the case of the
constraints (\ref{III1}), they coincide with the embeddings,
${\cal T}^{\mu}(\tau ,\vec \sigma ) = z^{\mu}(\tau ,\vec \sigma )$
\footnote{The interpretation of the embedding variables as
generalized c-number times is an extension of the
 multi-times formalism developed in Ref.
\cite{31} for the Todorov-Droz-Vincent-Komar relativistic two-body
problem \cite{38} and of the {\em multi-fingered time}
interpretation of the ADM metric gravity \cite{39}.}.

\medskip

The {\it physical states} are still defined by Eqs.(\ref{III5}),
but in the case of the constraints (\ref{III1}) these equations
and the physical wave functions are re-written in the form

\bea
 &&i\, \hbar\, {{\partial}\over {\partial \tau}}\, \tilde \psi
 (\tau , z^{\mu} | {\vec \eta}_i] = 0,\nonumber \\
 &&{}\nonumber \\
 &&i\, \hbar\, {{\delta}\over {\delta z^{\mu}(\tau ,\vec \sigma
 )}}\, \tilde \psi (\tau , z^{\mu} | {\vec \eta}_i] = {\hat
 {\cal G}}_{\mu}[z^{\alpha}(\tau ,\vec \sigma ), \eta^r_i(\tau ), i\,
 \hbar\, {{\partial}\over {\partial \eta^r_i(\tau )}}]\, \tilde
 \psi (\tau ,z^{\alpha} | {\vec \eta}_i],\nonumber \\
 &&{}\nonumber \\
 &&\Rightarrow\qquad \tilde \psi = \tilde \psi (z^{\mu}(\tau ,\vec
 \sigma ) | {\vec \eta}_i].
 \label{III7}
 \eea

The operator ordering in ${\hat {\cal G}}_{\mu}$ must imply the
validity of Eqs.(\ref{III3}), since they are the formal
integrability conditions of Eqs.(\ref{III7}).

\medskip

The {\it physical states} live in a (in general frame-dependent)
$N$-particle Hilbert space ${\tilde {\bf H}}$, whose wave
functions $\tilde \psi ({\vec \eta}_i,z)$  are square-integrable
with respect to the frame-dependent measure \footnote{The
frame-dependence is given by the terms $\sqrt{h(z^{\mu}, {\vec
\eta}_i)}$ which depend from the determinant of the 3-metric, i.e.
on the generalized times, on the simultaneity leaves
$\Sigma_{\tau}$ of the non-inertial frame. See Ref.\cite{40} for
examples of Hilbert spaces with a time-dependent measure of the
scalar product.} $d\mu_z(\vec{\eta}_i)= \prod_i\, \sqrt{h(z^{\mu},
\vec{\eta_i})}\, d^3\eta_i$ and depend on the generalized times
$z^{\mu}(\tau ,\vec \sigma )$. Its physical Hermitean scalar
product will be induced by the coupled functional
Schroedinger-like equations (\ref{III7}). The novel aspect is that
now the physical wave functions do not depend on a single time
variable but on a space of {\it generalized time parameters}
$\Big(\tau, {\cal T}(\vec \sigma ) = {\cal T}(., \vec \sigma
)\Big)$ with ${\cal T}(\vec \sigma ){|}_{\tau} = {\cal
T}^{\mu}(\tau ,\vec \sigma ) = z^{\mu}(\tau ,\vec \sigma )$
containing the gauge embedding variables. The physical
frame-dependent scalar product will be of the form $({\tilde
\psi}_1, {\tilde \psi}_2) = \int d \mu_z({\vec \eta}_i)\, {\bar
{\tilde \psi}}_1({\vec \eta}_i, z^{\mu})\, K({\vec \eta}_i, i\,
\hbar\, {{\partial}\over {\partial {\vec \eta}_i}}, z^{\mu})\,
{\tilde \psi}_2( {\vec \eta}_i, z^{\mu})$ with some kernel $K$
dictated by Eqs.(\ref{III7}) and it is independent from the {\it
generalized times}, ${{\delta}\over {\delta z^{\mu}(\tau ,\vec
\sigma )}}\, ({\tilde \psi}_1, {\tilde \psi}_2) = 0$, due to
Eqs.(\ref{III7}). Namely {\it this physical scalar product is
insensitive to the choice of the gauge time parameters}. As we
shall see explicitly in the next Subsection, this allows to
reformulate the theory in a frame-independent Hilbert space ${\bf
{\cal H}}$ with the standard measure $d \mu({\vec \eta}_i) =
\prod_i\, d^3\eta_i$, whose wave functions are $\Psi ({\vec
\eta}_i, z^{\mu}) = [h(z^{\mu}, {\vec \eta}_i)]^{1/4}\, \tilde
\psi ({\vec \eta}_i, z^{\mu})$ and whose scalar product is
$\Big(\Psi_1, \Psi_2\Big) = \int d \mu({\vec \eta}_i)\, {\bar
\Psi}_1({\vec \eta}_i, z^{\mu})\, \tilde K({\vec \eta}_i, i\,
\hbar\, {{\partial}\over {\partial {\vec \eta}_i}}, z^{\mu})\,
\Psi_2( {\vec \eta}_i, z^{\mu})$ with a suitably modified kernel.
Again we have ${{\delta}\over {\delta z^{\mu}(\tau ,\vec \sigma
)}}\, \Big(\Psi_1, \Psi_2\Big) = 0$, due to the induced
modification of the coupled Schroedinger equations. This procedure
is in general equivalent to a change of operator ordering
respecting Eqs.(\ref{III3}).

\bigskip

This construction shows that given a system with first class
constraints, for which, due to the Shanmugadhasan canonical
transformations, we know a canonical set of its Abelianized gauge
variables, {\it we can eliminate the traditional step of
introducing the non-physical Hilbert space}. Instead we can define
a {\it new quantization scheme} in a physical Hilbert space, in
which the wave functions depend on the true {\it time} $\tau$ and
on as many {\it generalized time parameters} as canonical gauge
variables. In this quantization, done in the Schroedinger (either
coordinate or momentum) representation, {\it only the physical
degrees of freedom} like the particle variables ${\vec
\eta}_i(\tau )$, ${\vec \kappa}_i(\tau )$ in our case (more in
general a canonical basis of gauge-invariant Dirac observables)
{\it are quantized}. Instead the gauge variables are not quantized
but considered as {\it c-number generalized times} in analogy to
treatment of time in the non-relativistic Schroedinger equation
$i\hbar\, {{\partial}\over {\partial t}}\, \psi (t,q) = \hat H(q,
\hat p)\, \psi (t,q)$. Like in this equation, where the classical
identification \footnote{$E$, the energy, is the generator of the
kinematical Poincare' group identified by the relativity
principle, while $H$ is the Hamiltonian governing the time
evolution.} $E = H$ is realized with $E\, \mapsto\, i\, \hbar
{{\partial}\over {\partial t}}$ and $H\, \mapsto\, \hat H(q, \hat
p)$, in our system the momenta conjugated to the gauge variables
are replaced with the functional derivatives with respect to the
time variables ($\rho_{\mu}(\tau ,\vec \sigma ) \mapsto i\hbar\,
{{\delta}\over {\delta z^{\mu}(\tau ,\vec \sigma )}}$ in our
example), and their action is identified with the action of the
generalized Hamiltonians ${\hat {\cal G}}_{\mu}$.
\medskip

All the topological problems connected to the description {\it in
the large} of the gauge system, which form the obstruction to do a
quantization of the reduced phase space after a complete canonical
reduction, are shifted to the global properties of the generalized
time parameter space, so that our new quantization is in general
well defined only locally in this parameter space. However, we
have that the restriction of the wave function to a {\it line} in
this parameter space, defined by putting ${\cal T}(\vec \sigma
){|}_{\tau} = z^{\mu}(\tau ,\vec \sigma ) = x^{\mu}(\tau ) +
F^{\mu}(\tau ,\vec \sigma )$ with $F^{\mu}$ given [see for
instance Eqs.(\ref{II2}) or (\ref{II23})] corresponds to the
quantum description of the classical reduced phase space
associated to the classical gauge fixings $z^{\mu}(\tau ,\vec
\sigma ) - x^{\mu}(\tau ) - F^{\mu}(\tau ,\vec \sigma ) \approx
0$.

\bigskip

Since, even in the case of $N$ free positive-energy particles, it
is difficult to find an ordering such that the quantization of the
constraints (\ref{III1}) satisfies Eqs.(\ref{III3}) for arbitrary
embeddings $z^{\mu}(\tau ,\vec \sigma )$, in the next Subsection
we shall apply this new type of quantization to the restricted
case of $N$ free particles described on the family of space-like
hyper-planes with admissible rotating 3-coordinates associated to
the embeddings studied in Subsection IIC. To preserve manifest
Lorentz covariance we must introduce the extra particle
$X^{\mu}(\tau )$, $U^{\mu}$ with ${\hat  U}^{\mu}$ orthogonal to
$\Sigma_{\tau}$. We assume to have introduced the gauge-fixing
(\ref{II45}), so that the Dirac Hamiltonian $H_D$ of
Eq.(\ref{II25}) has $\kappa (\tau ) = 0$ and the only remaining
extra canonical variables are $k^i$ and $z^i$ (the decoupled
non-covariant Newton-Wigner-like variable $\vec z$ is the only
quantity breaking Lorentz covariance).

\subsection{Quantization on Parallel Hyper-Planes with
Admissible Differentially Rotating 3-Coordinates.}

Let us restrict  ourselves to the family of foliations whose
leaves $\Sigma_{\tau}$ are the parallel hyper-planes with
admissible differentially rotating 3-coordinates discussed in the
Subsection IIC, namely the $\Sigma_{\tau}$'s are flat Riemannian
3-manifolds ${\bf R}^3$ with a non-Cartesian coordinate chart
where the 3-metric is given by $h_{rs}(\tau ,\vec \sigma )$ of
Eqs.(\ref{II23}). As shown there, these embeddings are
parametrized by the extra canonical variables $z^i$, $k^i$ of
Eqs.(\ref{II45}), $\theta (\tau )$, $M_U(\tau )$, ${\cal A}^a(\tau
,\vec \sigma )$, $\rho_{Ua}(\tau ,\vec \sigma )$. While $M_U(\tau
)$ is determined by the constraint $H_{\perp}(\tau ) \approx 0$ of
Eq.(\ref{II29}) and $\rho_{Ua}(\tau ,\vec \sigma )$ by the
constraints ${\cal H}_r(\tau ,\vec \sigma ) \approx 0$ of
Eqs.(\ref{II30}), the non-evolving extra variables $\vec k$ and
$\vec z$ are determined only when we add by hand the constraints
$\widehat{U}^{\mu} - l^{\mu} \approx 0$ and $\vec z \approx 0$ of
Eqs. (\ref{II46}) and (\ref{II47}), respectively. However, to
preserve manifest Lorentz (or better Wigner) covariance (with the
exception of the Newton-Wigner-like 3-position $\vec z$), we shall
not add the latter constraints at this preliminary stage.
Therefore $\vec z$ and $\vec k$ are {\it non-evolving spectator
variables} with $\widehat{U}^{\mu}(\vec k)$ describing the unit
normal to the arbitrary hyper-planes $\Sigma_{\tau}$.

\subsubsection{Quantization: Times, Operators and the Frame-Dependent Hilbert Space.}

i) We shall consider the gauge variables $\theta = \theta (.)$ and
${\cal A}^a(\vec \sigma ) = {\cal A}^a(., \vec \sigma )$ (we
suppress the $\tau$-dependence)  as {\it c-number generalized
times}, with the conjugate momenta replaced by the following {\it
time}-derivatives

\beq
 M_U \rightarrow i\hbar\,\frac{\partial}{\partial\theta},
\qquad \rho_{Ua}(\vec{\sigma})\rightarrow i\hbar\,
\frac{\delta}{\delta {\cal A}^a(\vec{\sigma})}.
 \label{III8}
\eeq

ii) The  positions and  momenta $\eta^r_i,\kappa_{i\,r}$ of the
particles are quantized in the usual way as operators on a
standard Hilbert space. We choose a coordinate representation
where $\eta^r_i$ are multiplicative operators and where the
self-adjoint momentum operators

\begin{eqnarray*}
 {\kappa}_{i\,r} &\mapsto& i\, \hbar\, \frac{\partial}{\partial\eta^r_i}
+ i\, \hbar\, {{\partial}\over {\partial \eta_i^r}}\, \ln\,
\sqrt{\det\, \left( \frac{\partial {\cal A}^a(\vec{\eta}_i
)}{\partial \eta_i^r} \right)} =\end{eqnarray*}

\bea
 &=& {1\over {\sqrt{\det\, \left( \frac{\partial {\cal A}^a(\vec{\eta}_i
)}{\partial \eta_i^r} \right)}}}\, i\, \hbar\,
\frac{\partial}{\partial\eta^r_i}\, \sqrt{\det\, \left(
\frac{\partial {\cal A}^a(\vec{\eta}_i )}{\partial \eta_i^r}
\right)},
 \label{III9}
  \eea

\noindent are derivative operators on a {\it frame-dependent
Hilbert space}

\beq
 \widetilde{\bf H}_{\cal A}=L^2(\mu_{\cal A},{\bf R}^{3N}),
 \label{III10}
  \eeq

\noindent whose states are the wave functions $\Psi(\vec{\eta_i})$
with the scalar product

\begin{equation}
(\Psi_1,\Psi_2)=\int d\mu_{\cal A}(\vec{\eta}_i)\, \overline{
\Psi}_1(\vec{\eta_i}) \Psi_2(\vec{\eta_i}),
 \label{III11}
\end{equation}

\noindent defined by a {\it ${\cal A}$-dependent measure} [use is
done of Eq.(\ref{c4})]

 \beq
 d\mu_{\vec {\cal
A}}(\vec{\eta}_i)= \prod_i\, \sqrt{h(\vec{\eta_i})}\, d^3\eta_i=
\prod_i\, \det\, \left( \frac{\partial {\cal A}^a(\vec{\eta}_i
)}{\partial \eta_i^r} \right)\, d^3\eta_i.
 \label{III12}
  \eeq

The scalar product depends on the gauge variables ${\vec {\cal
A}}(\vec \sigma )$ describing the inertial {\it translational and
rotational} effects as discussed after Eq.(\ref{II23}), so that a
different Hilbert space is associated to every non-inertial frame.
In a topologically trivial region of the generalized time space
${\cal M} = \{ \tau , \theta , {\cal A}^a(\vec \sigma )\}$, all
the (translationally and rotationally different) spaces
$\widetilde{\bf H}_{\cal A}$ are isomorphic and can be replaced with a
frame-independent Hilbert space ${\mbox{\boldmath{${\cal H}$}}}$, as it will be
shown in Subsections IIIB3.

iii) The canonical variables $z^i,k^i$, describing  a decoupled
point particle, will be quantized as operators. A convenient
choice is to realize them in the momentum representation: the
variables $k^i$ are multiplicative {\it c-number operators} and
the positions $z^i$ are represented as the folllowing self-adjoint
(non-covariant Newton-Wigner-like) operators

\beq
 z^i \rightarrow i\, \hbar\, \frac{\partial}{\partial k^i} -
 i\, \hbar\, {{k^i}\over {2\, \sqrt{1 + {\vec k}^2}}}.
 \label{III13}
 \eeq

\noindent on a Hilbert Space $L^2(\nu(\vec{k}),{\bf R}^3)$, whose
states are complex functions $F(\vec{k})$ with scalar product

\beq
 (F_1,F_2)=\int d\nu(\vec{k})\,\overline{F}_1(\vec{k})
\,F_2(\vec{k}),
 \label{III14}
  \eeq

\noindent whose {\em covariant} measure is

  \beq
d\nu(\vec{k})=\frac{d^3k}{2\sqrt{1+\vec{k}^2}}.
 \label{III15}
  \eeq

\medskip

From the steps (ii) and (iii), the states of the complete quantum
theory are wave functions $\Phi(\vec{\eta}_i,\vec{k})$ in the
Hilbert spaces

 \beq
  {\bf H}_{\cal A}=L^2(\nu(\vec{k}),{\bf
R}^3)\otimes \widetilde{\bf H}_{\cal A},
 \label{III16}
  \eeq

\noindent with scalar product

 \beq
  \langle \Phi_1,\Phi_2\rangle=
\int d\nu(\vec{k})\,\int d\mu_{\cal A}(\vec{\eta}_i)\,
\overline{\Phi}_1(\vec{\eta}_i,\vec{k})
\Phi_2(\vec{\eta}_i,\vec{k}).
 \label{III17}
  \eeq

\subsubsection{Generalized Temporal Evolution}

A state will evolve in the frame-dependent Hilbert space ${\bf
H}_{\cal A}$ as functional of the {\em time} $\tau$ and of the
{\em generalized times} $\theta$ and ${\cal A}^a(\vec{\sigma})$.
The evolution in these {\em generalized times} is determined by
the quantization of the Dirac constraints  $H_{\perp}(\tau )
\approx 0$, ${\cal H}_r(\tau ,\vec \sigma ) \approx 0$ of
Eqs.(\ref{II26}), (\ref{II28}) in the form \footnote{A quantity
$\psi (a; b]$ means a function of $a$ and a functional of $b$.}

\begin{eqnarray}
&&\widehat{H}_\perp\cdot\Phi\left(\vec{\eta}_i,\vec{k};\tau,\theta,
 {\cal A}^a\right]=0,\nonumber \\
&&\nonumber\\
&&\widehat{\cal
H}_r(\vec{\sigma})\cdot\Phi\left(\vec{\eta}_i,\vec{k};\tau,\theta,
 {\cal A}^a\right]=0,
 \label{III18}
\end{eqnarray}

\noindent whereas the evolution in the {\em time} $\tau$ is
determined by the Schroedinger equation

 \beq
i\hbar\frac{\partial}{\partial\tau}\Phi\left(\vec{\eta}_i,\vec{k};\tau,\theta,
 {\cal A}^a\right]=\widehat{H}_D\cdot\Phi\left(\vec{\eta}_i,\vec{k};\tau,\theta,
 {\cal A}^a\right] = 0,
 \label{III19}
  \eeq

\noindent since Eq.(\ref{II25}) with $\kappa (\tau ) = 0$ implies

\beq
 \widehat{H}_D=\mu(\tau)\cdot \widehat{H}_\perp+\int d^3\sigma
\lambda^r(\tau,\vec{\sigma})\cdot\widehat{\cal H}_r(\vec{\sigma}),
 \label{III20}
  \eeq

 Therefore we will use wave functions that satisfy this condition, that is
wave functions that do {\it not} depend explicitly on the {\em
time} $\tau$

 \beq
  \Phi\left(\vec{\eta}_i,\vec{k};\tau,\theta,
 {\cal A}^a\right]= \Phi\left(\vec{\eta}_i,\vec{k};\theta,
 {\cal A}^a\right].
 \label{III21}
 \eeq

\medskip

Using the rules i),ii),iii) we can obtain the explicit form of
Eqs.(\ref{III18}), (\ref{III19}). The only problem is to find
suitable pseudo-differential operators \cite{41,42} $\widehat{\cal
R}_i$ as the representative the square-roots  $\sqrt{m^2_i +
h^{rs}(\tau ,{\vec \eta}_i(\tau ))\, \kappa_{ir}(\tau )\,
\kappa_{is}(\tau ) }$ appearing in the constraint $H_\perp(\tau )
\approx 0$ in Eq.(\ref{II28})

\begin{equation}
\sqrt{m_i^2\,c^2+h^{rs}(\tau,\vec{\eta}_i)\kappa_{i\,r}(\tau)
\kappa_{i\,s}(\tau)}\rightarrow\,\widehat{\cal R}_i.
 \label{III22}
\end{equation}

Since this is a non trivial problem, let us start by replacing the
momenta $\kappa_{ir}(\tau )$ with the non-self-adjoint operators
$i \hbar\, {{\partial}\over {\partial \eta_i^r}}$ and by choosing
an ordering equivalent to replace the quadratic forms $
h^{rs}(\tau ,{\vec \eta}_i(\tau ))\, \kappa_{ir}(\tau )\,
\kappa_{is}(\tau )$ with the Laplace-Beltrami operator

 \beq
  \triangle_{\eta_i} = - \hbar^2\,
\frac{1}{\sqrt{h({\vec \eta}_i)}}\frac{\partial}{\partial\eta_i^r}
\left( h^{rs}({\vec \eta}_i)\sqrt{h({\vec \eta}_i)}
\frac{\partial}{\partial\eta_i^s}\right).
 \label{III23}
  \eeq

\noindent Therefore, we look for pseudo-differential operators
corresponding to the formal operators

\begin{equation}
\widehat{\cal R}_i\, f({\vec \eta}_i) = \sqrt{m_i^2\,c^2 - \hbar^2
\frac{1}{\sqrt{h({\vec \eta}_i)}}\,
\frac{\partial}{\partial\eta_i^r} \left( h^{rs}({\vec \eta}_i)\,
\sqrt{h({\vec \eta}_i)}
\frac{\partial}{\partial\eta_i^s}\right)}\, f({\vec \eta}_i).
 \label{III24}
\end{equation}

In Appendix C it is shown  that these formal operators can be
interpreted as pseudo-differential operators with the following
integral representation

\begin{eqnarray}
&&\widehat{\cal R}_i\, f({\vec \eta}_i) =\nonumber \\
 &&\nonumber\\
  &{\buildrel {def}\over =}&\frac{1}{(2\pi)^3}\,
  \int d^3K\, \sqrt{m_i^2\,c^2 + \vec{K}^2}\, \int
\sqrt{h({\vec \zeta}_i)}\,d^3\zeta_i \,f({\vec{\zeta}}_i)\,
e^{\frac{i}{\hbar}\vec{K}\cdot (\vec{\cal A }({\vec
\eta}_i)-\vec{\cal A}(\vec{\zeta}_i))}.\nonumber \\
 &&{}
 \label{III25}
 \end{eqnarray}

\hfill

With this definition, we can obtain the following explicit form of
Eqs.(\ref{III18}), that we will call {\em generalized Schroedinger
equations} (also in the second set of  these equations the
classical momenta have been replaced with the non-self-adjoint
operators $i \hbar\, {{\partial}\over {\partial \eta_i^r}}$)

\begin{eqnarray*}
 &&\widehat{H}_\perp\cdot\Phi\left(\vec{\eta}_i,\vec{k};\tau,\theta,
 {\cal A}^a\right]=
 \left( i\, \hbar\, {{\partial}\over {\partial \theta}} -
 \sum_{i=1}^N\, \widehat{\cal R}_i\right)\,
 \Phi\left(\vec{\eta}_i,\vec{k};\tau,\theta,
 {\cal A}^a\right]
 = 0, \end{eqnarray*}

\bea
 && \widehat{\cal H}_r(\vec{\sigma})\cdot\Phi\left(\vec{\eta}_i,\vec{k};\tau,\theta,
 {\cal A}^a\right]=\nonumber\\
&&\nonumber\\
&=&i\, \hbar\, \left( {{\partial {\cal A}^a(\vec \sigma )}\over {
 \partial \sigma^r}}\, {{\delta}\over {\delta {\cal A}^a(\vec \sigma )}} -
 \sgn\, \sum_{i=1}^N\, \delta^3(\vec \sigma - {\vec \eta}_i(\tau
 ))\, {{\partial}\over {\partial \eta^r_i}}\right)\,
 \Phi\left(\vec{\eta}_i,\vec{k};\tau,\theta,
 {\cal A}^a\right]= 0.
  \label{III26}
 \eea

\bigskip

The chosen ordering  and the definition given for the operators
$\widehat{\cal R}_i$ {\it guarantee the formal integrability} of
Eqs.(\ref{III26}), namely the validity of Eqs.(\ref{III3}) with
vanishing structure functions, since we have

\begin{eqnarray}
&&[\widehat{\cal H}_r(\vec{\sigma}) ,\widehat{\cal
H}_s(\vec{\sigma}')]=i\hbar\left[
\frac{\partial}{\partial\sigma^{\prime\,s}}
\delta^3(\vec{\sigma}-\vec{\sigma}')\, \widehat{\cal
H}_r(\vec{\sigma}') - \frac{\partial}{\partial\sigma^{s}}
\delta^3(\vec{\sigma}-\vec{\sigma}')\,
\widehat{\cal H}_s(\vec{\sigma})\right],\nonumber\\
 &&\nonumber\\
&&[\widehat{H}_\perp,\widehat{\cal H}_r(\vec{\sigma})]=0.
 \label{III27}
\end{eqnarray}

\bigskip

We can formalize the {\em generalized time evolution} introducing
a {\em space of generalized times}, parametrized with {\em time}
$\tau$ and with the {\em generalized times} $\theta$ and ${\cal
A}^a(\vec{\sigma})$. When topologically trivial, this {\em space
of generalized times} is the cartesian product ${\cal M}={\bf
R}\times{\bf R}\times C^\infty({\bf R}^3,{\bf R}^3)$ and its
points are represented by $(\tau,\theta,{\cal A}^a(\vec{\sigma}))$
[${\cal A}^a\in C^\infty({\bf R}^3,{\bf R}^3)$]. Then, the {\em
generalized temporal evolution} can be defined as a map from the
Cartesian product of the {\em space of generalized times} ${\cal
M}$ with the Hilbert space of the {\it initial states} ${\bf
H}_{{\cal A}_o}$ to the Hilbert space ${\bf H}_{\cal A}$

 \beq
 {\cal M}\times{\bf H}_{{\cal A}_o} \rightarrow{\bf H}_{\cal A},
\qquad \left[ (\tau,\theta,{\cal
A}^a(\vec{\sigma})),\Phi_o(\vec{\eta}_i,\vec{k})
\right]\rightarrow\Phi\left(\vec{\eta}_i,\vec{k};\tau,\theta,{\cal
A}^a\right],
 \label{III28}
  \eeq

\noindent with ($\theta_o = \theta{|}_{\tau =0}$, ${\cal
A}^a_o(\vec \sigma ) = {\cal A}^a(\vec \sigma ){|}_{\tau =0}$)

   \beq
\Phi\left(\vec{\eta}_i,\vec{k};0,\theta_o,{\cal
A}^a_o\right]=\Phi_o(\vec{\eta}_i,\vec{k}).
 \label{III29}
  \eeq
\medskip

As a consequence of the {\it zero curvature} condition
(\ref{III27}), the evolution from an initial time-configuration
$\Big(\tau_1, \theta_1, {\cal A}^a_1(\vec \sigma )\Big)$ to a
final one $\Big(\tau_2, \theta_2, {\cal A}^a_2(\vec \sigma )\Big)$
{\it does not depend upon the path in the generalized time space
joining the two time configurations} (see Refs.\cite{30,31}).

\medskip
Since the generalized Schroedinger equations (\ref{III26}) depend
neither on $\vec{k}$ nor on $\vec{k}$-derivatives, the {\em
generalized temporal evolution} is realized in the Hilbert space
$\widetilde{\bf H}_{\cal A}$. By construction any state in the
Hilbert space ${\bf H}_{{\cal A}_o}=L^2(\nu(\vec{k}),{\bf
R}^3)\times \widetilde{\bf H}_{{\cal A}_o}$ can be expanded on a
basis of factorized states

 \beq
\Phi\left(\vec{\eta}_i,\vec{k};\theta_o,
 {\cal A}_o^a\right]=\sum_{\lambda}
\,F_{\lambda}(\vec{k})\, \Psi_{\lambda}\left(\vec{\eta}_i;
\theta_o, {\cal A}_o^a\right].
 \label{III30}
  \eeq

Let us remark that $\lambda$ is a label for a basis of eigenstates
in $L^2(\nu (\vec k), R^3)$. In what follows for $\lambda$ we
always use the continuum basis labeled by $\vec k$
[$\sum_{\lambda} \mapsto \int d\nu (\vec k)$].

\medskip

Then the {\em generalized temporal evolution} in $\widetilde{\bf
H}_{\cal A}$ is determined by the following equation for
$\Psi_{\lambda}(\vec{\eta}_i; \theta , {\cal A}^a]$

 \bea
 &&\widehat{H}_\perp\cdot\Psi_\lambda\left(\vec{\eta}_i;\tau,\theta,
 {\cal A}^a\right]=
\left( i\, \hbar\, {{\partial}\over {\partial \theta}} -
\sum_{i=1}^N\, \widehat{\cal R}_i\right)\,
 \Psi_\lambda\left(\vec{\eta}_i;\theta,
 {\cal A}^a\right]
 = 0,\nonumber \\
 &&{}\nonumber \\
 && \widehat{\cal H}_r(\vec{\sigma})\cdot\Psi_\lambda\left(\vec{\eta}_i;\tau,\theta,
 {\cal A}^a\right]=\nonumber\\
&&\nonumber\\
&=&i\, \hbar\, \left( {{\partial {\cal A}^a(\vec \sigma )}\over {
 \partial \sigma^r}}\, {{\delta}\over {\delta {\cal A}^a}} -
 \sgn\, \sum_{i=1}^N\, \delta^3(\vec \sigma - {\vec \eta}_i(\tau
 ))\, {{\partial}\over {\partial \eta^r_i}}\right)\,
 \Psi_\lambda\left(\vec{\eta}_i;\theta,
 {\cal A}^a\right]= 0.
  \label{III31}
 \eea

\bigskip

By using Eqs.(\ref{III25}) and (\ref{III11}), (\ref{III12}) it can
be checked that the operator ${\widehat{\cal R}} = \sum_{i=1}^N\,
{\widehat{\cal R}}_i$ is {\it self-adjoint} in ${\widetilde{\bf
H}}_{\cal A}$, $(\Psi_1, {\widehat{\cal R}}\, \Psi_2) =
({\widehat{\cal R}}\, \Psi_1, \Psi_2)$. As a consequence the first
of Eqs.(\ref{III31}) implies ${{\partial}\over {\partial
\theta}}\, (\Psi_1, \Psi_2) = (\Psi_1, {\hat {\cal R}}\, \Psi_2) -
({\hat {\cal R}}\, \Psi_1, \Psi_2) = 0$, namely that the scalar
product is independent from the generalized time $\theta$.

Instead, if we rewrite the second of Eqs.(\ref{III31}) in the form
[${\cal A}^r_a(\vec \sigma )$ is defined in Eq.(\ref{c2})]

\beq
 {\cal A}^r_a(\vec{\sigma})\cdot\widehat{\cal
H}_r(\vec{\sigma})\cdot\Phi\left(\vec{\eta}_i,\vec{k};\tau,\theta,
 {\cal A}^a\right]
=i\, \hbar\, \left( {{\delta}\over {\delta {\cal A}^a}} -
 \sum_{i=1}^N\, \widehat{T}_{ia}(\vec \sigma )
\right)\,
 \Psi\left(\vec{\eta}_i;\theta,
 {\cal A}^a\right]= 0,
  \label{III32}
 \eeq

\noindent the operators

 \beq
  \widehat{T}_{ia}(\vec \sigma ) = i\,
\hbar\, {\cal A}^u_a({\vec \eta}_i)\, \delta^3(\vec \sigma - {\vec
\eta}_i)\, {{\partial}\over {\partial \eta_i^u}},
 \label{III33}
\eeq

\noindent {\it are not self-adjoint}, but satisfy

 \bea
 (\Psi_1, \widehat{T}_{ia}(\vec \sigma )\, \Psi_2) &=& (\widehat{
 T}_{ia}(\vec \sigma )\, \Psi_1, \Psi_2) + \nonumber\\
&&\nonumber\\
&+&i\, \hbar\,\int \prod_{j\neq i}
d^3\eta_j\,\sqrt{h(\vec{\eta}_j)} \int
 d^3\eta_i\, {{\delta \sqrt{h({\vec \eta}_i)}}\over {\delta {\cal
 A}^a(\vec \sigma )}}\,
\overline{\Psi}_1\left(\vec{\eta}_i;\theta ,{\cal A}^a\right] \,
\Psi_2 \left(\vec{\eta}_i;\theta ,{\cal A}^a\right], \nonumber \\
 &&{}
 \label{III34}
 \eea

\noindent since Eqs.(\ref{c1})-(\ref{c5}) imply ${{\delta
\sqrt{h({\vec \sigma}^{'})}}\over {\delta {\cal A}^a(\vec \sigma
)}} = - {\cal A}^u_a(\vec \sigma )\, \sqrt{h(\vec \sigma )}\,
{{\partial}\over {\partial \sigma^u}}\, \delta^3(\vec \sigma -
{\vec \sigma}^{'})$.

\medskip

As a consequence, in the generalized Schroedinger equations
(\ref{III32}) the effective Hamiltonians $\sum_{i=1}^N\, \widehat{
T}_{ia}(\vec \sigma )$ are not self-adjoint operators. But this is
not a problem, because Eqs.(\ref{III32}) and (\ref{III34}) imply
that {\it the scalar product is also independent from the
generalized times} ${\cal A}^a(\vec \sigma )$, because, due to the
time-dependent measure, we have

 \bea
 i\hbar\,{{\delta}\over {\delta {\cal A}^a(\vec \sigma )}}\, (\Psi_1,
 \Psi_2) &=& \sum_{i=1}^N\,(\Psi_1, \widehat{T}_{ia}(\vec \sigma )\,
 \Psi_2) -\sum_i\, (\widehat{
 T}_{ia}(\vec \sigma )\, \Psi_1, \Psi_2)+\label{III35}\\
&&\nonumber\\
 + i\, \hbar\,\sum_{i=1}^N\,\int&&
 \prod_{j\neq i} d^3\eta_j\,\sqrt{h(\vec{\eta}_j)} \int
 d^3\eta_i\, {{\delta \sqrt{h({\vec \eta}_i)}}\over {\delta {\cal
 A}^a(\vec \sigma )}}\,
\overline{\Psi}_1\left(\vec{\eta}_i;\theta ,{\cal A}^a\right] \,
\Psi_2 \left(\vec{\eta}_i;\theta ,{\cal A}^a\right]=0.\nonumber
\eea

As a consequence, we get a  scalar product independent from both
the time and the generalized times

\begin{equation}
 {{\partial}\over {\partial \tau}}\, (\Psi_1, \Psi_2) =
\frac{\partial}{\partial \theta}\, (\Psi_1,\Psi_2)=
\frac{\delta}{\delta {\cal A}^a(\vec{\sigma})} (\Psi_1,\Psi_2)=0,
 \label{III36}
\end{equation}

\noindent and, as it will be shown explicitly in Subsection IIIB5,
this implies that all the Hilbert spaces ${\widetilde{\bf
H}}_{\cal A}$ are isomorphic.

\subsubsection{A Frame-Independent Hilbert Space with all the Hamiltonians
Self-Adjoint.}

Eqs.(\ref{III36}) suggest that it must be possible to reformulate
the multi-temporal quantization scheme in a frame-independent
Hilbert space ${\mbox{\boldmath{${\cal H}$}}} = L^2(\nu (\vec k), R^3) \times
{\widetilde{\mbox{\boldmath{${\cal H}$}}}}$ with wave functions

\beq
 \widehat{\Phi} ({\vec \eta}_i, \vec k) = \prod_i\, \sqrt{\det\, \Big({{\partial
 {\cal A}^a({\vec \eta}_i)}\over {\partial \eta^r_i}}\Big)}\, \Phi ({\vec
 \eta}_i, \vec k),
 \label{III37}
 \eeq

 \noindent and with scalar product (now the classical momenta $\kappa_{ir}$ are replaced
 by the self-adjoint operators $i \hbar\, {{\partial}\over {\partial \eta^r_i}}$)

 \bea
  \langle \widehat{\Phi}_1, \widehat{\Phi}_2 \rangle &=& \int d \nu (\vec k)\, \int
  d \mu({\vec \eta}_i)\, \overline{\widehat{\Phi}}_1({\vec \eta}_i, \vec
  k)\, \widehat{\Phi}_2({\vec \eta}_i, \vec k),\nonumber \\
&&\nonumber\\
  &&d \mu ({\vec \eta}_i) = \prod_i\, d^3\eta_i.
  \label{III38}
  \eea
\medskip

As a consequence, as shown at the end of Appendix C, the coupled
Schroedinger equations (\ref{III31}) or (\ref{III32}) are replaced
by the following ones all containing {\it self-adjoint}
Hamiltonian operators ($[\hat A, \hat B]_{+} = \hat A\, \hat B +
\hat B\, \hat A$)

\medskip

\bea
 &&\widehat{H}^{'}_{\perp}\, \widehat{\Phi} ({\vec \eta}_i, \vec k; \tau ,
 \theta , {\cal A}^a] =\nonumber\\
&&\nonumber\\
  &&=\sqrt{\prod_i\,\det\, \Big({{\partial
 {\cal A}^a({\vec \eta}_i)}\over {\partial \eta^r_i}}\Big)}\,
 \widehat{H}_{\perp}\, {1\over { \sqrt{\prod_i\,\det\, \Big({{\partial
 {\cal A}^a({\vec \eta}_i)}\over {\partial \eta^r_i}}\Big)}\,}}\,
 \widehat{\Phi}
 ({\vec \eta}_i, \vec k; \tau , \theta ,
 {\cal A}^a] =\nonumber  \\
&&\nonumber\\
 &&=\Big( i\hbar\, {{\partial}\over {\partial  \theta}} -
 \sum_{i=1}^N\, {\widehat{\cal R}}^{'}_i \Big)\,
 \widehat{\Phi} ({\vec \eta}_i, \vec k; \tau , \theta ,
 {\cal A}^a],\nonumber \\
 &&{}\nonumber \\
&&\nonumber\\
 &&{\widehat{\cal H}}^{'}_a(\vec \sigma )\, \widehat{\Phi} ({\vec \eta}_i,
 \vec k; \tau , \theta ,{\cal A}^a] =\nonumber\\
&&\nonumber\\
  &&=\sqrt{\prod_i\,\det\, \Big({{\partial
 {\cal A}^a({\vec \eta}_i)}\over {\partial \eta^r_i}}\Big)}\,
 {\cal A}^r_a(\vec \sigma )\, {\widehat{\cal H}}_r(\vec \sigma )\,
 {1\over { \sqrt{\prod_i\,\det\, \Big({{\partial
 {\cal A}^a({\vec \eta}_i)}\over {\partial
 \eta^r_i}}\Big)}\,}}\widehat{\Phi}
 ({\vec \eta}_i, \vec k; \tau ,\theta , {\cal A}^a]
 =\nonumber \\
&&\nonumber\\
 &&=\Big(i\hbar\, {{\delta}\over {\delta {\cal A}^a(\vec \sigma )}}
 - {{\sgn}\over 2}\, \sum_{i=1}^N\, \Big[
 {\cal A}^r_a(\vec \sigma )\, \delta^3(\vec \sigma - {\vec \eta}_i),
 i\hbar\, {{\partial}\over {\partial \eta^r_i}}\Big]_{+}
 \Big) \widehat{\Phi} ({\vec \eta}_i, \vec k; \tau , \theta ,
 {\cal A}^a] = 0,
 \label{III39}
 \eea

\medskip

\noindent where ${\widehat{\cal R}}_i^{'}$ are new
pseudo-differential operators defined in Eqs. (\ref{c25}) and
(\ref{c26}).
\medskip

Eq.(\ref{III19}) is still satisfied with $\widehat{H}^{'}_D =
\mu(\tau )\, \widehat{H}^{'}_{\perp} + \int d^3\sigma\,
\lambda_a(\tau ,\vec \sigma )\, \widehat{\cal H}^{'}_a(\tau
,\vec \sigma )$ , so that we have $\widehat{\Phi}
= \widehat{\Phi}({\vec \eta}_i, \vec k; \theta , {\cal A}^a]$.

Instead Eq.(\ref{III27}) is replaced by

\beq
 [\widehat{H}^{'}_{\perp}(\tau ), {\widehat{\cal H}}^{'}_a(\tau ,\vec
 \sigma )] = [{\widehat{\cal H}}^{'}_a(\tau ,\vec \sigma ), {\widehat
 {\cal H}}^{'}_b(\tau ,{\vec \sigma}_1)] = 0.
 \label{III40}
 \eeq

\bigskip

As a consequence the transition from the frame-dependent  Hilbert
spaces ${\bf H}_{\cal A}$ to the frame-independent Hilbert space
${\mbox{\boldmath{${\cal H}$}}}$ is equivalent to replace the previous operator
ordering with a new one, which turns out to be the {\it
symmetrization ordering rule} for transforming classical
quantities in operators. Let us remark that the symmetrized
operators depend on the (now) self-adjoint operators $i \hbar\,
{{\partial}\over {\partial \eta_i^r}}$.
\medskip

Finally, by using the analogue  of the expansion (\ref{III30}),
Eqs.(\ref{III36}) become [$\Big(\widehat{\Psi}_1, \widehat {\Psi}_2\Big)
= \int d \mu({\vec \eta}_i)\,
\overline{\widehat{\Psi}}_1({\vec \eta}_i)
\widehat{\Psi}_2({\vec \eta}_i)$]

\beq
 {{\partial}\over {\partial \tau}}\, \Big(\widehat{\Psi}_1, \widehat
 {\Psi}_2\Big) = {{\partial}\over {\partial \theta}}\, \Big(\widehat
 {\Psi}_1, \widehat{\Psi}_2\Big) = {{\delta}\over {\delta {\cal
 A}^a(\vec \sigma )}}\, \Big(\widehat{\Psi}_1, \widehat{\Psi}_2\Big) =
 0.
 \label{III41}
 \eeq

\subsubsection{Representation of Poincar\'e Group}

Differently from the {\em generalized time evolution} the action
of the Poincar\'e group has to be analyzed in the complete space
${\bf H}_{\cal A}$ since the infinitesimal canonical generator of
this group depend of $\vec{k}$ and $\vec{z}$. If we follow the
quantization rules i),ii) and iii) we would have, from
Eqs.(\ref{II42}), the following expression of the quantum
Poincare' generators

\begin{eqnarray*}
 \widehat{p}^{\mu} &=& {\hat U}^{\mu}(\vec k)
\left[1 - i\hbar\,{{\partial}\over {\partial
 \theta}}\right]\,  + i\hbar\,
 \epsilon^{\mu}_a(\hat U(\vec k))\, \int d^3\sigma\,
 {{\delta}\over {\delta {\cal A}^a(\vec \sigma )}},
 \nonumber \\
  &&{}\nonumber \\
 \widehat{J}^{\mu\nu} &=& {\hat L}^{\mu\nu} - D^{\mu\nu}_{ab}(U(\vec k))\,
\int d^3\sigma\, \Big({\cal A}^a(\vec \sigma )\, i\, \hbar\,
 {{\delta}\over {\delta {\cal A}^b(\vec \sigma )}} -
{\cal A}^b(\vec \sigma )\, i\, \hbar\,
 {{\delta}\over {\delta {\cal A}^a(\vec \sigma )}}
 \Big),
\end{eqnarray*}

 \bea
 &&\widehat{L}^{ij} = i\, \hbar\, (k^i\, {{\partial}\over {\partial
 k^j}} - k^j\, {{\partial}\over {\partial k^i}}),\nonumber \\
&&\nonumber\\
 &&\widehat{L}^{oi} = - \widehat{L}^{io} = - {{i\hbar}\over 2}\, \Big[
 \sqrt{1 + {\vec k}^2}, {{\partial}\over {\partial k^i}}\Big]_{+}
 = - i\hbar\, \sqrt{1 + {\vec k}^2}\, {{\partial}\over {\partial
 k^i}} - {{i\hbar}\over 2}\, {{k^i}\over {\sqrt{1 + {\vec
 k}^2}}}.
 \label{III42}
 \eea

These expressions have the unpleasant feature of containing
explicitly (functional) derivative in the generalized times. This
fact can produces some difficulties: for example the (functional)
derivative in the generalized times are not operators on the
Hilbert space ${\bf H}_{\cal A}$ and then the previous
infinitesimal generator are not a representation  of the
Poincar\'e Lie algebra on ${\bf H}_{\cal A}$. To avoid these
difficulties we can observe that the physically important case is
that one where the wave function depend on the generalized times
as solutions of the generalized Schroedinger equations, that is
when they are evaluated {\em on shell}. In this case we can
substitute the (functional) derivative in the generalized times
with the corresponding {\it Hamiltonians} in the quantum
infinitesimal generators of Poincar\'e group. We obtain the
following expression for the self-adjoint generators

\begin{eqnarray}
 \widehat{p}^{\mu}_{(on)} &=& {\hat U}^{\mu}(\vec k)
\left[1 + i\hbar\,\sum_i\widehat{\cal R}_i\right]\,  - i\hbar\,
 \epsilon^{\mu}_a(\hat U(\vec k))\,\sum_i\,
 {\cal A}^r_a(\vec{\eta}_i)\, \frac{\partial}{\partial \eta^r_i},
 \nonumber \\
  &&{}\nonumber \\
 \widehat{J}^{\mu\nu}_{(on)} &=& \widehat{L}^{\mu\nu} - D^{\mu\nu}_{ab}(U(\vec k))\,
i\hbar\,\sum_i \left({\cal A}^a(\vec{\eta}_i )\,
 {\cal A}^r_b(\vec{\eta}_i)\frac{\partial}{\partial \eta^r_i}-
{\cal A}^b(\vec{\eta}_i )\, {\cal
A}^r_a(\vec{\eta}_i)\frac{\partial}{\partial \eta_i} \right).
\label{III43}
\end{eqnarray}

These operators are a  representation of the Poincar\'e Lie
algebra on the Hilbert space ${\bf H}_{\cal A}$.
\bigskip

Instead, in the frame-independent Hilbert space
${\mbox{\boldmath{${\cal H}$}}}$ the
form of the Poincare' generators as self-adjoint operators is

\bea
  \widehat{p}^{\mu}_{(on)} &=& {\hat U}^{\mu}(\vec k)
\left[1 + i\hbar\,\sum_i\widehat{\cal R}^{'}_i\right]\,  -
i\hbar\,
 \epsilon^{\mu}_a(\hat U(\vec k))\,\sum_i\, \Big(
 {\cal A}^r_a(\vec{\eta}_i)\, \frac{\partial}{\partial \eta^r_i} +
 {1\over 2}\, {{\partial\, {\cal A}^r_a(\vec{\eta}_i)}\over {\partial \eta^r_i}}\Big),
 \nonumber \\
  &&{}\nonumber \\
 \widehat{J}^{\mu\nu}_{(on)} &=& \widehat{L}^{\mu\nu} - D^{\mu\nu}_{ab}(U(\vec k))\,
i\hbar\cdot\nonumber \\
 &\cdot&\sum_i\, \Big({\cal A}^a(\vec{\eta}_i )\, \Big[
{\cal A}^r_b(\vec{\eta}_i)\, \frac{\partial}{\partial \eta^r_i} +
 {1\over 2}\, {{\partial\, {\cal A}^r_b(\vec{\eta}_i)}\over {\partial \eta^r_i}}
 \Big] - {\cal A}^b(\vec{\eta}_i )\,
\Big[ {\cal A}^r_a(\vec{\eta}_i)\, \frac{\partial}{\partial
\eta^r_i} + {1\over 2}\, {{\partial\, {\cal
A}^r_a(\vec{\eta}_i)}\over {\partial \eta^r_i}} \Big] \Big).\nonumber\\
 \label{III43b}
 \eea
\medskip

In both cases it can be checked that these generators satisfy the
Poincare' algebra.

\subsubsection{Generalized Temporal Evolution  and Plane Wave Solutions.}

Let us come back to the coupled generalized Schroedinger equations
(\ref{III31}).

We want show that any solution of Eqs.(\ref{III31}) can be
obtained by mapping an {\it initial state} $\Psi_o(\vec{\eta}_i)$,
at $\theta_o,\vec{\cal A}_o(\vec{\sigma})$ in the Hilbert space
$\widetilde{\bf H}_{{\cal A}_o}$, to a {\it final state}
$\Psi_\lambda\left(\vec{\eta}_i;\theta,{\cal A}^a\right]$, at
$\theta,\vec{\cal A}(\vec{\sigma})$ in the Hilbert space
$\widetilde{\bf H}_{\cal A}$ with a {\it isometry} ${\cal
J}[\theta,{\cal A};\theta_o,{\cal A}_o]$ from $\widetilde{\bf
H}_{{\cal A}_o}$ to $\widetilde{\bf H}_{\cal A}$, implying that
all the Hilbert spaces ${\bf H}_{\cal A}$ are isomorphic. In the
reformulation in the frame-independent Hilbert space
$\widetilde{\mbox{\boldmath{${\cal H}$}}}$ the isometry ${\cal J}[\theta ,{\vec {\cal A}};
\theta_o, {\vec {\cal A}}_o]$ becomes a {\it unitary operator}
${\widehat{\cal J}}[\theta ,{\vec {\cal A}}; \theta_o, {\vec {\cal
A}}_o]$.

\bigskip

We search a solution, $\Psi_\lambda\left(\vec{\eta}_i;\theta,{\cal
A}^a\right]$, of Eqs.(\ref{III31}), which satisfies the boundary
condition

 \beq
\Psi_\lambda\left(\vec{\eta}_i;\theta_o,{\cal
A}^a_o\right]=\Psi_o(\vec{\eta}_i),
 \label{III44}
  \eeq

\noindent when it is evaluated at the generalized times
$\theta=\theta_o,{\cal A}^a(\vec{\sigma})= {\cal
A}^a_o(\vec{\sigma})$. The second set of Eqs.(\ref{III31}) have a
general solution that restricts the dependence of the wave
functions on the variables ${\vec {\cal A}}(\vec \sigma )$ and
${\vec \eta}_i$ to the following form

\begin{equation}
\Psi_{\lambda}\left(\vec{\eta}_i;\theta , {\cal A}^a\right] =
\widetilde{\Psi}_{\lambda}(\theta ,{\cal A}
^a(\vec{\eta}_1),\,...\,,{\cal A} ^a(\vec{\eta}_N) ).
 \label{III45}
\end{equation}

Instead the first of Eqs. (\ref{III31}) can be solved using the
unitary operator on $\widetilde{\bf H}_{\cal A}$

\begin{equation}
{\cal U}(\theta,\theta_o)
=\exp\left[-\frac{i}{\hbar}\,(\theta-\theta_o)\sum_{i=1}^N\,
\widehat{\cal R}_i\right].
 \label{III46}
\end{equation}

In other words, we can formally write a solutions of the
generalized Schroedinger equations (\ref{III31}) as

 \beq
\Psi_\lambda\left(\vec{\eta}_i;\theta,{\cal A}^a\right]= {\cal
U}(\theta,\theta_o)
\cdot\Psi_\lambda\left(\vec{\eta}_i;\theta_o,{\cal A}^a\right],
 \label{III47}
  \eeq

\noindent  where

 \beq
\Psi_\lambda\left(\vec{\eta}_i;\theta_o,{\cal A}^a\right]
=\widetilde{\Psi}_\lambda(\theta_o, {\cal
A}^a(\vec{\eta}_1),\,...\,,{\cal A}^a(\vec{\eta}_N)).
 \label{III48}
  \eeq

The boundary condition (\ref{III44}) becomes

 \beq
\widetilde{\Psi}_\lambda (\theta_o,{\cal
A}_o^a(\vec{\eta}_1),\,...\,,{\cal A}_o^a(\vec{\eta}_N)
)=\Psi_o(\vec{\eta}_i).
 \label{III49}
  \eeq

Since the functions ${\cal A}^a = {\cal A}^a_o(\vec{\sigma})$ are
assumed to be invertible to $\sigma^r = S^r({\vec {\cal A}})$,
there is a unique function

 \beq
\widetilde{\Psi}_\lambda(\theta_o,\vec{X}_1,\,....\,,\vec{X}_N)=
\Psi_o(\vec{S}(\vec{X}_i)),
 \label{III50}
  \eeq

\noindent  so that we have the unique solution \footnote{Since
$S^r_o({\vec {\cal A}})$ is the inverse  of ${\cal
A}_o(\vec{\sigma})$ and not of ${\cal A}(\vec{\sigma})\neq{\cal
A}_o(\vec{\sigma})$, then in this equation  we have
$\vec{S}_o(\vec{\cal A}(\vec{\eta}_i))\neq \vec{\eta}_i$. }

 \beq
\Psi_\lambda\left(\vec{\eta}_i;\theta_o,{\cal A}^a\right]
=\Psi_o(\vec{S}_o(\vec{\cal A}(\vec{\eta}_i))).
 \label{III51}
  \eeq
\medskip

Eqs.(\ref{III51}) define a map between the Hilbert space
$\widetilde{\bf H}_{{\cal A}_o}$ and the Hilbert space
$\widetilde{\bf H}_{\cal A}$. Since such map is defined by solving
the generalized temporal evolution determined by the generalized
Schroedinger equations, Eqs.(\ref{III36}) imply that this map is a
isometry ${\cal I}[{\cal A},{\cal A}_o]$

 \beq
  {\cal I}[{\cal
A},{\cal A}_o]:\widetilde{\bf H}_{{\cal A}_o}\mapsto\widetilde{\bf
H}_{\cal A},\qquad \Psi_{\lambda}({\vec \eta}_i; \theta ,{\cal
A}^a) = {\cal I}[{\cal A}, {\cal A}_o]\, \Psi_o({\vec \eta}_i),
 \label{III52}
  \eeq

\noindent because we have

 \beq
  \Psi_1,\Psi_2\in {\tilde {\bf H}}_{{\cal
A}_o}\;\;\Rightarrow\;\; ({\cal I}[{\cal A},{\cal
A}_o]\cdot\Psi_1,{\cal I}[{\cal A},{\cal
A}_o]\cdot\Psi_2)=(\Psi_1,\Psi_2).
 \label{III53}
  \eeq

Since ${\cal U}(\theta,\theta_o)$ is a unitary transformation on
$\widetilde{\bf H}_{\cal A}$, also the product

 \beq
  {\cal
J}[\theta,{\cal A};\theta_o,{\cal A}_o]= {\cal
U}(\theta,\theta_o)\cdot {\cal I}[{\cal A},{\cal A}_o],
 \label{III54}
  \eeq

\noindent is an isometry

 \beq
  {\cal J}[\theta,{\cal
A};\theta_o,{\cal A}_o]: \widetilde{\bf H}_{{\cal
A}_o}\mapsto\widetilde{\bf H}_{\cal A}.
 \label{III55}
 \eeq

Then we can conclude that the general solution of
Eqs.(\ref{III31}) can be realized with the isometry (\ref{III55}):
$\Psi_\lambda\left(\vec{\eta}_i;\theta,{\cal A}^a\right]={\cal
J}[\theta,{\cal A};\theta_o,{\cal A}_o]\cdot\Psi_o(\vec{\eta}_i)$,
which explicitly realizes the isomorphism of these Hilbert spaces.

\hfill

In the frame-independent Hilbert space ${\widetilde{\mbox{\boldmath{${\cal H}$}}}}$
with wave functions $\widehat{\Psi}_{\lambda}({\vec \eta}_i; \theta ,
{\vec {\cal A}}]$, the same type of discussion leads to the
unitary operators

\bea
 &&\widehat{\cal J}[\theta ,{\vec {\cal A}}; \theta_o, {\vec {\cal
 A}}_o]: {\widetilde{\mbox{\boldmath{${\cal H}$}}}} \mapsto
 {\widetilde{\mbox{\boldmath{${\cal H}$}}}},\nonumber \\
 &&{}\nonumber \\
&&{\widehat{\cal J}}[\theta ,{\vec {\cal A}}; \theta_o, {\vec {\cal
A}}_o] = {\widehat{\cal U}}(\theta ,\theta_o)\, {\widehat{\cal I}}[{\vec
{\cal A}}; {\vec {\cal A}}_o],\qquad {\widehat{\cal I}}[{\vec {\cal
A}}, {\vec {\cal A}}_o]: {\widetilde{\mbox{\boldmath{${\cal H}$}}}} \mapsto
{\widetilde{\mbox{\boldmath{${\cal H}$}}}},\nonumber  \\
&&{}\nonumber \\
&&{\widehat{\cal U}}(\theta ,\theta_o) = \exp\, \Big[ - {i\over
{\hbar}}\, (\theta - \theta_o)\, \sum_{i=1}^N\, {\hat {\cal
R}}^{'}_i\Big] : {\widetilde{\mbox{\boldmath{${\cal H}$}}}}
\mapsto {\widetilde{\mbox{\boldmath{${\cal H}$}}}}.
 \label{III56}
 \eea

\bigskip

We can also observe that, using Eq.(\ref{III45}), the first of
Eqs.(\ref{III31}) can be written as a condition on the
$\widetilde{\Psi}$ in the form

 \bea
 i\hbar\frac{\partial}{\partial\theta}
 \widetilde{\Psi}_{\lambda}(\theta;{\cal A}
^a_1,\,...\,,{\cal A}^a_N)&=& \sum_{i=1}^N\,\sqrt{m_i^2c^2-\hbar^2
\frac{\partial}{\partial {\cal A}^a_i} \delta^{ab}
\frac{\partial}{\partial {\cal A}^b_i}}\;
\widetilde{\Psi}_{\lambda}(\theta;{\cal
A}^a_1,\,...\,,{\cal A}^a_N),\nonumber \\
 &&{}
 \label{III57}
  \eea

\noindent which imply the following {\em  plane wave} solutions of
the generalized Schroedinger equations (${\vec K}_1,..,{\vec K}_N$
are $N$ constant vectors; $A_{\lambda}$ is a normalization
constant)

\bea
 \widetilde{\Psi}_{\lambda| K_1,..,K_N}\left(\vec{\eta}_i;\theta
,{\cal A}^a\right] &=& A_{\lambda}\, \prod_i^{1..N}\,
\frac{1}{(2\pi)^{3/2}}\, \exp\left[ -\frac{i}{\hbar}\, \left(
\theta\sqrt{m_i^2c^2+\vec{K}_i^2}\pm \vec{K}_i\cdot\vec{\cal A
}(\vec{\eta}_i) \right) \right],\nonumber \\
&&\nonumber\\
 \widehat{\widetilde{\Psi}}_{\lambda |K_1,..,K_N}({\vec \eta}_i;
 \theta ,{\vec {\cal A}}] &=& \sqrt{\prod_i\, \det\, \Big({{\partial
 {\cal A}^a({\vec \eta}_i)}\over {\partial \eta^r_i}}\Big)}\,
 \widetilde{\Psi}_{\lambda |K_1,..,K_N}({\vec \eta}_i; \theta ,{\vec
 {\cal A}}].\nonumber \\
 &&{}
 \label{III58}
\eea

\subsubsection{The Choice of a Non-Inertial Frame}

In the classical theory the selection of a non-inertial frame,
i.e. of a congruence of non inertial observers, is done in  two
steps:

i) with a first class constraint we select a unique value $\vec{k}
\approx \vec{k}_o=const.$ for the momentum of the  extra particle,
because this selects a family of parallel hyperplane orthogonal to
${\hat U}^{\mu}({\vec k}_o)$; this amounts to eliminate the extra
particle (now a gauge object), eventually by adding the gauge
fixing $\vec z \approx 0$;

ii) then we fix the gauge variables $\theta(\tau),\vec{\cal
A}(\tau,\vec{\sigma})$ with suitable gauge fixings.
\medskip

In the quantum theory the first step corresponds to select a
eigenspace of ${\hat {\vec k}}$ [we choose the basis $\lambda =
\vec k$] in ${\bf H}_{\cal A}$ corresponding to the eigenvalue
$\vec{k}_o$: we call this eigenspace $\widetilde{\bf H}_{k_o,{\cal
A}}$ and its states have the form

 \beq
\Phi_{k_o}(\vec{\eta}_i,\vec{k})=\Delta(\vec{k},\vec{k}_o)
\cdot\Psi_{k_o}\left(\vec{\eta}_i;\theta,{\cal A}^a\right],
 \label{III59}
  \eeq

\noindent where $\Delta (\vec k, {\vec k}_o) = 2\, \sqrt{1 + {\vec
k}_o^2}\, \delta^3(\vec k - {\vec k}_o)$. This is the covariant
delta function satisfying $\int d\nu (\vec k)\, \Delta (\vec k,
{\vec k}_o)\, f(\vec k) = f({\vec k}_o)$ and $\Delta (\overrightarrow{\Lambda\,k}, {\vec k}_o) = \Delta
(\vec k, \overrightarrow{\Lambda^{-1}\, k_o})$,
where $\lambda$ is a Lorentz transformation.

\medskip

In the eigenspace $\widetilde{\bf H}_{k_o,{\cal A}}$ the scalar
product of ${\bf H}_{\cal A}$ diverges, because $\vec k = {\vec
k}_o$ is an eigenvalue of the continuous spectrum

 \beq
\langle \Phi_{k_o,1}, \Phi_{k_o,2}\rangle=
2\sqrt{1+\vec{k}_o^2}\,\delta(0)\,(\Psi_{k_o,1},\Psi_{k_o,2}).
\label{III60}
 \eeq

\noindent However, we have to go to the quotient (the reduced
phase space at the classical level) with respect to the extra
particle and this implies that we must use the well defined scalar
product of $\widetilde{\bf H}_{\cal A}$, that is
$(\Psi_1,\Psi_2)$. In conclusion we must restrict ourselves  to
the eingenspace $\widetilde{\bf H}_{k_o,{\cal A}}$, isomorphic to
$\widetilde{\bf H}_{\cal A}$.

This step {\it breaks the Lorentz covariance} of the quantum
theory. Actually,  the eingenspace $\widetilde{\bf H}_{k_o,{\cal
A}}$ is not invariant under Poincar\'e transformations and there
is not a representation of the Poincar\'e group on it. However we
can interpret the action of a Poincar\'e transformation as an
isometry from a eigenspace $\widetilde{\bf H}_{k_o,{\cal A}}$ to a
another eingenspace $\widetilde{\bf H}_{k'_o,{\cal A}}$.

\hfill

To realize the second step at quantum level, we must define a path
(labeled by an index $c$)

 \beq
  {\cal P}_c(\tau)=(\tau,\theta_c(\tau),{\cal
A}^a_c(\tau,\vec{\sigma})),
 \label{III61}
  \eeq

\noindent connecting two points $(\tau_o,\theta_o,{\cal
A}^a_o(\vec{\sigma}))$ and $(\tau_f,\theta_f,{\cal
A}^a_f(\vec{\sigma}))$ of the space ${\cal M}$ of generalized
times. The index $c$ means that we have restricted ourselves to
the evolution between $\tau_o$ and $\tau_f$ in a foliation with
hyper-planes whose normal is $\widehat{U}^{\mu}({\vec k}_o)$, i.e.
to a non-inertial frame where $\theta_o = \theta_c(\tau_o)$,
$\theta_f = \theta_c(\tau_f)$, ${\vec {\cal A}}_o(\vec \sigma ) =
{\vec {\cal A}}_c(\tau_o, \vec \sigma )$, ${\vec {\cal A}}_f(\vec
\sigma ) = {\vec {\cal A}}_c(\tau_f, \vec \sigma )$.

\medskip

For the non-inertial observer, whose world-line is the origin of
the observer-dependent coordinates $(\tau ,\vec \sigma )$ adapted
to the foliation with hyper-planes $z^\mu = {\hat
U}^\mu(\vec{k}_o)\, \theta_c(\tau) + \epsilon^\mu_a(\vec{k}_o)\,
{\cal A}^a_c(\tau,\vec{\sigma})$, the {\em effective} wave
function will be the wave function $\Psi_{{\vec
k}_o}\left(\vec{\eta}_i;\theta,{\cal A}^a\right]$ evaluated along
the path ${\cal P}(\tau)$

\begin{equation}
\psi_c(\tau,\vec{\eta}_i)= \Psi_{ {\vec
k}_o}\left(\vec{\eta_i};\theta_c(\tau),{\cal A}^a_c(\tau)\right].
 \label{III62}
\end{equation}

Since we have

\begin{eqnarray}
i\hbar\,\frac{\partial}{\partial\tau}\,
\psi_c(\tau,\vec{\eta}_i)&=& i\hbar\, \dot \theta (\tau )\, \left[
\frac {\partial\Psi_{{\vec k}_o}}
{\partial\theta}\right]\left(\vec{\eta_i},\theta_c(\tau );{\cal
A}_c^a(\tau)\right] +\nonumber\\
 &&\nonumber\\
  &+&i\hbar\, \int d^3\sigma\,
\frac{\partial {\cal A}^a_c(\tau,\vec{\sigma})}{\partial\tau}\,
 \left[ \frac{\delta\Psi_{{\vec k}_o}} {\delta {\cal A}
^a(\vec{\sigma})}\right] \left(\vec{\eta_i},\theta_c(\tau );{\cal
A} ^a_c(\tau)\right],
 \label{III63}
\end{eqnarray}

\noindent we see that Eqs.(\ref{III31}) {\it imply the following
effective Schroedinger equation along the path} ${\cal P}(\tau )$
in the generalized time parameter space

\begin{eqnarray*}
 i\hbar\,\frac{\partial}{\partial\tau}\, \psi_c(\tau,\vec{\eta}_i)&=&
\left[\dot \theta (\tau )\, \sum_{i=1}^N\, \widehat{\cal R}_i\, +
\sum_{i=1}^N\, V^r(\tau,\vec{\eta}_i)\, i\hbar\, \frac{\partial}
{\partial\eta^r_i}\right]\, \psi_c(\tau,\vec{\eta}_i)
=\nonumber \\
&&\nonumber\\
 &{\buildrel {def}\over =}& \widehat{H}_{ni}\cdot \psi_c(\tau ,{\vec
 \eta}_i),\end{eqnarray*}

\bea
 &&  V^r(\tau,\vec{\sigma})={\cal A}^r_{c\, a}(\tau,\vec{\sigma})
\frac{\partial {\cal A}^a_c(\tau,\vec{\sigma})}{\partial\tau}.
 \label{III64}
\end{eqnarray}

{\it The effective Hamiltonian operator $\widehat{H}_{ni}$ is just
the quantized version of the effective non-inertial Hamiltonian}
$H_{ni}$  of Eq.(\ref{II44}) and the {\it generalized inertial
forces} are generated by the potential $\sum_{i=1}^N\, V^r(\tau
,{\vec \eta}_i(\tau ))\, i \hbar\, {{\partial}\over {\partial
\eta_i^r}}$.

\hfill

For each value of $\tau$, $\psi_c(\tau,\vec{\eta}_i)$ is a state
in the Hilbert space

 \beq
  \widetilde{\bf H}_{k_o,\tau}=\widetilde{\bf
H}_{k_o,{\cal A}_c(\tau)},
 \label{III65}
  \eeq

\noindent with a scalar product with a $\tau$-dependent measure

 \bea
  &&d\mu_c(\tau,\vec{\eta}_i)=\prod_i\,d^3\eta_i\,\det\left(
\frac{\partial{\cal A}^a_c(\tau,\vec{\eta}_i)}{\partial\eta^r_i}
\right),\nonumber\\
&&\nonumber\\
&& (\psi_{1,c}(\tau),\psi_{2,c}(\tau))=\int
d\mu_c(\tau,\vec{\eta}_i)\,
\overline{\psi}_{1,c}(\tau,\vec{\eta}_i)\,\psi_{2,c}(\tau,\vec{\eta}_i).
\label{III66}
 \eea

We can see that the effective $\tau$-dependent non inertial
Hamiltonian defined in Eq.(\ref{III64}) is {\it not self-adjoint}
in $\widetilde{\bf H}_{k_o,\tau}$. However, it follows from the
results of Subsection IVB5 that, due to the $\tau$-dependent
measure, the effective $\tau$-evolution still defines an isometry
between the initial state $\psi_c(\tau_o,\vec{\eta}_i)\,
\in\,\widetilde{\bf H}_{k_o,\tau_o}$ and the final state
$\psi_c(\tau_f,\vec{\eta}_i)\,\in\,\widetilde{\bf H}_{k_o,\tau_f}$

\beq
 {\cal J}(\tau_f,\tau_o):\widetilde{\bf
H}_{k_o,\tau_o}\mapsto\widetilde{\bf H}_{k_o,\tau_1}.
 \label{III67}
  \eeq

\noindent Indeed we have

 \bea
  &&{\cal J}(\tau_f,\tau_o)=\exp\left[-\frac{i}{\hbar}\,
(\theta_c(\tau_f)-\theta_c(\tau_o))
\right]\cdot{\cal I}(\tau_f,\tau_o),\nonumber\\
&&\nonumber\\
&& {\cal I}(\tau_f,\tau_o)={\cal I}[{\cal A}_f,{\cal A}_o],
 \label{III68}
  \eea

\noindent and

\begin{equation}
\frac{d}{d\tau}(\psi_{1c},\psi_{2c})=0.
 \label{III69}
 \end{equation}

\bigskip

The discussion in the frame-independent Hilbert space
${\mbox{\boldmath{${\cal H}$}}}$ is analogous. We select $\vec k = {\vec k}_o$ and we restrict
ourselves to the Hilbert space ${\widetilde{\mbox{\boldmath{${\cal H}$}}}}_{\lambda =
{\vec k}_o}$ with the scalar product $\Big(\widehat{\Psi}_1, \widehat
{\Psi}_2\Big)$. Then we select a non-inertial frame with the path
(\ref{III61}). The effective wave function $\widehat{\psi}_c(\tau
,{\vec \eta}_i)$ will satisfy the following effective Schroedinger
equation replacing Eq.(\ref{III64}) (see the end of Appendix C)

\bea
 i\hbar\, {{\partial \widehat{\psi}_c(\tau ,{\vec \eta}_i)}\over
 {\partial \tau}} &=& \Big(\dot \theta (\tau )\, \sum_{i=1}^N\,
 {\widehat{\cal R}}^{'}_i + {1\over 2}\, \sum_{i=1}^N\, \Big[
 V^r(\tau ,{\vec \eta}_i), i\hbar\, {{\partial}\over
 {\partial \eta^r_i}}\Big]_{+}\Big)\, \widehat{\psi}_c(\tau ,{\vec
 \eta}_i) =\nonumber \\
&&\nonumber\\
 &{\buildrel {def}\over =}& \widehat{H}^{'}_{ni}\, \widehat
 {\psi}_c(\tau ,{\vec \eta}_i).
 \label{III70}
 \eea

 \medskip

 Now, due to the different inertial potentials, $\widehat{
 H}^{'}_{ni}$ is a self-adjoint operator on
 ${\widetilde{\mbox{\boldmath{${\cal H}$}}}}_{{\vec k}_o}$, the isometry (\ref{III67}) becomes a unitary
 operator ${\widehat{\cal J}}(\tau_f, \tau_o):
 {\widetilde{\mbox{\boldmath{${\cal H}$}}}}_{{\vec k}_o} \mapsto {\widetilde{\mbox{\boldmath{${\cal H}$}}}}_{{\vec k}_o}$
 and we have ${d\over {d \tau}}\, \Big(\widehat{\psi}_{c1}, \widehat
 {\psi}_{c2}\Big) = 0$.

\newpage

\section{Center of Mass, Relative Variables and Bound States in
Non-Inertial Frames.}

In this Section we consider N positive-energy particles with
relativistic action-at-a-distance interactions (Subsection A).
Then, in Subsection B, we discuss a definition of bound states and
of their spectra, to be applied to atoms in the approximation of
replacing the electro-magnetic field with an effective (either
Coulomb or Darwin \cite{9}) action-at-a-distance potential. Then
we show that in relativistic non-rigid non-inertial frames there
exist suitable frame-adapted relative variables, whose use implies
that {\it the levels of atoms can be labeled by the same quantum
numbers used in inertial frames.}

\subsection{Relativistic Action-at-a-Distance Interactions.}

As shown in Ref.\cite{34,9} and their bibliography, the
relativistic action-at-a-distance interactions inside the Wigner
hyperplane of the rest-frame instant form may be introduced either
under the square roots (scalar and vector potentials) appearing in
the free Hamiltonian (\ref{II22}) or outside them (scalar
potential like the Coulomb one) .

In the rest-frame instant form the most general Hamiltonian with
action-at-a-distance interactions is

\beq
 H = \sum_{i=1}^N \sqrt{ m_i^2+U_i+[{\vec \kappa}_i-{\vec V}_i]^2}
+ V + \vec \lambda (\tau ) \cdot \sum_{i=1}^N\, {\vec
\kappa}_i(\tau ),
 \label{IV1}
  \eeq

\noindent where $U=U({\vec \kappa}_k, {\vec \eta}_h-{\vec
\eta}_k)$, ${\vec V}_i={\vec V}_i({\vec \kappa}_{j\not= i}, {\vec
\eta}_i-{\vec \eta}_{j\not= i})$, $V=V_o(|{\vec \eta}_i-{\vec
\eta}_j|)+V^{'}({\vec \kappa}_i, {\vec \eta}_i-{\vec \eta}_j)$.
\medskip

If we use the canonical transformation \footnote{See Ref.\cite{33}
and Ref.\cite{34} for its explicit construction. It is a {\it
point} canonical transformation {\it only in the momenta}.
Instead, the corresponding non-relativistic canonical
transformation is {\it point both in the coordinates and in the
momenta}.} defining the relativistic center of mass and relative
variables on $\Sigma_{\tau}$ (see Subsection B for the case N=2)

\beq
 {\vec \eta}_i, {\vec \kappa}_i\,\, \mapsto\,\, {\vec {\cal X}}, \vec
 \kappa = \sum_{i=1}^N\, {\vec \kappa}_i, {\vec \rho}_{qa}, {\vec
 \pi}_{qa},\quad (a=1,..,N-1),
 \label{IV2}
 \eeq

\noindent the rest frame Hamiltonian for the relative motion
becomes

\beq
 H_{rel} = \sum_{i=1}^N\sqrt{m_i^2+{\tilde
U}_i+[\sqrt{N}\sum_{a=1}^{N-1}\gamma_{ai}{\vec \pi}_{qa}- {\tilde
{\vec V}}_i]^2} +\tilde V,
 \label{IV3}
  \eeq

\noindent where

\bea
 {\tilde U}_i&=&U\Big(\sqrt{N}\sum_{a=1}^{N-1}\gamma_{ak}{\vec
\pi}_{qa},
{1\over{\sqrt{N}}}\sum_{a=1}^{N-1}(\gamma_{ah}-\gamma_{ak}){\vec
\rho}_{qa}\Big),\nonumber \\
 {\tilde {\vec V}}_i&=&{\vec V}_i\Big(\sqrt{N}\sum_{a=1}^{N-1}
\gamma_{aj\not= i}{\vec
\pi}_{qa},{1\over{\sqrt{N}}}\sum_{a=1}^{N-1}
(\gamma_{ai}-\gamma_{aj\not= i}){\vec \rho}_{qa}\Big),\nonumber \\
 \tilde V &=& V_o\Big(|{1\over
{\sqrt{N}}}\sum_{a=1}^{N-1}(\gamma_{ai}-\gamma_{aj}){\vec
\rho}_{qa}|\Big) + V^{'}\Big(\sqrt{N}\sum_{a=1}^{N-1}\gamma_{ai}{\vec
\pi}_{qa},
{1\over{\sqrt{N}}}\sum_{a=1}^{N-1}(\gamma_{ai}-\gamma_{aj}){\vec
\rho}_{qa}\Big).
 \label{IV4}
  \eea

\bigskip

Since a Lagrangian density, replacing Eq.(\ref{II11}) in presence
of action-at-a-distance mutual interactions, is not known, we must
introduce the potentials {\it by hand} in the constraints
(\ref{II9}), but {\it only} into the constraint ${\cal
H}_{\perp}(\tau ,\vec \sigma ) \approx 0$ restricted to
hyper-planes, since we are working in an instant form of dynamics.
The only restriction is that the Poisson brackets of the modified
constraints must generate the same algebra of the free ones. When
this happens, the restriction to the embeddings (\ref{II23}) will
produce only a modification of ${\cal H}_{\perp}(\tau ,\vec \sigma
) \approx 0$ of Eq.(\ref{II28}), namely only of ${\cal E}[{\cal
A}^r]$ of Eq.(\ref{II31}) and of the effective non-inertial
Hamiltonian $H_{ni}$ of Eq.(\ref{II44}).
\medskip

The observation that the quantum result (\ref{III45}), namely the
dependence of the wave functions only upon the variables ${\vec
{\cal A}}_i = {\vec {\cal A}}({\vec \eta}_i)$ after the solution
of the second set of Eqs.(\ref{III31}), is already present at the
classical level in Eq.(\ref{II44}), i.e. in the fact that the
effective non-inertial Hamiltonian $H_{ni}$ depends only on ${\cal
A}^a_i(\tau ) = {\cal A}^a(\tau ,{\vec \eta}_i(\tau ))$ and ${\cal
A}^s_a(\tau ,{\vec \eta}_i(\tau ))\, \kappa_{is}(\tau )$, suggests
to introduce the following Shanmugadhasan {\it $\tau$-dependent
point canonical transformation} adapted to the constraints ${\cal
H}_r(\tau ,\vec \sigma ) \approx 0$ \footnote{The existence of
this canonical transformation explains the second set of the
quantum Eqs. (\ref{III40}).} when we are in non-inertial frames of
the type (\ref{II42}) [see Eq.(\ref{c2}) for ${\cal A}^s_a(\tau
,\vec \sigma )$]

\begin{eqnarray*}
 \begin{minipage}[t]{3cm}
\begin{tabular}{|l|l|} \hline
${\vec \eta}_i(\tau )$ & ${\vec {\cal A}}(\tau ,\vec \sigma )$  \\
\hline ${\vec \kappa}_i(\tau )$ &
${\vec \rho}_U(\tau ,\vec \sigma )$  \\
\hline
\end{tabular}
\end{minipage} &&\hspace{.2cm} {\longmapsto \hspace{.2cm}} \
\begin{minipage}[t]{4 cm}
\begin{tabular}{|l|l|} \hline
${\vec \eta}_i{}^{'}(\tau )$ & ${\vec {\cal A}}(\tau ,\vec \sigma )$ \\
\hline
 ${\vec \kappa}_i{}^{'}(\tau )$ & ${\vec \rho}_U{}^{'}(\tau ,\vec \sigma )$
 \\ \hline
\end{tabular}
\end{minipage},
 \end{eqnarray*}

\bea
 {\vec \eta}^{'}_i(\tau ) &=& {\vec {\cal A}}(\tau ,{\vec
 \eta}_i(\tau )),\nonumber  \\
 \kappa^{'}_{ia}(\tau ) &=& {\cal A}_a^s(\tau ,{\vec \eta}_i(\tau
 ))\, \kappa_{is}(\tau ),\nonumber \\
 \rho^{'}_{Ua}(\tau ,\vec \sigma ) &=& {\cal A}^s_a(\tau ,\vec
 \sigma )\, {\cal H}_a(\tau ,\vec \sigma ) = \rho_{Ua}(\tau ,\vec
 \sigma ) - \sgn\, \sum_{i=1}^N\, \delta^3(\vec \sigma - {\vec
 \eta}_i(\tau ))\, \kappa^{'}_{ia}(\tau ) \approx 0.\nonumber \\
 &&{}
 \label{IV5}
 \eea

\medskip

Then, given the embedding (\ref{II23}), we can rewrite the
particle positions $x^{\mu}_i(\tau ) = z^{\mu}(\tau ,{\vec
\eta}_i(\tau ))$ in the form $x^{\mu}_i(\tau ) = \theta (\tau )\,
{\hat U}^{\mu}(\tau ) + \epsilon^{\mu}_a(\hat U(\tau ))\,
\eta^{{'}\, a}_i(\tau )$, i.e. {\it in a form similar to the one
given in the inertial systems on the Wigner hyper-planes} [see
before Eq.(\ref{II14})].
\medskip

But this implies that at the classical level to {\it introduce
mutual interactions in non-inertial frames} is equivalent to
replace the square root term in Eq.(\ref{II44}) with the term

\beq
 \sum_{i=1}^N\, \sqrt{m^2_i + U_i + [{\cal A}^r_a(\tau
 ,{\vec \eta}_i(\tau ))\, \kappa_{ir}(\tau ) - V_{ia}]\,
 \delta^{ab}\, [{\cal A}^s_b(\tau ,{\vec \eta}_i(\tau ))\,
 \kappa_{is}(\tau ) - V_{ib}]} + V,
 \label{IV6}
 \eeq

\noindent with $U_i$, ${\vec V}_i$, $V$ the same functions
appearing in Eq.(\ref{IV1}) but with the replacement ${\vec
\eta}_i, {\vec \kappa}_i\, \mapsto\, {\vec \eta}^{'}_i, {\vec
\kappa}^{'}_i$. Then, the {\it time-dependent canonical
transformation (\ref{IV5}) sends the Hamiltonian} (\ref{II44}),
modified according to Eq.(\ref{IV6}), {\it into the inertial
Hamiltonian}

\bea
 H_{inertial} &=& H_{ni} - \sum_{i=1}^N\, V^r(\tau ,{\vec \eta}_i(\tau
 ))\, \kappa_{ir}(\tau ) =
 \dot \theta (\tau )\, \Big[\sum_{i=1}^N \sqrt{ m_i^2+U_i+[{\vec
\kappa}^{'}_i-{\vec V}_i]^2} + V\Big],\nonumber \\
 &&{}\nonumber \\
 &&V^r(\tau ,{\vec \eta}_i(\tau ) )\, \kappa_{ir}(\tau ) \approx \sum_{i=1}^N\,
 [\vec v(\tau  ) + \vec \Omega (\tau ,{\vec \eta}_i(\tau ) ) \times {\vec
 \eta}_i(\tau )] \cdot {\vec \kappa}^{'}_i(\tau ),\qquad
 \mbox{if (\ref{II42}) holds.}\nonumber \\
 &&{}
 \label{IV7}
 \eea

\bigskip

\subsection{Bound States in Relativistic Non-Inertial Reference
Frames.}

In the relativistic case the effective classical non-inertial
Hamiltonian $H_{ni}$ is given in Eq.(\ref{II44}) with the
admissible class of functions ${\vec {\cal A}}(\tau ,\vec \sigma
)$, given in Eq.(\ref{II42}). As it will be shown in paper II,
these ${\vec A}(\tau ,\vec \sigma )$ and also the constraints
${\cal H}_r(\tau ,\vec \sigma ) \approx 0$ have the same form of
the non-relativistic ones. In presence of action-at-a-distance
interactions the square roots in Eq.(\ref{II44}) have to be
modified to the form of Eq.(\ref{IV6}). The quantum version $\widehat{
H}_{ni}$ is the (self-adjoint on ${\mbox{\boldmath{${\cal H}$}}}$) Hamiltonian
operator in the Schroedinger equation (\ref{III70}) along the path
${\cal P}_c(\tau ) = [\theta_c(\tau ), {\vec {\cal A}}_c(\tau
,\vec \sigma )]$.

\bigskip

Let us now consider the canonical transformation realizing the
separation of the center of mass from the relative variables on
$\Sigma_{\tau}$. If we make the sequence of two canonical
transformations, first Eq.(\ref{IV5}) followed by Eq.(\ref{IV2})
applied to ${\vec \eta}_i^{'}$, ${\vec \kappa}_i^{'}$, {\it the
inverse total canonical transformation allows to define a
non-inertial notion of center of mass and relative variables on
$\Sigma_{\tau}$}.

As shown in paper II, in the non-relativistic case such a
transformation is {\it point} both in the coordinates and the
momenta. Instead in the relativistic case this canonical
transformation, defined in Ref.\cite{33} and given explicitly in
Ref.\cite{34}, is very complicated and it is {\it point only in
the momenta}.

For the sake of simplicity let us consider only the case $N=2$ in
absence of action-at-a-distance interactions, when we have
$H_{inertial} = \dot \theta (\tau ) \, M$ with $M = \sqrt{m^2_1 +
{\vec \kappa}^{{'}\, 2}_1} + \sqrt{m^2_2 +  {\vec \kappa}_2^{{'}\,
2}}$. Then the canonical transformation sending the canonical
basis ${\vec \eta}_i^{'}$, ${\vec \kappa}_i^{'}$ in the canonical
basis ${\vec {\cal X}}$ (the relativistic 3-center of mass on
$\Sigma_{\tau}$), $\vec \kappa$ (the total 3-momentum on
$\Sigma_{\tau}$), $\vec \rho$ and $\vec \pi$ (the relativistic
relative variables on $\Sigma_{\tau}$) is

\begin{eqnarray*}
 {\vec {\cal X}} &=& {{\sqrt{m^2_1 + {\vec \kappa}_1^{{'}\, 2}}\,
 {\vec \eta}^{'}_1 + \sqrt{m^2_2 + {\vec \kappa}_2^{{'}\, 2}}\,
 {\vec \eta}^{'}_2}\over {\sqrt{m^2_1 + {\vec \kappa}_1^{{'}}} +
 \sqrt{m^2_2 + {\vec \kappa}_2^{{'}}}}} + {{{\vec S}_{rel} \times \vec \kappa}\over
 {M\, (M + \sqrt{M^2 - {\vec \kappa}^2})}},\nonumber \\
 \vec \kappa &=& {\vec \kappa}_1^{'} + {\vec
 \kappa}_2^{'},\nonumber \\
 \vec \rho &=& {\vec \eta}^{'}_1 - {\vec \eta}^{'}_2 +
 \Big({{\sqrt{m^2_1 + {\vec \kappa}_1^{{'}}}}\over {\sqrt{m^2_2 + {\vec \kappa}_2^{{'}}}}}
  + {{\sqrt{m^2_2 + {\vec \kappa}_2^{{'}}}}\over
  {\sqrt{m^2_1 + {\vec \kappa}_1^{{'}}}}}\Big)\, {{({\vec \eta}_1^{'} - {\vec \eta}_2^{'})
  \cdot \vec \kappa}\over {M\, \sqrt{M^2 - {\vec \kappa}^2}}}\, \vec
  \pi, \nonumber \\
 \vec \pi &=& {1\over 2}\, ({\vec \kappa}_1^{'} - {\vec
 \kappa}_2^{'}) - {{\vec \kappa}\over {\sqrt{M^2 - {\vec
 \kappa}^2}}}\nonumber \\
  &&\Big({1\over 2}\, (\sqrt{m^2_1 + {\vec \kappa}_1^{{'}}} -
  \sqrt{m^2_2 + {\vec \kappa}_2^{{'}}}) - {{\vec \kappa \cdot ({\vec \kappa}_1^{'} -
  {\vec \kappa}_2^{'})}\over {2 {\vec \kappa}^2}}\,
  (M - \sqrt{M^2 - {\vec \kappa}^2})\Big), \nonumber \\
 &&{}\nonumber \\
 {\vec S}_{rel} &=& \vec \rho \times \vec \pi,
 \end{eqnarray*}

\bea
 M &=& \sqrt{m^2_1 + {\vec \kappa}_1^{{'}\, 2}} + \sqrt{m^2_2 + {\vec \kappa}_2^{{'}\, 2}}
 = \sqrt{{\cal M}^2 + {\vec \kappa}^2},\qquad M = {\cal M}\,\mbox{ in
 the rest frame } \vec \kappa = 0,\nonumber \\
&&\nonumber\\
 {\cal M} &=& \sqrt{m^2_1 + {\vec \pi}^2} + \sqrt{m^2_2 + {\vec \pi}^2}.
 \label{IV8}
 \eea
\medskip

In the case of action-at-a-distance interactions (see Ref.\cite{9}
for the semi-classical Coulomb and Darwin potentials between two
charged particles and Ref.\cite{31} for an older treatment) the
modification is $M(\vec \kappa ,\vec \pi ) \mapsto M_{int}(\vec
\kappa, \vec \rho , \vec \pi )$ and ${\cal M}(\vec \pi ) \mapsto
{\cal M}_{int}(\vec \rho ,\vec \pi )$ ($M_{int} = {\cal M}_{int}$
for $\vec \kappa = 0$).
\medskip

Therefore, in the relativistic case the isolated particle systems
has the 3-center of mass ${\vec {\cal X}}$ on $\Sigma_{\tau}$
describing an {\it effective free particle} with 3-momentum $\vec
\kappa$ and with {\it effective mass} ${\cal M}_{int}$ determined
by the relative motion.
\medskip

{\it The quantization in an inertial system must be done in the
canonical variables (\ref{IV8}) and not in the individual particle
variables}, because it is not clear how to build a unitary
implementation of the canonical transformation (\ref{IV8}) and it
is also strongly suggested by older results \footnote{See
Ref.\cite{31} for a treatment of a two-body system with mutual
action-at-a-distance interaction described by the canonical
variables $x^{\mu}_i(\tau )$, $p^{\mu}_i(\tau )$ and by two first
class constraints. The only way to arrive at an {\it equal time}
description of the two-body system with a well defined equal-time
physical scalar product was to quantize a set of center-of-mass
and relative variables adapted to the gauge fixing $[p_{1\mu} +
p_{2\mu}]\, [x^{\mu}_1 - x^{\mu}_2] \approx 0$ (elimination of the
relative time to get simultaneity in the rest frame). The standard
use of a Hilbert space tensor product of two free particle Hilbert
spaces does not allow to incorporate a notion of {\it equal time}
(nothing forbids to a {\it in state} to be in the future of
another {\it state}). As a consequence, the {\it equal time}
quantization of the 3-center of mass and 3-relative variables is
unavoidable and this leads to a Hilbert space, which is always
(also in the free limit) the tensor product of a center-of-mass
Hilbert space with a relative motion Hilbert space. Let us remark
that the understanding of the gauge nature of the  relative times
was the starting point to develop parametrized Minkowski
theories.}.

Moreover the {\it point-in-the-momenta} nature of the canonical
transformation (\ref{IV8}) forces to use {\it the momentum
representation}. As a consequence in an {\it inertial frame} we
get the Schroedinger equation (the positive energy square root of
a Klein-Gordon type equation; ${\hat {\vec \rho}} = - i\hbar\,
{{\partial}\over {\partial \vec \pi}}$)

\bea
 i\hbar\, {{\partial}\over {\partial \tau}}\, {\widetilde{\Psi}}_{in}(\tau ,
 {\vec \kappa}, \vec \pi ) &=& \dot \theta (\tau )\,
 \sqrt{{\widehat{\cal M}}_{int}^2({\hat {\vec \rho}}, \vec \pi )
 + {\vec \kappa}^2}\,
 {\widetilde{\Psi}}_{in}(\tau , {\vec \kappa}, \vec \pi ),\nonumber \\
 &&{}\nonumber \\
 \Psi_{in}(\tau , {\vec {\cal X}}, \vec \rho ) &=& \int d^3k
 d^3\pi\, e^{-i\, ({\vec {\cal X}} \cdot \vec \kappa + \vec \rho
 \cdot \vec \pi )}\, {\widetilde{\Psi}}_{in}(\tau ,\vec \kappa , \vec
 \pi ),
 \label{IV9}
 \eea
 \medskip

If the bound states are defined as the solutions of the
 stationary equation

 \beq
  {\widehat{\cal M}}_{int}\, \widetilde{\psi}_n(\vec \pi ) = B_n\, {\widetilde
  {\psi}}_n(\vec  \pi ),
  \label{IV10}
  \eeq

 \noindent then we can consider the following factorized
 solution of Eq.(\ref{IV9})

 \bea
 {\widetilde{\Psi}}_{in,\, n\, {\vec \kappa}_o}(\tau , {\vec \kappa}, \vec \pi ) &=&
 {\widetilde{\Phi}}_{n, {\vec \kappa}_o}(\tau , \vec \kappa)\, {\widetilde{\psi}}_n(\vec
 \pi ),\nonumber \\
 &&{}\nonumber \\
 i\hbar\, {{\partial}\over {\partial \tau}}\, {\widetilde{\Phi}}_{n, {\vec
 \kappa}_o}(\tau ,{\vec \kappa}) &=& \dot \theta (\tau )\,
 \sqrt{B^2_n + {\vec \kappa}^2}\, {\widetilde{\Phi}}_{n, {\vec \kappa}_o}(\tau
 ,{\vec \kappa}),\nonumber \\
 &&{}\nonumber \\
 \Rightarrow&& {\widetilde{\Phi}}_{n, {\vec \kappa}_o}(\tau ,\vec
 \kappa ) = e^{- {i\over {\hbar}}\, \theta (\tau )\, \sqrt{B^2_n
 + {\vec \kappa}^2}}\, \delta^3(\vec \kappa - {\vec
 \kappa}_o),\nonumber \\
 \Rightarrow&& \Phi_{n, {\vec \kappa}_o}(\tau , {\vec {\cal X}}) =
 const.\,
 e^{- {i\over {\hbar}}\, (\theta (\tau )\, \sqrt{B^2_n + {\vec \kappa}_o^2}
 - {\vec \kappa}_o \cdot {\vec {\cal X}}}.
 \label{IV11}
 \eea

 \bigskip

The quantization in {\it non-inertial frames} follows the same
pattern, if we work in the momentum representation. From
Eq.(\ref{IV7}) we get $H_{ni} = H_{inertial} + (inertial\,
potentials)$ and the inversion of Eqs.(\ref{IV8}) allows to get
$H_{ni}$ in terms of the center of mass and the relative variable.
The final quantum Hamiltonian $\widehat{H}_{ni}$ will contain a term
of the type $\sqrt{{\widehat{\cal M}}^2 +[\sum_i\, {\vec
\kappa}_i^{'}]^2}$ with a {\it self-adjoint effective mass} plus a
self-adjoint (on ${\widetilde{\mbox{\boldmath{${\cal H}$}}}}_{{\vec k}_o}$) term
containing the potential of the inertial forces. Again we can
define the bound states by means of Eq.(\ref{IV10}) so that {\it
we get the same eigenvalues (i.e. spectral lines) as in the
inertial system}. Like for an atom in presence  of external
time-dependent electro-magnetic fields, the self-adjoint operator
$\widehat{H}_{ni}$ is in general time-dependent and does not admit a
unique associated eigenvalue equation except in special cases (for
instance when the inertial potentials are time-independent).
\medskip

Since the canonical transformation (\ref{IV8}) is a point one in
the momenta and the time-dependent canonical transformation
(\ref{IV5}) is a  point one in the coordinates, their combination
is unitarily implementable, so that {\it the Hamiltonian operators
$\widehat{H}_{ni}$ and $\widehat{H}_{inertial}$ are connected by a
time-dependent unitary transformation} (see the Introduction),
like it happens in the non-relativistic case (see II).

\bigskip

Let us remark that all these results are Lorentz invariant because
the 3-indices are internal indices inside $\Sigma_{\tau}$.

\newpage

\section{Conclusions.}

The main result of this paper is the definition of a quantization
scheme for isolated systems of relativistic  mutually interacting
particles in a sufficiently general class of non-rigid
non-inertial frames. This result allows to show that the only
possible definition of bound states by means of a stationary
eigenvalue equation is based on the analogous definition in
inertial frames, i.e. by using the self-adjoint relative energy
operator (the invariant mass after the decoupling of the center of
mass, which produces the over-imposed continuum spectrum of a free
particle). In general non-rigid non-inertial frames the
time-dependent potential of the inertial forces, appearing in the
effective self-adjoint non-inertial Hamiltonian, acts as a
time-dependent c-number external field. As a consequence, except
in special cases (for instance with time-independent inertial
potentials) it is not possible to find a unique eigenvalue
equation for the effective non-inertial Hamiltonian, which,
instead, governs the unitary evolution, allows to evaluate the
scattering matrix and produces the interferometric effects
signalling the non-inertiality of the frame.

Let us remark that, as said in Subsection IVB, at the relativistic
level the multi-temporal quantization scheme has to be applied
only after the separation of the relativistic center of mass from
the relativistic relative variables, because only in this way we
can get a satisfactory description of bound states on {\it
equal-time} Cauchy surfaces.

The fact that the effective Hamiltonian becomes frame-dependent
due to the potentials of the inertial forces, in analogy to the
energy density in general relativity where only non-inertial
frames are allowed, makes us hope that this quantization scheme
can be also useful for a future attempt to reopen the canonical
quantization of gravity with a softened ordering problem as a
consequence of the c-number nature of the gauge variables.

\bigskip

Since we will show  in paper II that non-relativistic quantum
mechanics in non-inertial frames follows the same pattern of the
relativistic, let us add here a remark on the applicability of the
equivalence principle to quantum mechanics in {\it non-rigid}
non-inertial frames. Since our approach to non-inertial frames is
originally defined in Minkowski space-time, where there is no
accepted formulation of action-at-a-distance gravity, and then
restricted to Galilei space-time by means of the non-relativistic
limit (see the next paper II), there is no space for a
reinterpretation of the inertial forces in non-inertial frames as
gravitational fields. Only the formulation of the equivalence
principle stating the equality of inertial and gravitational
masses (free fall along a geodesics) retains its validity.

\medskip

The Hamiltonian treatment of general relativity and of its initial
value problem \cite{3} shows that in presence of matter the
gravitational field gives rise to three different types of
quantities:

A) deterministically predictable tidal-like effects (Dirac
observables for the gravitational field; they are absent in
Newtonian gravity), whose functional form is in general
coordinate-dependent;

B) action-at-a-distance potentials between elements of matter (in
the non-relativistic limit they go into the Newton potential),
whose functional form is in general coordinate-dependent;

C) inertial-like effects (the gauge variables) which change from a
4-coordinate system to another one \footnote{Remember that a
completely fixed Hamiltonian gauge is equivalent to the choice of
4-coordinate system on the solutions of Einstein's equations.
Since this corresponds to the choice of a non-rigid non-inertial
frame (an extended space-time laboratory), in the non-relativistic
limit we get a either non-rigid or rigid non-inertial frame with
its  local or global  inertial effects.}). The gauge variables
describe how the {\it appearances} of the phenomena change locally
from a point to another one due to the absence of a global
inertial reference frame in general relativity: they have nothing
to do with the action-at-a-distance potentials.

\medskip

In particular {\it uniform} accelerations in a sufficiently small
neighborhood of a test particle in free fall are {\it not
equivalent to non-inertial frames} but are locally connected to
gauge variables, because the {\it uniform} gravitational fields,
equivalent to them according to Einstein, do not exist on finite
regions according to Synge \cite{18} but are replaced by physical
and action-at-a-distance tidal effects.  At the Newtonian level
the physical tidal effects do not exist and only
action-at-a-distance tidal effects induced by the Earth on nearby
particles exist.

\medskip

Moreover in general relativity the concept of energy is
coordinate-dependent and strictly speaking we do not know how to
define the energy of an atom except in the post-Newtonian
approximation after the introduction of a background Minkowski
4-metric.

\medskip

Another problem is whether or not we consider special relativity,
i.e. flat Minkowski space-time, as a limiting case of general
relativity.

i) If we consider flat Minkowski space-time as the limit of
general relativity for vanishing 4-Riemann tensor (special
solution of Einstein equations), then in absence of matter (no
action-at-a-distance potentials) this limit implies the vanishing
of the physical tidal effects, namely of the Dirac observables of
the gravitational field. This leads to a description of Minkowski
space-time as a {\it void space-time} (see the first paper in
Ref.\cite{3}) solution of Einstein's equations in absence of
matter. This fact puts {\it restrictions} on the leaves of the
allowed 3+1 splittings with Cauchy simultaneity space-like
hyper-surfaces: in absence of matter  the simultaneity 3-surfaces
must be {\it 3-conformally flat} (this is a restriction on the
admissible non-inertial frames). It is not yet clear which type of
restrictions (more complicated of 3-conformal flatness) are
introduced by the presence of matter on the admissible 3+1
splittings in the zero 4-curvature limit, in which the Dirac
observables have to be expressed only in terms of the Dirac
observables for the matter. One should solve Einstein's equations
to get the Dirac observables of the gravitational field in terms
of the matter's ones and then put the solution in the matter
equations in analogy to what can be done to go from the Coulomb to
the Darwin potential in electro-magnetism \cite{9}. If this is
possible, an action-at-a-distance formulation of gravity in
Minkowski space-time would emergge and then the non-relativistic
limit should allow to recover Newtonian gravity for the given
matter.

ii) If, on the contrary, we consider special relativity as an
autonomous theory, i.e. not as a solution of Einstein equations,
there is no such limitation: every 3+1 splitting of Minkowski
space-time  is admissible like in parametrized Minkowski theories
for any kind of isolated system.

\bigskip

Therefore the determination of gravitational potentials is a
problem much more difficult than the determination of the inertial
forces appearing in non-inertial frames and the use of the
equivalence principle, usually done in non-relativistic quantum
mechanics, does not seem acceptable.

\medskip

Even if there is no accepted formulation of quantum gravity, the
multi-temporal quantization developed in this paper suggests to
look for a quantization scheme of the gravitational field based on
the following prescriptions:

A) action-at-a-distance potentials between elements of matter will
be quantized once matter is quantized;

B) tidal effects (Dirac observables for the gravitational field)
will be quantized;

C) inertial effects connected with the gauge variables  must not
be quantized (they should become {\it c-number generalized
times}), since they describe only the appearances of phenomena
seen by local non-rigid non-inertial frames. In this way it is
hoped to arrive to a background independent quantization of
canonical gravity in a way respecting relativistic causality.

Besides the necessity of arriving to replace the Dirac observables
of the gravitational field with a canonical set of Bergmann
observables (coordinate-independent Dirac observables) \cite{1},
to implement this program we have first of all to find a
ultraviolet regularization for the Tomonaga-Schwinger formalism
\cite{36} (the Torre-Varadarajan no-go theorem \cite{37}) emerging
from the future attempt to extend the results of this paper to the
quantization of fields on arbitrary foliations of Minkowski
space-time in the framework of parametrized Minkowski theories.

\bigskip

Finally, the formalism developed in this paper should be useful to
try to define relativistic Bel inequalities in a way compatible
with the gauge nature of the notion of relativistic simultaneity.
In any case, the  need of a convention on the synchronization of
clocks to define an instantaneous 3-space together with the
related necessity to factorize a many-particle wave function in a
center-of-mass part and in a relative motion one (see the
discussion about relative times in Section IV) show that at the
relativistic level there is an {\it extra non-locality} besides
the standard quantum  one connected with the non-relativistic
entangled states.

\vfill\eject

\appendix

\section{The Wigner Standard Boost and  Wigner Rotations.}

The standard Wigner boost $L(\hat U,{\hat U}_o)$, mapping a
standard unit time-like four-vector ${\hat U}_o^\nu=(1,0,0,0)$
onto ${\hat U}^\mu$, i.e. such that ${\hat U}^\mu = L^\mu_\nu(\hat
U,{\hat U}_o)\,{\hat U}^\nu_o=L^\mu_o(\hat U,{\hat U}_o)$, can be
parametrized as

\begin{equation}
L^\mu{}_\nu(\hat U, {\hat U}_o) =
L^\mu{}_\nu(\vec{\beta})=\frac{1}{\sqrt{1-{\vec \beta}^2}} \left(
\begin{array}{cc}
1&\beta^i\\
&\\
\beta^j&N^{ij}(\vec{\beta})
\end{array}
\right),
 \label{a1}
\end{equation}

\noindent where

\begin{equation}
N^{ij}(\vec{\beta})=\delta^{ij}+\frac{\beta^i\beta^j}{\beta^2}
\left(\frac{1-\sqrt{1-{\vec \beta}^2}}{\sqrt{1-{\vec
\beta}^2}}\right).
 \label{a2}
\end{equation}

Then we can define the tetrads $\epsilon^{\mu}_A(\hat U)$

\begin{equation}
 \epsilon^{\mu}_o(\hat U) = {\hat U}^\mu = L^\mu{}_o(\vec{\beta})
 = \frac{1}{\sqrt{1-{\vec
\beta}^2}}\left(1,\vec{\beta}\right),
 \label{a3}
\end{equation}

\begin{equation}
\epsilon^\mu_a(\hat U)=L^\mu{}_a(\vec{\beta}),
 \label{a4}
\end{equation}

\noindent whose associated cotetrads $\epsilon^A_{\mu}(\hat U)$
are defined by $\epsilon^A_{\mu}(\hat U)\, \epsilon^{\mu}_B(\hat
U) = \delta^A_B$.

We also have

\beq
 \eta^{\mu\nu} = \sgn\, \Big[\epsilon^{\mu}_o(\hat U)\, \epsilon^{\nu}_o(\hat U)
 + \sum_a\, \epsilon^{a\mu}(\hat U)\, \epsilon^{\nu}_a(\hat U)\Big]
 = \sgn\, \Big[\epsilon^{\mu}_o(\hat U)\, \epsilon^{\nu}_o(\hat U)
 - \sum_a\, \epsilon^{\mu}_a(\hat U)\, \epsilon^{\nu}_a(\hat U)\Big].
 \label{a5}
 \eeq

\medskip

The {\em Wigner Rotation} $R^b{}_a(\Lambda,\hat U)$ associated to
a Lorentz transformation $\Lambda$ is defined by

\begin{equation}
R^{a=i}{}_{b=j}(\Lambda,\hat U)= [L(\hat U,{\hat
U}_o)\,\Lambda\,L^{-1}(\hat U,{\hat U}_o)]^i{}_j.
 \label{a6}
\end{equation}

The definition (\ref{a4}) implies that the $\epsilon^\mu_a(\hat
U)$ are a triad of space-like four-vector such that \cite{7,31}

\begin{eqnarray}
&& {\hat U}_\mu\,\epsilon^\mu_a(\hat U)=0,\qquad
\epsilon^\mu_a(\hat U)\epsilon_{\mu\,b}(\hat U)=\eta_{ab},\qquad
{\hat U}_\mu\,\frac{\partial\epsilon^\lambda_a(\hat U)}{\partial
{\hat U}_\mu}=0, \nonumber\\
 &&\nonumber\\
  &&\epsilon^\mu_a(\Lambda\,\hat U)=\Lambda^\mu{}_\nu\,
\epsilon^\nu_b(\hat U)R^b{}_a(\Lambda,\hat U).
 \label{a7}
\end{eqnarray}

\medskip

In the relativistic canonical theory of Subsection IIC, the
parameter $\vec{\beta}(\tau)$ in $U^\mu(\tau)$ is equivalent to
the canonical variable $\vec{k}(\tau)$, since we have

\begin{equation}
\vec{\beta}(\tau)=
\frac{\vec{k}(\tau)}{\sqrt{1+\vec{k}^2(\tau)}},\;\;\;\;\;
\Leftrightarrow\;\;\;\;\; \vec{k}(\tau)=
\frac{\vec{\beta}(\tau)}{\sqrt{1-\vec{\beta}^2(\tau)}}.
 \label{a8}
\end{equation}

This relation can be viewed as half of a canonical transformation,
whose generating function  is

\begin{equation}
G(\vec{\beta},\vec{z})=\frac{\vec{\beta}}{\sqrt{1-\vec{\beta}^2}}\cdot\vec{z}.
 \label{a9}
\end{equation}

Then if we define

\beq
 k^i(\tau) = \frac{\partial G}{\partial
z^i}(\vec{\beta}(\tau),\vec{z}(\tau)), \qquad \xi^i(\tau) =
\frac{\partial G}{\partial
\beta^i}(\vec{\beta}(\tau),\vec{z}(\tau)),
 \label{a10}
\eeq

\noindent we obtain

\begin{eqnarray}
k^i(\tau)&=&\frac{\vec{\beta}(\tau)}{\sqrt{1-\vec{\beta}^2(\tau)}},\nonumber\\
 &&\nonumber\\
\xi^i(\tau)&=&\sqrt{1+\vec{k}^2(\tau)}\,
\left[\,z^i(\tau)+(\vec{k}(\tau)\cdot\vec{z}(\tau))\,k^i(\tau)\,
\right].
 \label{a11}
\end{eqnarray}

By construction the variables $\xi^i,\beta^i$ are three pairs of
canonical variables

\begin{equation}
\{\xi^i(\tau),\beta^j(\tau)\}=\delta^{ij}.
 \label{a12}
\end{equation}

\newpage

\section{Calculations for Subsection IIC}

\subsection{Dirac Brackets.}

Let $F({\cal I})$ be a function of the canonical variables ${\cal
I} = {\vec \eta}_i, {\vec \kappa}_i$ only.
\medskip

By explicit computation it can be shown that $ F({\cal I}),{\cal
A}^a(\tau,\vec{\sigma}),
\rho_{U\,a}(\tau,\vec{\sigma}),\theta(\tau),M_U(\tau),U_\mu(\tau)$
and $\widetilde{X}^\mu(\tau)$ have null Poisson brackets with the
gauge fixing (\ref{II22}), $S(\tau,\vec{\sigma})= {\cal A}^o(\tau
,\vec \sigma ) - {\cal A}^o(\tau ,\vec 0) \approx 0$.

\medskip

Then it is easy to verify the following Dirac brackets

\begin{eqnarray*}
 &&\{F_1({\cal I}),F_2({\cal I})\}^*= \{F_1({\cal I}),F_2({\cal
I})\} ,\nonumber \\
 &&\nonumber\\
 &&\{{\cal A}^a(\tau,\vec{\sigma}),{\cal A}^b(\tau,\vec{\sigma}')\}^*=
\{\rho_{U\,a}(\tau,\vec{\sigma}),\rho_{U\,b}(\tau,\vec{\sigma}')\}^*=0,
\nonumber\\
 &&\nonumber\\
 &&\{{\cal A}^a(\tau,\vec{\sigma}),\rho_{U\,b}(\tau,\vec{\sigma}')\}^*=
 - \sgn\, \delta^a_b\,\delta(\vec{\sigma}-\vec{\sigma}'),
 \end{eqnarray*}

\bea
  &&\{{\cal A}^a(\tau,\vec{\sigma}),F({\cal I})\}^*=
\{\rho_{U\,b}(\tau,\vec{\sigma}),F({\cal I})\}^*=0,\nonumber \\
 &&\nonumber\\
 &&\{M_U(\tau),\theta(\tau)\}^*=\{M_U(\tau),\theta(\tau)\}= \sgn,\nonumber\\
 &&\nonumber\\
 &&\{M_U(\tau),M_U(\tau)\}^*=\{\theta(\tau),\theta(\tau)\}=0,\nonumber \\
 &&\nonumber\\
  &&\{M_U(\tau),F({\cal I})\}^*=
\{M_U(\tau),{\cal A}^a(\tau,\vec{\sigma})\}^*=
\{M_U(\tau),\rho_{U\,a}(\tau,\vec{\sigma})\}^*=0,\nonumber\\
 &&\nonumber\\
  &&\{\theta(\tau),F({\cal I})\}^*=
\{\theta(\tau),{\cal A}^a(\tau,\vec{\sigma})\}^*=
\{\theta(\tau),\rho_{U\,a}(\tau,\vec{\sigma})\}^*=0.
 \label{b1}
 \end{eqnarray}

\begin{eqnarray*}
&&\{U^\mu(\tau),F({\cal I})\}^*= \{U^\mu(\tau),{\cal A}
^a(\tau,\vec{\sigma})\}^*=
\{U^\mu(\tau),\rho_{U\,a}(\tau,\vec{\sigma})\}^*=\nonumber\\
&&\nonumber\\ &=& \{U^\mu(\tau),M_U(\tau)\}^*=
\{U^\mu(\tau),\theta(\tau)\}^* =0,
\end{eqnarray*}

\bea
 &&\{{\tilde X}^\mu(\tau),F({\cal I})\}^*= \{{\tilde X}
^\mu(\tau),{\cal A}^r(\tau,\vec{\sigma})\}^*= \{{\tilde X}
^\mu(\tau),\rho_{Ur}(\tau,\vec{\sigma})\}^*=\nonumber \\
&&\nonumber\\
 &=& \{{\tilde X}^\mu(\tau),M_U(\tau)\}^*=
\{{\tilde X}^\mu(\tau),\theta(\tau)\}^* =0\nonumber\\
&&\nonumber\\
&&\{{\tilde X}^\mu(\tau),{\tilde X}^\nu(\tau)\}^*=0,\qquad
 \{{\tilde X}^\mu(\tau),U^\nu(\tau)\}^* = -\eta^{\mu\nu},
 \label{b2}
\end{eqnarray}

\bigskip

All these brackets show us that the pairs $\theta(\tau)$, $
M_U(\tau)$, ${\cal A} ^a(\tau,\vec{\sigma})$, $
\rho_{Ua}(\tau,\vec{\sigma})$, ${\tilde X}^{\mu}(\tau )$,
$U^{\mu}(\tau )$ together with the particle variables ${\vec
\eta}_i(\tau )$, ${\vec \kappa}_i(\tau )$ are a canonical basis
for the reduced phase space.

\subsection{Lorentz Covariance of the Final Canonical Basis}

We must now study the Lorentz covariance of the new variables in
the reduced phase space.

\medskip

It is easy to check that the variables $F({\cal I})$, $M_U(\tau
)$, $\theta (\tau )$ are Lorentz scalars

\begin{equation}
\{J^{\mu\nu}(\tau),F({\cal
I})\}^*=\{J^{\mu\nu}(\tau),M_U(\tau)\}^*=
\{J^{\mu\nu}(\tau),\theta (\tau )\}^*=0.
 \label{b3}
\end{equation}
\medskip

On the contrary the variables ${\cal A}^a(\tau,\vec{\sigma})$,
$\rho^a_U(\tau,\vec{\sigma})= \eta^{ab}\,
\rho_{U\,b}(\tau,\vec{\sigma})$ are not scalar, but they transform
as {\em Wigner spin-1 3-vectors}, because we get

\begin{eqnarray}
\{{\cal A}^a(\tau,\vec{\sigma}),J^{\sigma\rho}(\tau)\}^*&=&
-2\,D_{ab}{}^{ \sigma\rho}(\hat U)\,{\cal A}
^b(\tau,\vec{\sigma}),\nonumber\\
 &&\nonumber\\
 \{\rho^a_U(\tau,\vec{\sigma}),J^{\sigma\rho}(\tau)\}^*&=&
-2\,D_{ab}{}^{ \sigma\rho}(\hat U)\,\rho^b_U(\tau,\vec{\sigma}).
 \label{b4}
\end{eqnarray}

\noindent where the matrix $D^{\alpha\beta}_{ab}(\hat U)$ turns
out to be the one given in Eqs.(\ref{II37}).

Indeed, by construction, under a Lorentz transformation a Wigner
3-vector $W^a$ transforms  as $W^a\rightarrow
W^b\,R_{ba}(\Lambda,U)$ [the Wigner rotation $R(\Lambda ,U)$ is
defined in Eq.(\ref{a6})].

\noindent Let us parametrize  the associated infinitesimal Wigner
rotation in terms of a D matrix

\begin{equation}
R_{ba}(\Lambda,\hat U)=\delta_{ba} +D_{ba}{}^{ \mu\nu}(\hat
U)\,\delta\omega_{\mu\nu}, \qquad D_{ba}{}^{ \mu\nu}(\hat U)=
-D_{ab}{}^{ \mu\nu}(\hat U)= -D_{ba}{}^{ \nu\mu}(\hat U).
 \label{b7}
\end{equation}

 so that at the infinitesimal level we
get $\delta W^a = W^b\, D_{ba}{}^{\sigma\rho}\, \delta
\omega_{\sigma\rho}$ under the infinitesimal Lorentz
transformation

\begin{equation}
\Lambda_{\mu\nu}=\eta_{\mu\nu}+\delta\omega_{\mu\nu},\qquad
\delta\omega_{\mu\nu}=-\delta\omega_{\nu\mu}.
 \label{b6}
\end{equation}

Then from the last of Eqs.(\ref{a7}) we obtain

\begin{equation}
\left(\eta_{\mu\nu}+\delta\omega_{\mu\nu}\right)\,\epsilon^\nu_s(\hat
U)\, \left(\delta_{ba}+D_{ba}{}^{ \sigma\rho}(\hat
U)\,\delta\omega_{\sigma\rho}\right)= \epsilon^\mu_r(\hat U)+
\frac{\partial \epsilon^\mu_r(\hat U)}{\partial
{U}_\sigma}\,U^\rho\,\delta\omega_{\sigma\rho},
 \label{b8}
\end{equation}
\medskip

\noindent and this implies that the D matrix has the form given in
Eq.(\ref{II37}).

\medskip

As a consequence, the behavior of ${\cal A}^a$ under an
infinitesimal Lorentz transformation is

\bea
&& \delta{\cal
A}^a(\tau,\vec{\sigma})=\frac{1}{2}\,\delta\omega_{\sigma\rho}
\{{\cal A}^a(\tau,\vec{\sigma}),J^{\sigma\rho}(\tau)\}^*= {\cal
A}^b(\tau,\vec{\sigma})\,D_{ba}{}^{\sigma\rho}\delta\omega_{\sigma\rho}
\nonumber \\
&& \nonumber \\
&\Rightarrow& {\cal A}^a(\tau,\vec{\sigma})\rightarrow{\cal
A}^b(\tau,\vec{\sigma})\,R_{ba}(\Lambda,U).
 \label{b9}
\eea

Since we can also show that $\rho^a_U = \sgn\, \rho_{Ua}
\rightarrow \rho^b_U\, R_{ba}(\Lambda  ,U)$, we get that both
${\cal A}^a$, $\rho_{Ua}$ are Wigner spin 1 3-vectors.

\bigskip

While  $U^\mu(\tau)$ is a true four-vector

\begin{equation}
\{J^{\mu\nu}(\tau),U^\sigma(\tau)\}^*
=\eta^{\nu\sigma}U^\mu(\tau)-\eta^{\mu\sigma}U^\nu(\tau),
 \label{b10}
\end{equation}

\noindent on the contrary ${\tilde X}^{\mu}$ is {\it not a Lorentz
four-vector}, since we have

\begin{eqnarray}
\{J^{\mu\nu}(\tau),{\tilde X}^\sigma(\tau)\}^*
&=&\eta^{\nu\sigma}{\tilde X}^\mu(\tau)-\eta^{\mu\sigma}{\tilde X}
^\nu(\tau)+ \nonumber\\
 &&\nonumber\\
  &+&\frac{\partial D_{ab}{}^{
\mu\nu}(\hat U)} {\partial {\hat U}_\sigma} \,\int d^3\sigma\,
\left[ {\cal A}
^a(\tau,\vec{\sigma})\,\rho^b_U(\tau,\vec{\sigma})- {\cal A}
^b(\tau,\vec{\sigma})\,\rho^a_U(\tau,\vec{\sigma}) \right].
 \label{b11}
\end{eqnarray}

\newpage

\section{A Pseudo-Differential Operator.}

Let us consider the 3-metric (\ref{II23}) on a fixed space-like
hyper-plane $\Sigma_{\tau}$ (we omit the $\tau$-deppendence)

\begin{equation}
h_{rs}(\vec{\sigma})= \frac{\partial {\cal A}^a}{\partial
\sigma^r}(\vec{\sigma}) \,\delta_{ab}\, \frac{\partial {\cal A}
^b}{\partial \sigma^s}(\vec{\sigma}).
 \label{c1}
 \end{equation}

\noindent  If ${\cal A}^r_a(\vec \sigma )$ is the inverse  of the
matrix $\frac{\partial {\cal
A}^a(\vec{\sigma})}{\partial\sigma^s}$, we have

\beq
 {\cal A}^r_a(\vec{\sigma}) \cdot \frac{\partial
{\cal A}^a(\vec{\sigma})}{\partial\sigma^s} = \delta^r_s,\qquad
{\cal A}^r_a(\vec{\sigma}) \cdot \frac{\partial {\cal A}
^b(\vec{\sigma})}{\partial\sigma^r} = \delta^b_a,
 \label{c2}
 \eeq

\noindent and then the inverse 3-metric $h^{rs}(\vec{\sigma})$ is
given by

\begin{equation}
h^{rs}(\vec{\sigma})={\cal A}^r_a(\vec{\sigma}) \,\delta^{ab}\,
{\cal A}^s_b(\vec{\sigma}).
 \label{c3}
 \end{equation}

Since we have

\begin{equation}
\sqrt{h(\vec{\sigma})}=\det\left( \frac{\partial {\cal A}
^a(\vec{\sigma})}{\partial\sigma^r} \right),
 \label{c4}
 \end{equation}

\noindent we get

\bea
 \sqrt{h(\vec{\sigma})}\, {\cal A}^r_a(\vec{\sigma}) &=& \frac{1}{2!}\,
\varepsilon^{ruv}\, \varepsilon_{abc}\, \frac{\partial {\cal A}
^b(\vec{\sigma})}{\partial\sigma^u} \frac{\partial {\cal A}
^c(\vec{\sigma})}{\partial\sigma^v},\nonumber \\
 &&{}\nonumber \\
 \Rightarrow && \frac{\partial}{\partial\sigma^r} \left[\,
\sqrt{h(\vec{\sigma})}\,{\cal A}^r_a(\vec{\sigma})\, \right]=0.
 \label{c5}
 \eea

\bigskip

If we use the notation [Eq.(\ref{c5}) is used; the operators
${\hat k}_{ia}$ and $\triangle_{\eta_i}$ are self-adjoint with
respect to the scalar product (\ref{III11}), but not with respect
to the one (\ref{III38})]

\bea
 \Delta_{\eta_i} &=&
\frac{1}{\sqrt{h({\vec{\eta}}_i)}}\frac{\partial}{\partial\eta_i^r}
\left( h^{rs}({\vec{\eta}}_i)\sqrt{h({\vec{\eta}}_i)}
\frac{\partial}{\partial\eta^s_i}\right) = - {1\over {\hbar^2}}\,
\sum_a\, {\hat k}_{ia}\, {\hat k}_{ia},\nonumber \\
&&{\hat k}_{ia} = i\hbar\, {\cal A}^r_a(\tau ,{\vec \eta}_i)\,
{{\partial}\over {\partial \eta^r_i}},
 \label{c6}
 \eea

\noindent for the Laplace-Beltrami operator on $\Sigma_{\tau}$, we
may define the operator (\ref{III24}) in the following way

\begin{equation}
\sqrt{m^2c^2-\hbar^2\,\Delta_{\eta_i}}=
mc\,\sum_{n=0}^\infty\,c_n\,\left(
-\frac{\hbar^2}{mc}\,\Delta_{\eta_i}\right)^n,
 \label{c7}
 \end{equation}

\noindent where the $c_n$'s are the coefficients of the Taylor
expansion

\begin{equation}
\sqrt{1+x}=\sum_{n=0}^\infty\,c_n\,x^n.
 \label{c8}
 \end{equation}

\bigskip

I) By doing the following calculation

\begin{eqnarray*}
&& \Delta_\eta\,\exp\left(\, \frac{i}{\hbar}\,\vec{K}\cdot { \vec
{\cal A}}(\vec{\eta})\right)= \nonumber\\
 &=&\frac{1}{\sqrt{h(\vec{\eta})}}\frac{\partial}{\partial\eta^r}
\left[ h^{rs}(\vec{\eta})\sqrt{h(\vec{\eta})}
\frac{\partial}{\partial\eta^s} \,\exp\left(\,
\frac{i}{\hbar}\,\vec{K}\cdot {\vec {\cal A}}(\vec{\eta})\right)
\right]= \nonumber\\
 &&\nonumber\\
  &=&\frac{i}{\hbar}\,
\frac{1}{\sqrt{h(\vec{\eta})}}\frac{\partial}{\partial\eta^r}
\left[ h^{rs}(\vec{\eta})\sqrt{h(\vec{\eta})} \frac{\partial {\cal
A} ^c(\vec{\eta})}{\partial\eta^s}\,K_c \,\exp\left(\,
\frac{i}{\hbar}\,\vec{K}\cdot {\vec {\cal A}}(\vec{\eta})\right)
\right]= \nonumber\\
 &&\nonumber\\
  &&\mbox{ by\, using\, Eq.(\ref{c3}))} \nonumber\\
 &&\nonumber\\
 &=&\frac{i}{\hbar}\,
\frac{1}{\sqrt{h(\vec{\eta})}}\frac{\partial}{\partial\eta^r}
\left[ \sqrt{h(\vec{\eta})}\, {\cal A}^r_a(\vec{\sigma})
\,\delta^{ab}\, {\cal A}^s_b(\vec{\sigma})\, \frac{\partial {\cal
A} ^c(\vec{\eta})}{\partial\eta^s}\,K_c \,\exp\left(\,
\frac{i}{\hbar}\,\vec{K}\cdot {\vec {\cal A}}(\vec{\eta})\right)
\right]=\nonumber\\
 &&\nonumber\\
  &&\mbox{ by\, using\, Eq.(\ref{c2}) }\nonumber\\
   &&\nonumber\\
    &=&\frac{i}{\hbar}\,
\frac{1}{\sqrt{h(\vec{\eta})}}\frac{\partial}{\partial\eta^r}
\left[ \sqrt{h(\vec{\eta})}\, {\cal A}^r_a(\vec{\sigma})
\,\delta^{ab}\, \,K_b \,\exp\left(\, \frac{i}{\hbar}\,\vec{K}\cdot
{\vec {\cal A}}(\vec{\eta})\right) \right]=\nonumber\\
 &&\nonumber\\
 &&\mbox{ by\, using\, Eq.(\ref{c5}) }\nonumber\\
  &&\nonumber\\
 &=&\frac{i}{\hbar}\, {\cal A}^r_a(\vec{\sigma}) \,\delta^{ab}\, \,K_b
\frac{\partial}{\partial\eta^r} \,\exp\left(\,
\frac{i}{\hbar}\,\vec{K}\cdot {\vec {\cal A}}(\vec{\eta})\right)
=\nonumber\\
 &&\nonumber\\
  &=&-\frac{1}{\hbar^2}\,
{\cal A}^r_a(\vec{\sigma}) \,\delta^{ab}\, \,K_b \frac{\partial
{\cal A}^c(\vec{\eta})}{\partial\eta^r}\,K_c \,\exp\left(\,
\frac{i}{\hbar}\,\vec{K}\cdot {\vec {\cal A}}(\vec{\eta})\right)
=\nonumber\\
 &&\nonumber\\
  &&\mbox{ by\, using\, Eq.(\ref{c2}) }
  \end{eqnarray*}

\bea
 &=&-\frac{K_a\,\delta^{ab}\,
\,K_b}{\hbar^2}\,\exp\left(\, \frac{i}{\hbar}\,\vec{K}\cdot {\vec
{\cal A}}(\vec{\eta})\right) \nonumber\\
 &&\nonumber\\
 &=&-\frac{\vec{K}^2}{\hbar^2}\,\exp\left(\,
\frac{i}{\hbar}\,\vec{K}\cdot {\vec {\cal A}}(\vec{\eta})\right),
 \label{c9}
 \end{eqnarray}

\noindent we arrive at the result

\begin{eqnarray}
&&\sqrt{m^2c^2-\hbar^2\,\Delta_\eta}\,\exp\left(\,
\frac{i}{\hbar}\,\vec{K}\cdot {\vec {\cal A}}(\vec{\eta})\right)=
\nonumber\\
 &&\nonumber\\
  &=&mc\,\sum_{n=0}^\infty\,c_n\,\left(
-\frac{\hbar^2}{mc}\,\Delta_\eta\right)^n\,\exp\left(\,
\frac{i}{\hbar}\,\vec{K}\cdot {\vec {\cal A}}(\vec{\eta})\right)=
\nonumber\\
 &&\nonumber\\
  &=&mc\,\sum_{n=0}^\infty\,c_n\,\Big(
\frac{\vec{K}^2}{mc}\Big)^n\,\exp\left(\,
\frac{i}{\hbar}\,\vec{K}\cdot {\vec {\cal A}}(\vec{\eta})\right)=
\nonumber\\
 &&\nonumber\\
  &=&\sqrt{m^2c^2+\vec{K}^2}\,\exp\left(\,
\frac{i}{\hbar}\,\vec{K}\cdot {\vec {\cal A}}(\vec{\eta})\right).
 \label{c10}
 \end{eqnarray}

\bigskip

II) Given a function  $f(\vec{\eta})$ let us introduce its {\em
transform}

\begin{equation}
F(\vec{K})=\frac{1}{(2\pi)^{3/2}} \int
d^3\eta'\,\sqrt{h(\vec{\eta}^{\,\prime})}\,
f(\vec{\eta}^{\,\prime})\,\exp\left(\,-
\frac{i}{\hbar}\,\vec{K}\cdot {\vec {\cal
A}}(\vec{\eta}^{\,\prime}) \right),
 \label{c11}
 \end{equation}

\noindent with {\em anti-transform}

\begin{equation}
f(\vec{\eta})=\frac{1}{(2\pi)^{3/2}} \int d^3K\,
F(\vec{K})\,\exp\left(\, \frac{i}{\hbar}\,\vec{K}\cdot { \vec
{\cal A}}(\vec{\eta}) \right).
 \label{c12}
 \end{equation}

By substituting Eq.(\ref{c11}) into Eq.(\ref{c12}) we get

\begin{eqnarray}
f(\vec{\eta})&=&\frac{1}{(2\pi)^3} \int d^3K\,\int
d^3\eta'\,\sqrt{h(\vec{\eta}^{\,\prime})}\,
f(\vec{\eta}^{\,\prime})\, \exp\left[\,
\frac{i}{\hbar}\,\vec{K}\cdot \left({\vec {\cal A}}(\vec{\eta})-
{\vec {\cal A}}(\vec{\eta}^{\,\prime}) \right)\right]=\nonumber\\
 &&\nonumber\\
  &=& \int d^3\eta'\,\sqrt{h(\vec{\eta}^{\,\prime})}\,
f(\vec{\eta}^{\,\prime})\, \delta\left({\vec {\cal
A}}(\vec{\eta})- {\vec {\cal A}}(\vec{\eta}^{\,\prime}) \right),
 \label{c13}
 \end{eqnarray}

\noindent which is an identity due to

\begin{equation}
\sqrt{h(\vec{\eta}^{\,\prime})}\, \delta\left({\vec {\cal A}
}(\vec{\eta})- {\vec {\cal A}}(\vec{\eta}^{\,\prime})\right)=
\delta(\vec{\eta}^{\,\prime}-\vec{\eta}).
 \label{c14}
 \end{equation}

\bigskip

III) Finally by combining Eq.(\ref{c10}) with Eq.(\ref{c13}) we
get

\begin{eqnarray}
&&\sqrt{m^2c^2-\hbar^2\,\Delta_\eta}\,f(\vec{\eta})= \nonumber\\
 &&\nonumber\\
 &=& \frac{1}{(2\pi)^{3/2}} \int
d^3K\,F(\vec{K})\,\sqrt{m^2c^2+\vec{K}^2}\; \exp\left(\,
\frac{i}{\hbar}\,\vec{K}\cdot {\vec {\cal A}}(\vec{\eta}) \right),
\label{c15}
 \end{eqnarray}

\noindent so that, by substituting $F(\vec{K})$ with its
expression (\ref{c11}), we get the integral representation
(\ref{III25}).

\bigskip

Let us now look for a general solution of Eqs.(\ref{III31}). To
this end let us introduce a new function $\widetilde{\Psi}(\theta,
\vec{x}_1,...,\vec{x}_N, \vec k)$, which is completely arbitrary
at this stage. Let us assume, as it is done in Eq.(\ref{III45}),
that a general solution of Eq.(\ref{III31}) may be written in the
form

\begin{equation}
\Psi({\vec \eta}_i;\theta,{\cal A}^a]=
\widetilde{\Psi}(\theta,{\vec {\cal A} }(\vec{\eta}_1),...,{\vec
{\cal A}}(\vec{\eta}_N)),
 \label{c16}
 \end{equation}

\noindent so that we get

\begin{eqnarray}
&& \left[ i\hbar\frac{\partial {\cal A}
^a(\vec{\sigma})}{\partial\sigma^r}\, \frac{\delta}{\delta {\cal
A} ^a(\vec{\sigma})}-i\hbar\,
\sum_i\,\delta(\vec{\sigma}-\vec{\eta}_i)\,
\frac{\partial}{\partial\eta^r_i}\right]\; \Psi
({\vec \eta}_i;\theta,{\cal A}^a]=\nonumber
\\
 &&\nonumber\\
  &=&\left[ i\hbar\frac{\partial {\cal A}^a(\vec{\sigma})}{\partial\sigma^r}\,
\frac{\delta}{\delta {\cal A}^a(\vec{\sigma})}-i\hbar\,
\sum_i\,\delta(\vec{\sigma}-\vec{\eta}_i)\,
\frac{\partial}{\partial\eta^r_i}\right]\;
\widetilde{\Psi}(\theta,{\vec {\cal A} }(\vec{\eta}_1),...,{\vec
{\cal A}}(\vec{\eta}_N)) =0.
 \label{c17}
 \end{eqnarray}

Now the chain-rule gives the following result

\begin{eqnarray}
\frac{\partial}{\partial\eta^r_i}\, \Psi({\vec \eta}_i;\theta,{\cal A}^a] &=&
\frac{\partial}{\partial\eta^r_i}\, \widetilde{\Psi}(\theta,{\vec
{\cal A} }(\vec{\eta}_1),...,{\vec {\cal A}}(\vec{\eta}_N)) =\nonumber\\
 &&\nonumber\\
  &=& \frac{\partial
{\cal A}^a(\vec{\eta}_i)}{\partial\eta^r_i}\, \left[
\frac{\partial \widetilde{\Psi}(\theta,\vec{x}_1,...,\vec{x}_N)}
{\partial x^a_i} \right]_{\vec{x}_i={\vec {\cal
A}}(\vec{\eta}_i)}.
 \label{c18}
 \end{eqnarray}

Moreover by definition we get

\begin{eqnarray*}
&&\delta \Psi({\vec \eta}_i;\theta,{\cal A}^a]=
\delta \widetilde{\Psi}(\theta, {\vec {\cal A}
}(\vec{\eta}_1),...,{\vec {\cal A}}(\vec{\eta}_N))=\nonumber\\
 &&\nonumber\\
 &=&\Psi({\vec \eta}_i;\theta,{\cal A}^a+\delta{\cal A}^a]-
\Psi(({\vec \eta}_i;\theta,{\cal A}^a]=\nonumber\\
 &&\nonumber\\
  &=&\widetilde{\Psi}(\theta, {\vec {\cal A}}(\vec{\eta}_1)+\delta
{\vec {\cal A}}(\vec{\eta}_1),..., {\vec {\cal A} }(\vec{\eta}_N )
+ \delta{\vec {\cal A}}(\vec{\eta}_N))-
\widetilde{\Psi}(\theta,{\vec {\cal A} }(\vec{\eta}_1),...,{\vec
{\cal A}}(\vec{\eta}_N))=
 \end{eqnarray*}

\bea
  &=& \sum_{i=1}^N\,\left[ \frac{\partial
\widetilde{\Psi}(\theta,\vec{x}_1,...,\vec{x}_N)}
{\partial x^a_i} \right]_{\vec{x}_i={\vec {\cal
A}}(\vec{\eta}_i)}\;\delta {\cal A} ^a(\vec{\eta}_i)=\nonumber\\
 &&\nonumber\\
  &=& \int
d^3\sigma\,\frac{\delta\Psi({\vec \eta}_i;\theta,{\cal A}^a]}{\delta {\cal
A}^a(\vec{\sigma})}\cdot\delta {\cal A}^a(\vec{\sigma})=
\nonumber\\
 &&\nonumber\\
  &=& \int
d^3\sigma\,\frac{\delta\widetilde{\Psi}(\theta, {\vec {\cal A}
}(\vec{\eta}_1),...,{\vec {\cal A}}(\vec{\eta}_N))}{\delta
{\cal A} ^a(\vec{\sigma})}\cdot\delta {\cal A}^a(\vec{\sigma}),
 \label{c19}
\end{eqnarray}

\noindent so that

\begin{eqnarray}
\frac{\delta\Psi({\vec \eta}_i;\theta,{\cal A}^a]}{\delta {\cal A}^a(\vec{\sigma})}&=&
\frac{\delta\widetilde{\Psi}(\theta,{\vec {\cal A}
}(\vec{\eta}_1),...,{\vec {\cal A}}(\vec{\eta}_N))}{\delta
{\cal A} ^a(\vec{\sigma})}=\nonumber\\
 &&\nonumber\\
  &=&\sum_{i=1}^N\,\left[ \frac{\partial
\widetilde{\Psi}(\theta,\vec{x}_1,...,\vec{x}_N)}
{\partial x^a_i} \right]_{\vec{x}_i={\vec {\cal
A}}(\vec{\eta}_i)}\; \delta(\vec{\sigma}-\vec{\eta}_i).
 \label{c20}
 \end{eqnarray}

Eq.(\ref{c17}) is a direct consequence of Eqs.(\ref{c18}) and
(\ref{c20}). As a consequence, functionals of the variables ${\cal
A}^a(\vec{\sigma})$ of the form (\ref{III45}), namely
Eq.(\ref{c16}), {\it solve three of the four constraints}. Let us
now look for the condition to be imposed on the function
$\widetilde{\Psi}$ to satisfy the equation

\begin{eqnarray}
&& \left[ i\hbar\,\frac{\partial}{\partial \theta}
-\sum_i\,\widehat{\cal R}_i\right]\; \Psi({\vec \eta}_i;\theta,{\cal A}^a]=\nonumber\\
 &&\nonumber\\
 &=&\left[ i\hbar\,\frac{\partial}{\partial \theta}
-\sum_i\,\widehat{\cal R}_i\right]\; \widetilde{\Psi}(\theta,{\vec
{\cal A}}(\vec{\eta}_1),...,{\vec {\cal A}}(\vec{\eta}_N))
=0.
 \label{c21}
\end{eqnarray}

Since we have

\begin{eqnarray*}
&&\widehat{\cal R}_i\;\widetilde{\Psi}(\theta,{\vec {\cal A}
}(\vec{\eta}_1),...,{\vec {\cal A}}(\vec{\eta}_N))
=\nonumber\\
 &&\nonumber\\
   &=& \frac{1}{(2\pi)^3}\,
\int d^3K\sqrt{m_i^2\,c^2+\vec{K}^2}\cdot\nonumber\\
 &&\nonumber\\
 &&\cdot\int \sqrt{h(\vec{\eta}')}\,d^3\eta' \; \widetilde{\Psi}
(\theta,{\vec {\cal A} }(\vec{\eta}_1),..., {\vec {\cal A}
}(\vec{\eta}_i^{\,\prime}),...,{\vec {\cal A}}(\vec{\eta}_N))
\;e^{\;\frac{i}{\hbar}\vec{K}\cdot ({\vec {\cal A} }({\vec
\eta}_i)-{\vec {\cal A}}(\vec{\eta}{}^{'}))}=
 \end{eqnarray*}

 \bea
   &=& \left[ \frac{1}{(2\pi)^3}\, \int
d^3K\sqrt{m_i^2\,c^2+\vec{K}^2} \int d^3x'_i\,
\widetilde{\Psi}(\theta,\vec{x}_1,...,{x}_i^{\,\prime},...,\vec{x}_N)
e^{\;\frac{i}{\hbar}\vec{K}\cdot
(\vec{x_i}-\vec{x}_i^{\,\prime})}\right]_{\vec{x}_i={\vec {\cal A}
}(\vec{\eta}_i)}= \nonumber\\
 &&\nonumber\\
  &&\nonumber\\
   &=&\left[
\sqrt{m_i^2c^2-\hbar^2\,\frac{\partial}{\partial x^a_i}\,
\delta^{ab}\,\frac{\partial}{\partial x^b_i}}
\;\widetilde{\Psi}(\theta,\vec{x}_1,...,{x}_i^{\,\prime},...,\vec{x}_N)
\right]_{\vec{x}_i={\vec {\cal A}}(\vec{\eta}_i)}.
 \label{c22}
\end{eqnarray}

\noindent then, by using Eq.(\ref{c21}), the condition turn out to
be

\begin{eqnarray}
&& \left[ i\hbar\,\frac{\partial}{\partial \theta}
-\sum_i\,\widehat{\cal R}_i\right]\; \Psi({\vec \eta}_i;\theta,{\cal A}^a]
=\nonumber \\
 &&\nonumber\\
  &=&\left[ i\hbar\,\frac{\partial}{\partial \theta}
-\sum_i\,\widehat{\cal R}_i\right]\; \widetilde{\Psi}(\theta,{\vec
{\cal A}}(\vec{\eta}_1),...,{\vec {\cal A}}(\vec{\eta}_N))
=\nonumber\\
 &&\nonumber\\
  &=& \left[\left( i\hbar\,\frac{\partial
}{\partial \theta} -\sum_i
\sqrt{m_i^2c^2-\hbar^2\,\frac{\partial}{\partial x^a_i}\,
\delta^{ab}\,\frac{\partial}{\partial x^b_i}} \;\right)
\widetilde{\Psi}(\theta,\vec{x}_1,...,{x}_i^{\,\prime},...,\vec{x}_N)
\right]_{\vec{x}_j={\vec {\cal
A}}(\vec{\eta}_j)}=0,\nonumber \\
&&{}
 \label{c23}
\end{eqnarray}

Since the variables ${\vec {\cal A}}(\vec{\sigma})$ may be chosen
arbitrarily, Eq.(\ref{III57}) is automatically implied.

\bigskip

Let us now consider the frame-independent Hilbert space
${\mbox{\boldmath{${\cal H}$}}}$ of Subsection IIIB3. Its wave functions are defined in
Eq.(\ref{III37}) and the scalar product is given in
Eq.(\ref{III38}). If we put $\Phi = \widehat{\Phi}/\sqrt{\prod_i\det\,
\Big({{\partial {\cal A}^a({\vec \eta}_i)}\over {\partial
\eta^r_i}}\Big)}$ in Eq.(\ref{III26}), they can be rewritten in
the form

\bea
 \widehat{H}^{'}_{\perp}\, \widehat{\Phi} &=& \sqrt{\prod_i\det\, \Big({{\partial
 {\cal A}^a({\vec \eta}_i)}\over {\partial \eta^r_i}}\Big)}\,
 \widehat{H}_{\perp}\, {1\over {\sqrt{\prod_i\det\, \Big({{\partial
 {\cal A}^a({\vec \eta}_i)}\over {\partial \eta^r_i}}\Big)}\,}}\,
 \widehat{\Phi} = 0,\nonumber \\
  &&{}\nonumber \\
 {\widehat{\cal H}}^{'}_a(\vec \sigma )\, \widehat{\Phi} &=& \sqrt{\prod_i\det\,
 \Big({{\partial
 {\cal A}^a({\vec \eta}_i)}\over {\partial \eta^r_i}}\Big)}\,
 {\cal A}^r_a(\vec \sigma )\, {\widehat{\cal H}}_r\, {1\over {\sqrt{\prod_i\det\, \Big({{\partial
 {\cal A}^a({\vec \eta}_i)}\over {\partial \eta^r_i}}\Big)}\,}}\,
 \widehat{\Phi} = 0.
 \label{c24}
 \eea
\medskip

The explicit evaluation of $\widehat{H}^{'}_{\perp}$ leads to define
the following new pseudo-differential operator

\beq
 {\widehat{\cal R}}_i^{'} = \sqrt{\prod_i\det\, \Big({{\partial
 {\cal A}^a({\vec \eta}_i)}\over {\partial \eta^r_i}}\Big)}\,
 {\widehat{\cal R}}_i\, {1\over {\sqrt{\prod_i\det\, \Big({{\partial
 {\cal A}^a({\vec \eta}_i)}\over {\partial \eta^r_i}}\Big)}\,}} =
 \sqrt{m^2 - \hbar^2\, \triangle^{'}_{\eta_i}},
 \label{c25}
 \eeq

\noindent where the covariant Laplace-Beltrami operator
$\triangle_{\eta_i}$ of Eq.(\ref{c9}) has been replaced with the
new Laplacian

\bea
 \triangle_{\eta_i}^{'} &=& \sqrt{\prod_i\det\, \Big({{\partial
 {\cal A}^a({\vec \eta}_i)}\over {\partial \eta^r_i}}\Big)}\,
 \triangle_{\eta_i}\, {1\over {\sqrt{\prod_i\det\, \Big({{\partial
 {\cal A}^a({\vec \eta}_i)}\over {\partial \eta^r_i}}
 \Big)}\,}},\nonumber  \\
 &&{}\nonumber \\
 - \hbar^2\, \triangle^{'}_{\eta_i} &=& \sum_a\, {\hat k}_{ia}^{'}\,
 {\hat k}_{ia}^{'},\nonumber \\
&&\nonumber\\
  {\hat k}_{ia}^{'} &=&
\sqrt{\prod_i\det\, \Big({{\partial
 {\cal A}^a({\vec \eta}_i)}\over {\partial \eta^r_i}}\Big)}\,
 {\hat k}_a\, {1\over {\sqrt{\prod_i\det\, \Big({{\partial
 {\cal A}^a({\vec \eta}_i)}\over {\partial \eta^r_i}}\Big)}\,}} =
 {1\over 2}\, \Big[{\cal A}^r_a(\tau ,{\vec \eta}_i), i\hbar\,
 {{\partial}\over {\partial \eta^r_i}}\Big]_{+},
 \label{c26}
 \eea

\noindent where ${\hat k}_{ia}$ is defined in Eq.(\ref{c9}). The
operators ${\hat k}_{ia}^{'}$ are obtained from the corresponding
classical quantities ${\cal A}^r_a(\tau ,{\vec \eta}_i)\,
\kappa_{ir}$ by means  of the symmetrization ordering rule. They
and $\triangle^{'}_{\eta_i}$ are self-adjoint  operators with
respect to the scalar product (\ref{III38}), but not with respect
to the one (\ref{III11}).
\bigskip

The explicit evaluation of ${\widehat{\cal H}}^{'}_a(\vec \sigma )$
leads to the result

\beq
 {\widehat{\cal H}}^{'}_a(\vec \sigma ) = i\hbar\, {{\delta}\over
 {\delta {\cal A}^a(\vec \sigma )}} - {1\over 2}\, \sum_{i=1}^N\,
 \Big[{\cal A}^r_a({\vec \eta}_i)\, \delta^3(\vec \sigma - {\vec \eta}_i),
 i\hbar\, {{\partial}\over {\partial \eta^r_i}}\Big]_{+},
 \label{c27}
 \eeq

 \noindent containing the symmetrization ordering rule for the
 classical quantities $\delta^3(\vec \sigma - {\vec \eta}_i)\,
 {\cal A}^r_a({\vec \eta}_i)\, \kappa_{ir}$. In this way we get
 Eqs.(\ref{III39}).
 \bigskip

 Finally, Eqs.(\ref{III70}) can be checked by putting $\widehat
 {\psi}_c = \sqrt{\prod_i\det\, \Big({{\partial
 {\cal A}_c^a(\tau,{\vec \eta}_i)}\over {\partial \eta^r_i}}\Big)}\,
 \psi_c$ in Eq.(\ref{III64}) and by defining
 \begin{eqnarray*}
 \widehat{H}^{'}_{ni} &=&
 \sqrt{\prod_i\det\, \Big({{\partial
 {\cal A}^a_c(\tau,{\vec \eta}_i)}\over {\partial \eta^r_i}}\Big)}\,
 \widehat{H}_{ni}\,
 {1\over {\sqrt{\prod_i\det\, \Big({{\partial
 {\cal A}^a_c(\tau,{\vec \eta}_i)}\over {\partial \eta^r_i}}
 \Big)}\,}}+\\
 &&\\
 &+&i\hbar{1\over {\sqrt{\prod_i\det\, \Big({{\partial
 {\cal A}^a_c(\tau,{\vec \eta}_i)}\over {\partial \eta^r_i}}
 \Big)}\,}}\,\frac{\partial}{\partial\tau}\,
 \sqrt{\prod_i\det\, \Big({{\partial
 {\cal A}^a_c(\tau,{\vec \eta}_i)}\over {\partial
 \eta^r_i}}\Big)}.
 \end{eqnarray*}
Again we obtain that the classical inertial potentials are
replaced by operators determined with the symmetrization ordering
rule.

\vfill\eject

\section{Relativistic Positive-Energy Spinning Particles.}

In this Appendix we delineate the treatment of N free
positive-energy spinning particles, in the semi-classical
approximation of describing the spin with Grassmann variables
\cite{43}, in the non-inertial frames of Subsection IIB. In
Subsection A we adapt the treatment of positive-energy spinning
particles on the Wigner hyper-planes of the rest-frame instant
form given in Ref.\cite{44} to the generic hyper-planes of our
non-inertial frames. Then in Subsection B we make the
multi-temporal quantization.

\subsection{N Free Semi-Classical Spinning Particles.}

Since in parametrized Minkowski theories we can describe only
particles with a definite sign of the energy, the position of the
spinning particles are described by the configuration variables
${\vec \eta}_i(\tau )$ defined by the embedding by means of
Eq.(\ref{II3}). Having positive energy the spin of the particle
has to be described by three Grassmann variables, which after
quantization will become two-by-two Pauli matrices ($\xi^a \mapsto
\sqrt{{{\hbar}\over 2}}\, \sigma_a$), like it happens for the
Dirac particle after the Foldy-Wouthuysen transformation. To
preserve manifest Lorentz covariance we associate to each spinning
particle a Grassmann 4-vector $\xi^{\mu}_i(\tau )$ [$\xi^{\mu}_i\,
\xi^{\nu}_j + \xi^{\nu}_j\, \xi^{\mu}_i = 0$] and then we will
introduce  suitable constraints to eliminate one of its
components. In the rest-frame instant form, where $p_{\mu} = \int
d^3\sigma\, \rho_{\mu}(\tau ,\vec \sigma )$ of Eqs.(\ref{II11}) is
equal to the conserved total particle 4-momentum and is the normal
to the Wigner hyper-plane, these constraints are $\phi_i \approx
i\, \xi^{\mu}_i\, p_{\mu} \approx 0$ \cite{44}. Instead here,
where the relevant embeddings are given by Eqs.(\ref{II23}), the
normal to the hyper-planes is the extra dynamical variable ${\hat
U}^{\mu}(\tau )$. As a consequence we shall introduce the
constraints $\phi_i \approx i\, \xi^{\mu}_i(\tau )\, {\hat
U}_{\mu}(\tau ) \approx 0$.

\medskip

Therefore, instead of the Lagrangian given by Eqs.(\ref{II17}) and
(\ref{II4}), we introduce the new action

\bea
  S &=& \int d\tau\, L(\tau ) = \int d\tau d^3\sigma\, {\cal
 L}(\tau ,\vec \sigma ) =\nonumber \\
 &=& \int d\tau \Big(- \sqrt{\sgn\, {\dot X}^2(\tau )}
 - {1\over 2}\, \sum_{i=1}^N\, \xi_{i\mu}(\tau )\,
 {\dot \xi}^{\mu}_i(\tau )
 + \sum_{i=1}^N\, \lambda_i(\tau )\, i\, \xi_{i\, \mu}(\tau )\,
 {{{\dot X}^{\mu}(\tau )}\over {\sqrt{\sgn\, {\dot X}^2(\tau )}}}
-\nonumber \\
 &-& \int d^3\sigma\, \sum_i
m_i\,\delta^3(\vec{\sigma}-\vec{\eta}_i(\tau)) \sqrt{\sgn\, [
g_{\tau\tau}(\tau,\vec{\sigma}) +2g_{\tau
{r}}(\tau,\vec{\sigma})\dot{\eta}^{{r}}_i(\tau)
+g_{{r}{s}}(\tau,\vec{\sigma})
\dot{\eta}^{{r}}_i(\tau)\dot{\eta}^{{s}}_i(\tau)]}\Big),\nonumber
\\
 &&{}
  \label{d1}
 \end{eqnarray}

\noindent where the second term in the second line is the kinetic
term for the Grassmann variables. The configuration variables
$\lambda_i(\tau )$ are Lagrange multipliers to implement the
constraints $\phi_i \approx 0$.
\medskip

The momenta $\rho_{\mu}(\tau ,\vec \sigma )$ and $\kappa_{ir}(\tau
)$ are still given by Eqs.(\ref{II5}) and Eqs.(\ref{II6}),
(\ref{II7}) and (\ref{II8}) are still valid. The momenta
$\pi_{\lambda_i}(\tau )$ conjugate to the Lagrange multipliers
vanish and satisfy the Poisson brackets $\{ \lambda_i(\tau ),
\pi_{\lambda_j}(\tau )\} = \delta_{ij}$. In Eqs.(\ref{II18}) the
momentum of the extra particle has the following modification

\beq
 U^{\mu}(\tau ) = {{{\dot X}^{\mu}(\tau )}\over {\sqrt{\sgn\,
 {\dot X}^2(\tau )}}} - \Big(\eta^{\mu\nu} - {{{\dot X}^{\mu}(\tau )\,
 {\dot X}^{\nu}(\tau )}\over {{\dot X}^2(\tau )}}\Big)\,
 {{\sum_{i=1}^N\, \lambda_i(\tau )\, \xi_{i\nu}(\tau )}\over
 {\sqrt{\sgn\, {\dot X}^2(\tau )}}}.
 \label{d2}
 \eeq

The momenta conjugate to the Grassmann variables are
$\pi_i^{\mu}(\tau ) = {i\over 2}\, \xi^{\mu}_i(\tau )$ and satisfy
the Poisson brackets $\{ \xi^{\mu}_i(\tau ), \pi^{\nu}_j(\tau )\}
= - \delta_{ij}\, \eta^{\mu\nu}$.

\bigskip

The primary constraints are ${\cal H}_{\mu}(\tau ,\vec \sigma )
\approx 0$ of Eqs.(\ref{II9}), $\chi^{\mu}_i = \pi^{\mu}_i(\tau )
- {i\over 2}\, \xi^{\mu}_i(\tau ) \approx 0$,
$\pi_{\lambda_i}(\tau ) \approx 0$ and $\chi = \sgn\, U^2(\tau ) -
1 \approx 0$ of Eqs.(\ref{II18}).

Since the canonical Hamiltonian is $H_c = - \sum_{i=1}^N\,
\lambda_i(\tau )\, \xi_{i \mu}(\tau )\, U^{\mu}(\tau )$, the
time-preservation of the constraints $\pi_{\lambda_i}(\tau )
\approx 0$ induces the secondary constraints $\phi_i = i\, \xi_{i
\mu}(\tau )\, U^{\mu}(\tau ) \approx (\pi^{\mu}_i(\tau ) + {i\over
2}\, \xi^{\mu}_i(\tau ))\, {\hat U}^{\mu}(\tau ) \approx 0$.

\medskip

Since we have $\{ \chi^{\mu}_i(\tau ), \chi^{\nu}_j(\tau )\} = i\,
\delta_{ij}\, \eta^{\mu\nu}$, $\{\chi^{\mu}_i(\tau ), \phi_j(\tau
)\} = 0$, $\{ \phi_i(\tau ), \phi_j(\tau )\} = - i\, \delta_{ij}$,
the Grassmann constraints $\chi^{\mu}_i \approx 0$ and
$\phi_i(\tau ) \approx 0$ are second class. All the other
constraints are first class. By eliminating the gauge variables
$\lambda_i(\tau )$ with the gauge fixing constraints
$\lambda_i(\tau ) \approx 0$, the extra particle momentum
$U^{\mu}(\tau )$ assumes the form of Eqs.(\ref{II18}).

\medskip

The Poincare' generators (\ref{II21}) have now the form

\bea
 p^{\mu} &=& U^{\mu}(\tau ) + \int d^3\sigma\, \rho^{\mu}(\tau
 ,\vec \sigma ),\nonumber \\
 &&{}\nonumber \\
 J^{\mu\nu} &=&  X^{\mu}(\tau )\, U^{\nu}(\tau ) - X^{\nu}(\tau
 )\, U^{\mu}(\tau ) + \int d^3\sigma\, [z^{\mu}\, \rho^{\nu} -
 z^{\nu}\, \rho^{\mu}](\tau ,\vec \sigma ) -\nonumber \\
 &-& \sum_{i=1}^N\, [\xi^{\mu}_i(\tau )\, \pi^{\nu}_i(\tau ) -
 \xi^{\nu}_i(\tau )\, \pi^{\mu}_i(\tau )] \approx\nonumber \\
 &\approx& X^{\mu}(\tau )\, {\hat U}^{\nu}(\tau ) - X^{\nu}(\tau
 )\, {\hat U}^{\mu}(\tau ) + \int d^3\sigma\, [z^{\mu}\, \rho^{\nu} -
 z^{\nu}\, \rho^{\mu}](\tau ,\vec \sigma ) - i\, \sum_{i=1}^N\,
 \xi^{\mu}_i(\tau )\, \xi^{\nu}_i(\tau ) =\nonumber \\
 &{\buildrel {def}\over =}& L^{\mu\nu} + S^{\mu\nu}_z +
 S^{\mu\nu}_{\xi}.
 \label{d3}
 \eea
\medskip

The second class constraints can be eliminated with the Dirac
brackets

\bea
 \{ A, B\}^* &=& \{ A, B\} + i\, \sum_{i=1}^N\, \Big[ \{A,
 \chi^{\mu}_i\}\, \{\chi_{i \mu}, B\} - \{ A, \phi_i\}\,
 \{ \phi_i, B\}\Big],\nonumber \\
 &&{}\nonumber \\
 &&\Downarrow\nonumber \\
 &&{}\nonumber \\
 &&\xi_{i \mu}(\tau )\, {\hat U}^{\mu}(\tau ) \equiv 0,\qquad
 S^{\mu\nu}_{\xi} \equiv - i\, \sum_{i=1}^N\, \xi^{\mu}_i\,
 \xi^{\nu}_i,\nonumber \\
 &&\{ \xi^{\mu}_i(\tau ), \xi^{\nu}_j(\tau )\}^* = i\,
 \delta_{ij}\, (\eta^{\mu\nu} - \sgn\, {\hat U}^{\mu}(\tau )\, {\hat
 U}^{\nu}(\tau )) = - \sgn\, \delta_{ij}\, \sum_a\,
 \epsilon^{\mu}_a(\hat U)\, \epsilon^{\nu}_a(\hat U),
 \label{d4}
 \eea

 \noindent where we used the notations of Appendix A.

 \bigskip

 Therefore, like in Ref.\cite{44}, only the following three
 Grassmann variables (a Wigner spin 1 3-vector) of each spinning
 particle survive

\bea
 \xi^a_i(\tau ) &=& \epsilon^a_{\mu}(\hat U)\, \xi^{\mu}_i(\tau
 ),\qquad \xi^{\mu}_i \equiv \epsilon^{\mu}_a(\hat U)\, \xi^a_i,\nonumber \\
 &&{}\nonumber  \\
 &&\{ \xi^a_i(\tau ), \xi^b_j(\tau )\}^* = - \sgn\, \delta_{ij}\,
 \delta^{ab},\nonumber \\
&&\nonumber\\
 &&\{ X^{\mu}(\tau ), \xi^a_i(\tau )\}^* = - {{\partial
 \epsilon^a_{\nu}(\hat U)}\over {\partial U_{\mu}}}\,
 \xi^{\nu}_i(\tau ),\nonumber \\
 &&{}\nonumber \\
 &&S^{\mu\nu}_{\xi} \equiv \epsilon^{\mu}_a(\hat U)\,
 \epsilon^{\nu}_b(\hat U)\, {\bar S}^{ab}_{\xi},\qquad
 {\bar S}^{ab}_{\xi} = - i\, \sum_{i=1}^N\, \xi^a_i\, \xi^b_i,
 \nonumber \\
 &&{\bar S}^a_{\xi} = {1\over 2}\, \epsilon^{abc}\, {\bar
 S}_{\xi}^{bc} = \sum_{i=1}^N\, {\bar S}^a_{\xi\, i},\qquad
 {\bar S}^a_{\xi\, i} = - {i\over 2}\, \epsilon^{abc}\, \xi^b_i\,
 \xi^c_i.
 \label{d5}
 \eea

As in Ref.\cite{44}, a Darboux basis for the Dirac brackets
requires to replace the 4-vector $X^{\mu}$ [$\{ X^{\mu}(\tau ),
X^{\nu}(\tau )\}^* = S^{\mu\nu}_{\xi}$] with a canonical
non-covariant Newton-Wigner-like variable ${\hat X}^{\mu}$

\bea
 {\hat X}^{\mu} &=& X^{\mu} - {1\over 2}\, \epsilon^A_{\nu}(\hat
 U)\, \eta_{AB}\, {{\partial \epsilon^B_{\rho}(\hat U)}\over
 {\partial U_{\mu}}}\, S^{\mu\nu}_{\xi},\nonumber \\
 &&{}\nonumber \\
 &&\{ {\hat X}^{\mu}(\tau ), {\hat X}^{\nu}(\tau )\}^* =
 \{ {\hat X}^{\mu}(\tau ), \xi^a_i(\tau )\}^* = 0,\qquad
 \{{\hat X}^{\mu}(\tau ), U^{\nu}(\tau )\}^* = - \eta^{\mu\nu}.
 \label{d6}
 \eea

As a consequence, by using the same $D$-matrix of
Eqs.(\ref{II37}), we get

\bea
 J^{\mu\nu} &=& {\hat L}^{\mu\nu} +
 S^{\mu\nu}_z + {\hat S}^{\mu\nu}_{\xi} ,\nonumber \\
 &&{}\nonumber \\
 {\hat L}^{\mu\nu} &=& {\hat X}^{\mu}\, {\hat U}^{\nu} -
 {\hat  X}^{\nu}\, {\hat U}^{\mu},\nonumber \\
&&\nonumber\\
 {\hat S}^{\mu\nu}_{\xi} &=& S^{\mu\nu}_{\xi} + {1\over 2}\,
 \epsilon^A_{\rho}(\hat U)\, \eta_{AB}\, \Big({{\partial
 \epsilon^B_{\sigma}(\hat U)}\over {\partial U_{\mu}}}\, U^{\nu}
 - {{\partial \epsilon^B_{\sigma}(\hat U)}\over {\partial
 U_{\nu}}}\, U^{\mu}\Big)\, S^{\rho\sigma}_{\xi} =\nonumber \\
 &{\buildrel {def}\over =}&
 \Big[\epsilon^{\mu}_C(\hat U)\, \epsilon^{\nu}_D(\hat U) +
 {1\over 2}\, \epsilon^A_{\rho}(\hat U)\, \eta_{AB}\, \Big({{\partial
 \epsilon^B_{\sigma}(\hat U)}\over {\partial U_{\mu}}}\, U^{\nu}
 - {{\partial \epsilon^B_{\sigma}(\hat U)}\over {\partial
 U_{\nu}}}\, U^{\mu}\Big)\, \epsilon^{\rho}_C(\hat U)\,
 \epsilon^{\sigma}_D(\hat U)\Big]\, {\bar S}^{CD}_{\xi}=\nonumber \\
 &=&D^{\mu\nu}_{ab}(\hat U)\,\bar{S}^{ab}
 \label{d7}
 \eea

\bigskip

We can now add the gauge fixing $S(\tau ,\vec \sigma ) \approx 0$
of Eqs.(\ref{II22}) and restrict the spinning particles to the
embeddings (\ref{II23}). All the equations from (\ref{II24}) till
(\ref{II32}) remain valid. The new Dirac brackets (\ref{II33}) now
impose the following modification of Eq.(\ref{II35})

\bea
 {\tilde X}^\mu(\tau ) &=& {\hat X}^{\mu}(\tau ) +
 \frac{\epsilon^\mu_a(\hat U(\tau ))}{\sqrt{\sgn\, U^2(\tau)}}\,\int
d^3\sigma\,\left(\theta(\tau)\,  \rho^a_U(\tau,\vec{\sigma})-
{\cal A}^a(\tau,\vec{\sigma})\, \rho_U(\tau,\vec{\sigma}) \right)
+\nonumber \\
 &+& \frac{\partial \epsilon^\alpha_a(\hat U(\tau ))}{\partial
{\hat U}_\mu} \,\epsilon_{b\alpha}(\hat U(\tau )) \int
d^3\sigma\,{\cal A}
^a(\tau,\vec{\sigma})\,\rho^b_U(\tau,\vec{\sigma}).
 \label{d8}
\eea

While  Eqs.(\ref{II36}) remain valid, Eqs.({\ref{II37}) are
modified in the following way

  \begin{eqnarray*}
 J^{\mu\nu} &=& {\tilde L}^{\mu\nu} + {\tilde S}_z^{\mu\nu} +
 {\hat S}^{\mu\nu}_{\xi},\nonumber \\
 &&{}\nonumber \\
 {\tilde S}^{\mu\nu}_z &=& D_{ab}{}^{\mu\nu}(\hat U)\,
 \int d^3\sigma\, [{\cal A}^a\, \rho^b_U - {\cal A}^b\,
 \rho^a_U](\tau ,\vec \sigma ), \nonumber\\
&&\nonumber\\
 {\tilde L}^{\mu\nu}&=&
{\tilde X}^{\mu}(\tau )\, U^{\nu}(\tau ) - {\tilde X}^{\nu}(\tau
)\, U^{\mu}(\tau ), \nonumber \\
&&\nonumber\\
 &&\{ {\tilde L}^{\mu\nu}, {\tilde S}_z^{\alpha\beta}\} \not=
 0,\nonumber \\
 &&{}\nonumber \\
  {\tilde X}^\mu(\tau)
 &=&(\hat{U}^\sigma(\tau)\,{\hat X}_\sigma(\tau))\,\hat{U}^\mu(\tau)+
J^{\mu\rho}(\tau)\hat{U}_\rho(\tau)\frac{1}{\sqrt{\sgn\,
U^2(\tau)}}-
 \frac{\partial
\epsilon^\alpha_a(\hat U(\tau )}{\partial {\hat U}_\nu}
\,\epsilon_{b\alpha}(\hat U(\tau ))\, S_z^{ab}(\tau ),
 \end{eqnarray*}

\bea
  \{ {\tilde S}_z^{\mu\nu}, {\tilde S}_z^{\alpha\beta} \} &=&
 C^{\mu\nu\alpha\beta}_{\rho\sigma}\, {\tilde S}_z^{\rho\sigma} +
 \Big({{\partial D_{ab}{}^{\mu\nu}(\hat U)}\over {\partial {\hat U}_{\beta}}}\,
 U^{\alpha} - {{\partial D_{ab}{}^{\mu\nu}(\hat U)}\over {\partial {\hat U}_{\alpha}}}\,
 U^{\beta} -\nonumber \\
 &&\quad - {{\partial D_{ab}{}^{\alpha\beta}(\hat U)}\over {\partial {\hat U}_{\nu}}}\,
 U^{\mu} + {{\partial D_{ab}{}^{\alpha\beta}(\hat U)}\over {\partial {\hat U}_{\mu}}}\,
 U^{\nu}\Big)\, S^{ab}_z,\nonumber \\
 &&{}\nonumber \\
 &&S^{ab}_z(\tau ) = \int d^3\sigma\, ({\cal A}^a\, \rho^b_U - {\cal A}^b\,
 \rho^a_U)(\tau ,\vec \sigma ).
 \label{d9}
 \eea

Eqs.(\ref{II38}) and (\ref{II39}) remain valid as well as
Eqs.(\ref{II40}), (\ref{II41}) and (\ref{II42}). In
Eq.(\ref{II43}), where $S^{rs}$ must be replaced with $S^{rs}_z$,
we must add $ D^{\mu\nu}_{ab}(\hat U)\, {\bar S}^{ab}_{\xi}$ to
$J^{\mu\nu}$. Finally the effective Hamiltonian (\ref{II44}) and
the gauge fixings (\ref{II45}), (\ref{II46}) and (\ref{II47}) are
not modified.

\subsection{The Multi-Temporal Quantization.}

In absence of interactions on the hyper-planes (\ref{II23}) and
with the Dirac brackets (\ref{d4}) a system of positive energy
spinning particles is described by the same first class
constraints (\ref{II28}) and (\ref{II31}) valid for spinless
particles. As a consequence the multi-temporal quantization
follows the same pattern of Section III.
\medskip

With the quantization rule

\beq
 \xi^a_i\, \mapsto\, {{\hbar}\over {\sqrt{2}}}\, \sigma^a_i,
 \label{d10}
 \eeq

 \noindent where $\sigma^a_i$ are Pauli matrices,
we obtain that the wave functions of Section III are now
two-component spinors $\Psi = \left(
\begin{array}{c} \Psi_{(+)}\\ \Psi_{(-)}\end{array}\right)$
belonging to the $({1\over 2}, 0)$ representation of the Poincare'
group. Therefore Eqs.(\ref{III64}) and (\ref{III70}) must be
called effective Pauli equations.

\medskip

Finally we must add a term ${\hat S}^{\mu\nu}_{\xi} = {1\over 2}\,
D_{ab}{}^{\mu\nu}(\hat U)\, \epsilon^{abc}\, \sum_{i=1}^N\,
\sigma_i^c$ to the angular momentum generator of
Eqs.(\ref{III43}).

\end{document}